\documentclass{aa}  

\usepackage{graphicx}

\usepackage{txfonts}
\DeclareMathOperator*{\E}{\mathbb{E}}

\usepackage{natbib,twoopt}
\usepackage[breaklinks=true]{hyperref} 

\bibpunct{(}{)}{;}{a}{}{,}             
\makeatletter
  \newcommandtwoopt{\citeads}[3][][]{\href{http://adsabs.harvard.edu/abs/#3}%
    {\def\hyper@linkstart##1##2{}%
     \let\hyper@linkend\@empty\citealp[#1][#2]{#3}}}
  \newcommandtwoopt{\citepads}[3][][]{\href{http://adsabs.harvard.edu/abs/#3}%
    {\def\hyper@linkstart##1##2{}%
     \let\hyper@linkend\@empty\citep[#1][#2]{#3}}}
      \newcommandtwoopt{\citeyearads}[3][][]%
    {\href{http://adsabs.harvard.edu/abs/#3}
    {\def\hyper@linkstart##1##2{}%
    \let\hyper@linkend\@empty\citeyear[#1][#2]{#3}}}
\makeatother

\begin{document} 

\title{Description of turbulent dynamics in the interstellar medium: multifractal/microcanonical analysis}
\subtitle{I. Application to {\sl Herschel} observations of the Musca filament}
\author{ H. Yahia\inst{1}
  \and N. Schneider\inst{2}
  \and S. Bontemps\inst{3}
  \and L. Bonne\inst{3} \and G. Attuel\inst{1}  
\and S. Dib\inst{3}  \and V. Ossenkopf\inst{2} 
  \and  A. Turiel\inst{4}
  \and A. Zebadua\inst{1}  \and D. Elia\inst{5} \and S. K. Maji\inst{6}
 \and F. G. Schmitt\inst{7} \and J.-F. Robitaille\inst{8}}

\institute{INRIA, Geostat team, France  
\email{hussein.yahia@inria.fr} 
\and I. Physik. Institut, University of Cologne, Z\"ulpicher Str. 77, 50937 Cologne, Germany
\and CNRS LAB, UMR 5804, Bordeaux University
\and ICM CSIC, Barcelona,  Spain
\and INAF, Roma, Italy
\and Indian Institute of Technology, Patna, India  
\and CNRS, Univ. Lille, Univ. Littoral Cote d'Opale, UMR 8187, LOG, 62930 Wimereux, France
\and Univ. Grenoble Alpes, CNRS, IPAG, 38000 Grenoble, France }

\date{Draft of \today}

\abstract{Observations of the interstellar medium (ISM) show a complex density and velocity structure which is in part attributed to turbulence. Consequently, the multifractal formalism is required to be applied to observation maps of the ISM in order to characterize its turbulent and multiplicative cascade properties. However, the multifractal formalism, even in its more advanced and recent {\it canonical} versions, requires a large number of realizations of the system which usually cannot be obtained in astronomy. We here present a self-contained introduction to the multifractal formalism in a {\it microcanonical} version which allows us for the first time to compute precise turbulence characteristic parameters from a single observational map without the need for averages in a grand ensemble of statistical observables (temporal sequence of images for instance). We compute the singularity exponents and the singularity spectrum 
for both observations and magnetohydrodynamic simulations, which include key parameters to describe turbulence in the ISM. For the observations we focus on studying the 250 $\mu$m {\sl Herschel} map of the Musca filament. 
Scaling properties are investigated using spatial 2D structure functions, and we apply a two-point $\log$-correlation magnitude analysis over various lines of the spatial observation which is known to be directly related to the existence of a multiplicative cascade under precise conditions. It reveals a clear signature of a multiplicative cascade in Musca with an inertial range from 0.05 to 0.65$\,$pc. We show that the proposed microcanonical  approach provides singularity spectra which are truly scale invariant as required to validate any method to analyze multifractality. The obtained, for the first time precise enough, singularity spectrum of Musca is clearly not as symmetric as usually observed in $\log$-normal behavior. We claim that the ISM towards Musca features more a $\log$-Poisson shape of its singularity spectrum. Since $\log$-Poisson behavior is claimed to exist when dissipation is stronger for rare events in turbulent flows in contrast to more homogeneous (in volume and time) dissipation events, we suggest that this deviation from $\log$-normality could trace enhanced dissipation in rare events at small scales, which may explain or is at least consistent with the dominant filamentary structure in Musca. Moreover we find that sub-regions in Musca tends to show different multifractal properties: while a few regions can be described by a $\log$-normal model other regions have singularity spectra better fitted by a $\log$-Poisson model.  It strongly suggests that different types of dynamics exist inside the Musca cloud. We stress that this deviation from $\log$-normality and these differences between sub-regions appear only after eliminating noise features - using a sparse edge-aware algorithm - which have the tendency to "$\log$-normalize" an observational map. Implications on the star formation process are discussed.  Our study sets up fundamental tools which will be applied to other galactic clouds and simulations in forthcoming studies. }
\maketitle

\section{Introduction}
Stars form in the cold ($\sim$10~K) and dense ($>10^4\,$cm$^{-3}$) gas phase of the interstellar medium (ISM) which represents only a fraction of the order of a few \% in mass \citep[e.g.][]{Tielens2005,McKeeOstriker2007,BerginTafalla2007,Lada2010,Andre2014,Shimajiri2017}. This rare occurrence of {\it fertile} gas in the ISM explains the low star formation rates of galaxies and gets its origin from the global physics of the ISM from galactic down to star formation scales, including heating, cooling, magnetic fields,  cosmic rays and gravity. It is only by understanding the origin of this rare dense gas from the bulk of the predominantly diffuse and warm ISM that initial conditions for star formation will be fully constrained and understood. 

In the 80's and 90's the star-forming ISM was generally described as driven by a quasi-static evolution of clouds/clumps which could condense into dense collapsing cores \citep[e.g.][]{Mouschovias1976,Shu1987}. The importance of supersonic motions and dynamical processes have since then been recognized and better understood \citep[e.g.][]{Elmegreen2000,Koyama2000,MacLowKlessen2004} and a gravo-turbulent paradigm has emerged to explain how islands of quiet dense gas can emerge from a sea of turbulent low density gas. More recently it has been realized, notably thanks the {\sl Herschel} Space Telescope \citep{Pilbratt2010}, that the dense star forming gas is located in filamentary structures \citep[e.g.][]{Myers2009,Andre2010,Schneider2010,Bontemps2010,Molinari2010,Schneider2012,Schisano2014,Koenyves2015,Marsh2016,Rayner2017,Schisano2014,Motte2018} which complicates the usually implied simple view of a mostly isotropic and thermal pressure like turbulence. In parallel to these observational results, the formation of coherent structures such as filaments and hubs (where filaments merge) has also been discerned in numerical simulations by many groups. These simulations take into account the role of one or several physical processes such as gravity, turbulence, magnetic fields, radiation and thermodynamics; they also play a fundamental role in the evaluation of the statistics of intensive variables displaying a multifractal behaviour \citep{MacLow2000,Heitsch2005,Chappell2001,Klessen2010,Dib_2005,Krumholz2005,KRUMHOLZ201449,Hull2017,Elia2018}. 

The turbulent nature of the ISM is well established by the extremely high values of Reynolds  number  \citep{Elmegreen2004,Kowal2007,Burkhart2009a,Burkhart2009b,Schneider2011,Seifried2015,Kritsuk2017,Mocz2017,Elia2018,Lee2019}. 
The need to progress towards a more precise way to measure the properties of the turbulent motions led to fractal approaches. Self-similarity of the ISM was observed and described first with monofractal descriptors by \citep[e.g.][]{Stutzki1998,Falgarone1998,Bensch2001,Sanchez_2006,Elia_2014} but it was
realized that the observations can not be characterized as pure
monofractals, which is in accordance with most advanced phenomenological descriptions of turbulence \citep{frisch1995}. Characteristic scales, mostly on the sub-parsec scale up to
a few parsec, were found \citep{Padoan2003,Hartmann2002,Sun2006,Brunt2010,Schneider2011,Elia2018,Dib_2020},
breaking the premise of pure self-similarity. Moreover, the transition from incoherent to coherent structures is usually assumed to be related to the formation of dense structures and to the dissipation of turbulence. 
If coherent structures display a significant scale invariance, they are of multifractal nature, expressed and observed in power law statistics for spatial and time correlation functions. This leads to
the existence of critical manifolds as predicted in dynamical systems, especially those of  turbulence
\citep{Arneodo1995,Venugopal2006b,Khalil2006,Turiel2008,Robitaille2019}.
In order to understand the turbulent mechanisms present in the ISM and to be able to discriminate between the effects of magnetic pressure  and gravity on structure evolution, it is required
to make use of most recent advances in the analysis of multiscale and multifractal signals.
 
Computational tools have been developed in order to obtain the most important fingerprints of multifractality from the data and to establish links with multiplicative processes and energy cascades in the case of turbulence  \citep{Chhabra1989,Meneveau1991,Bacry1993,Muzy1993,Arneodo1995,Hosokwa1997,Delour2001,Venugopal2006a,Venugopal2006b,Turiel2006,Turiel2008,Muzy2016,Leonarduzzi2016,Salat2017,Muzy2019}. In
physical systems, multifractality is closely related to the existence of singular mesures defined from the observables; for instance in a turbulent 3-dimensional medium, if we denote by $\dot{\varepsilon}({\bf x})$ the rate of kinetic energy dissipation per unit volume, the total dissipation of energy $E$ is a measure of density $\dot{\varepsilon}({\bf x})\mbox{d}^3{\bf x}$ which has a
singular behaviour around each point ${\bf x}$ with a particular singularity exponent ${\bf h}({\bf x})$;  this means that $E({\cal B}({\bf x},{\bf r}))$, the energy dissipation in a ball ${\cal B}({\bf x},{\bf r})$ centered at point ${\bf x}$ and of radius ${\bf r}$, behaves  like ${\bf r}^{{\bf h}({\bf x})} + o({\bf r}^{{\bf h}({\bf x})} )$ when ${\bf r} \rightarrow 0$.  The spatial intermittence of the support of energy transfer is directly related to
the existence of critical manifolds defined by the geometrical distribution of the singularity exponents ${\bf h}({\bf x})$, as it implies a complex partition of the energy at different scales, and the
power law behaviour observed in the statistics of physical variables \citep{frisch1995}.

To apply multifractal analysis to acquired signals, a canonical formalism is generally used \citep{Arneodo1995,Venugopal2006a,Venugopal2006b,Turiel2006,Turiel2008}. 
This formalism is based on statistical averages (moments of different orders, correlation and structure functions) computed on grand ensembles of realisations, from which the singularity spectrum and singular values are obtained as mean values, through log-regression, of quantities usually defined out of partition functions.
In astronomy, this procedure is usually only applicable for MHD simulations since multi-epoch data are usually rare or not possible due to long timescales of evolution. The canonical approach is ruled out to characterize turbulent behaviours from single realisation observations, i.e. for typical astronomical images.

In this study, we go beyond the limitations of the canonical approach by using a microcanonical formulation of the multifractal signal analysis, in which individual microstates are evaluated in a single realisation \citep{Turiel2006,Turiel2008}. As a direct application of the method, we examine the turbulence properties of gas associated with the Musca filament observed with {\sl Herschel}. This source (see Sec.~\ref{musca} for details) is a prototypical example of a rather isolated filamentary structure in the Chamaeleon-Musca cloud complex \citep{Cox2016}, not affected by stellar feedback. Thanks to its sensitive access from space to the bulk of dust emission of nearby cloud complexes, the {\sl Herschel} mission provides us with excellent datasets with unprecedented spatial  and flux dynamical ranges to provide the required statistical significance to study the properties of the turbulent flows presumably leading to dense gas and star formation \citep[e.g.][]{Andre2010,Molinari2010,Elia2018}. The spectral coverage of the Herschel photometric surveys is 70 - 500 $\mu$m at an angular resolution of 6$''$ to 36$''$ covering several square degrees per regions.

Section~\ref{musca} describes the {\sl Herschel} data used in this study. The multifractal formalism, its relations with turbulence, intermittency and the multiplicative cascade is presented in
Sect.~\ref{multifractal}. We also explain in Sect.~\ref{multifractal} the canonical and microcanonical approaches to multifractality. The scale invariance of an observation map is studied with some detail in Sect.~\ref{h-loglog}. In section~\ref{musca-appli} we introduce a sparse filtering methodology designed to reveal the filaments hidden by background noise, and the 2D structure function approach. In Sect.~\ref{simu} we present the magneto-hydrodynamic simulations used in this work. Sect.~\ref{results} presents the results obtained in applying the multifractal analysis to our available data: the  {\sl Herschel} observation map of Musca and the MHD simulations. These results are discussed in Sect.~\ref{discussion}, followed by the conclusions  of this study in Sect.~\ref{conclusion}.
\section{The Musca cloud: {\sl  Herschel} flux density map}
\label{musca}
\begin{figure*}[h]
  \centering
  \includegraphics[width=1.0\textwidth]{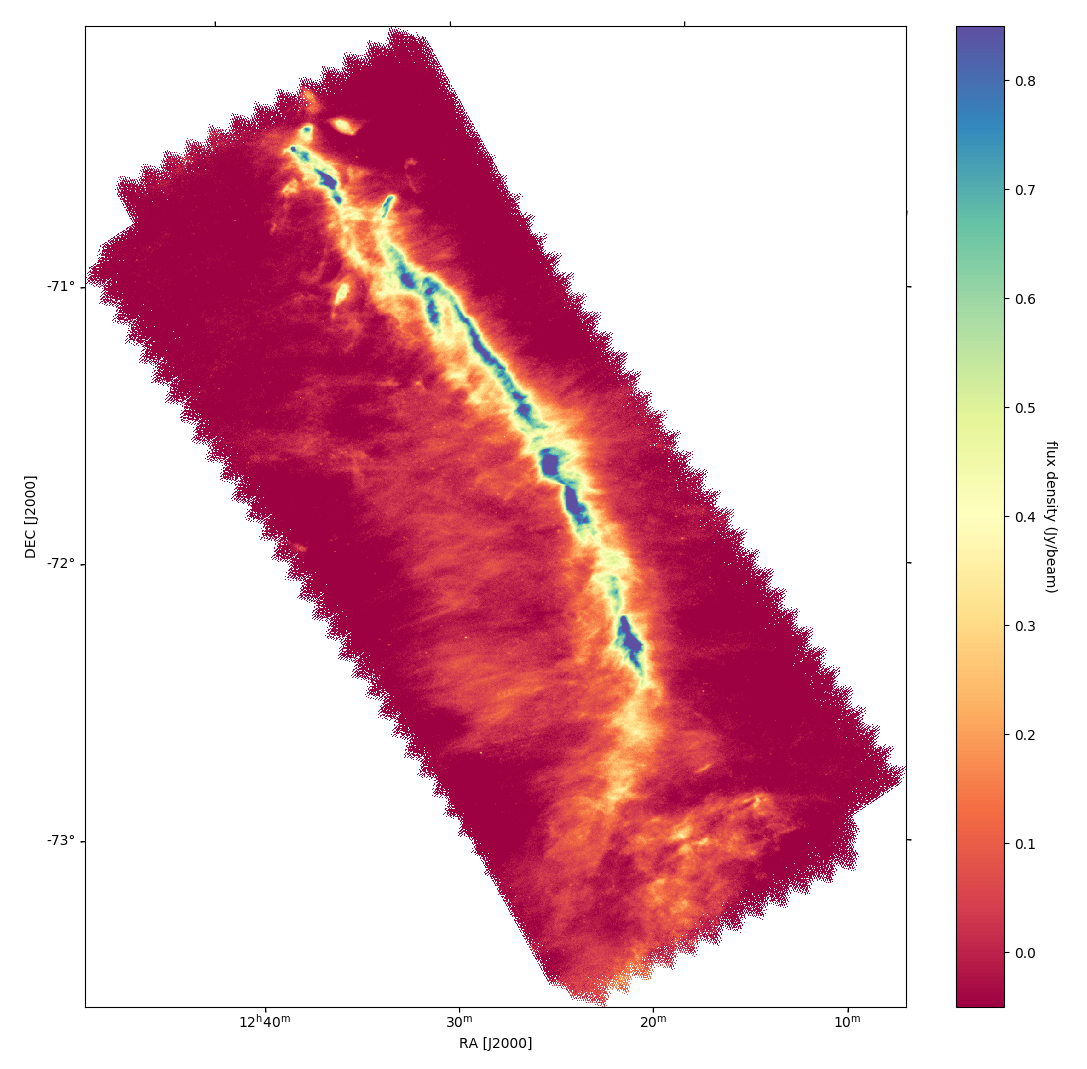} 

  \caption{Musca flux density map from {\sl Herschel} at 250 $\mu$m. The filamentary structure with
    a high aspect ratio is obvious. Perpendicular to the main ridge of emission are fainter
    hair-like structures (called striations in \citep{Cox2016}) that are mostly attached to the
    main filament. \label{muscaPSW}}
\end{figure*}
Musca is a prominent 6 pc long filamentary structure with a high
aspect ratio at a distance of only 140-150 pc
\citep{Hacar2016,Cox2016,2016A&A...586A..27K,GaiaDR2}.  It has a low
average column density $N$, with $N$(max)$\sim$4-8 10$^{21}$ cm$^{-2}$
\citep{Bonne2020a}, and shows only one protostellar source located at
the northern end of the filament \citep{Juvela2012,Machaieie2017}.
Furthermore, {\sl Herschel} observations reveal a network of
striations orthogonal to the filament, which are suggested to be
indications of mass inflow along the magnetic field
\citep{Cox2016}. Indication of continuous mass accretion from inflow
towards the Musca filament was found by observed presence of
low-velocity filament accretion shocks around the Musca filament
\citep{Bonne2020a}. \cite{Hacar2016} conclude from individual
$^{13}$CO and C$^{18}$O pointings that the crest of the Musca filament
is a single velocity coherent structure. It was
proposed~\citep{Tritsis635} that the Musca filament is only a sheet
viewed edge-on, but our detailed study using APEX and SOFIA
spectroscopic observations \citep{Bonne2020b} showed that the data
best fit with the concept that the crest of Musca is a dense
($n_{H_{2}}\sim$10$^4$ cm$^{-3}$), cylindrical structure and not a
low-density sheet and that the mass input for building up the crest is
provided by large-scale flows but there is no evidence that this
inflow appears in the form of striations. The striations are then only
the result of gas compression by magnetosonic waves, which could fit
with what was recently suggested by theoretical studies
\citep{2016MNRAS.462.3602T}.

The {\sl Herschel} SPIRE 250~$\mu$m data of Musca
(Fig.~\ref{muscaPSW}) used in this paper are part of the Herschel
Gould Belt Survey \citep{Andre2010} and are published in
\citep{Cox2016}. We use here the same dataset that was obtained in
Parallel Mode at a sampling at 10Hz and at high speed (60$''$/s), and
reduced using modified pipeline scripts of HIPE version 10. The
resulting Level1 contexts for each scan direction (two maps were taken
with orthogonal directions) were then combined using the 'naive' map
maker in the destriper module. The conversion of the maps into surface
brightness (from Jy/beam into MJy/sr) was done using the beam-areas
obtained from measurements of Neptune (March 2013). The
zeroPointCorrection task was used to absolutly calibrate the maps
using information from the Planck satellite. The angular resolution of
the maps at 250~$\mu$m is 18$''$. We here use the 250 $\mu$m flux map
for our studies and not the dust column density map used in
\citep{Cox2016,Bonne2020a,Bonne2020b} because the latter has only an
angular resolution of 36$''$. The 250 $\mu$m map can be considered as
a good proxy for the column density, however, in regions that are not
strongly affected by stellar feedback \citep{Miville2010} because
it traces mostly cold dust that is mixed with the cold molecular
gas. 

\section{Canonical and microcanonical approach to multifractality}
\label{multifractal}

A number of Galactic studies using continuum data or emission lines of atomic hydrogen or molecules showed that the Fourier power spectrum of the observed line intensity was well fitted by a power-law - at least over certain size scales - and which was commonly interpreted as a consequence of the scale-free and turbulent nature of the ISM~\citep{Scalo1987,Green1993,Elmegreen2004,Stutzki1998,Miville2010,Schneider2011,Robitaille2020}. As mentioned since the introduction of the first phenomenological descriptions of hydrodynamic turbulence  \citep{frisch1995}, the observed scale invariance suggests the existence of a energy cascade in which energy injected at large scales is transferred into smaller ones hence providing a natural explanation for the complex structure of the ISM at each scale.  The Fourier power spectrum with its single descriptive parameter (the slope of the spectrum), however, turned out to be a too coarse descriptor to encode all turbulent and scale-free phenomena observed in the ISM and notably filamentary structures~\citep{Roy2015,Roy2019,Arzoumanian2019}. The simplest models able to describe the partition of energy across the scales of a turbulent medium are monofractal: they rely on a single parameter, the fractal dimension, which characterizes the geometry of the sets between which the energy transfer is operated. 

For modeling purposes in astronomy, monofractal signals can be generated using stochastic processes such as the fractional Brownian motion (fBm)~\citep{Mandelbrot1968} and, more specifically,
2-dimensional fBMs~\citep{Heneghan1996,Stutzki1998,Robitaille2020}. Such fBms are parametrized by their Hurst exponent which defines their monofractal
properties. In \citep{Khalil2006} a monofractal signature is found in atomic hydrogen data from the Canadian Galactic Plane Survey, while an anisotropic signature is also detected. The complex organization of the ISM can be fully described by the multifractal formalism which  is as necessary in magneto-hydrodynamic turbulence as it is needed in hydrodynamic turbulence~\citep{frisch1995,Robitaille2020}.

The multifractal formalism is already used in astronomy, e.g., in
heliophysics turbulence
\citep{Movahed2006,McAteer_2007,Salem_2009,Kestener_2010,Macek_2014,WU20151,Cadavid2016,MARUYAMA20171363},
the ISM \citep{Elia2018}, the large scale structure of the
universe \citep{Gaite_2007a,Gaite_2007b}, galaxy mergers
\citep{DelaFuente2006}, and gravitational waves detection
\citep{Eghdami_2018}. This formalism comes into two distinct
presentations: canonical and microcanonical. The first one is the most
popular \citep{Arneodo1995};  it includes the Wavelet Transform Modulus Maxima method (WTMM) \citep{Arneodo1995,Turiel2006} and the cumulant approach 
\citep{Brillinger1994,Delour2001, wendt:ensl-00144568,4740319,Venugopal2006a}. Microcanonical
approaches  where developed first in poorly accurate versions such as the counting box and histogram methods, then in an efficient geometrico-statistical formulation \citep{Turiel2008} 
which we will use in this work. In the following, we review the most
important aspects of the multifractal approach without being
exhaustive.

In hydrodynamics or magneto-hydrodynamic turbulence, one observes, as explained before, a
very complex partition of the energy according to the scale
\citep{She1990226,frisch1995,SHIVAMOGGI20151887} and it was understood
that this complexity was an effect of intermittency, i.e., a
consequence of the extremely complex geometrical organization, or
partition, of the support of the energy at small
scales. Observationally, this was supported by the detection of strong
velocity-shears at sub-parsec scale
\citep{Falgarone2009}. Multifractal models were introduced to describe
this intermittency because they allow to relate the fluctuations of a
physical variable (such as the projection of velocity on a given fixed
axis) to the complex geometry of the points where such a physical
variable changes abruptly.
First, we consider a fluid evolving in a two or three dimensional medium, and denote by ${\bf v}({\bf x})$ the velocity vector at  a given point ${\bf x}$ inside the medium. One can then measure the values taken by ${\bf v}({\bf x})$ either in time at a fixed point
${\bf x}$ or at a fixed time at many different points ${\bf x}$. Experimentally, some decades ago, it was much more easier to measure the projection of the turbulent velocity field ${\bf v}$ in a specific fixed direction defined by a unitary vector ${\bf u}$ instead
of the vector ${\bf v}({\bf x})$ itself. Nowadays, laser-mounted optical devices allow a direct measurement of the velocity vector field. Nevertheless, following the experimental setup described in \citet{frisch1995}, measuring the velocity field in a fixed direction
allows to evaluate the statistical moments
\begin{equation}
 \label{expturbulence}
 \langle \left | \left( {\bf v}({\bf x} + {\bf r}{\bf u}) - {\bf v}({\bf x} ) \right) \cdot \displaystyle {\bf u} \right |^q  \rangle
 \end{equation} 
of the turbulent velocity field ${\bf v}$; depending on the
experimental setup, in equation~\ref{expturbulence} the averages
$\langle ~\cdot ~\rangle$ can be computed either in time at a given
point ${\bf x}$ or spatially over the domain. Then it was observed
that the moments in equation~\ref{expturbulence} behaves as ${\bf
  r}^{\xi(q)}$ when ${\bf r}$ is inside a well defined interval of
scale values $[ {\bf r} _1, {\bf r} _2]$ called the inertial range:
\begin{equation}
 \label{scaling-turbulence}
  \langle \left | \left( {\bf v}({\bf x} + {\bf r}{\bf u}) - {\bf
    v}({\bf x} ) \right) \cdot \displaystyle {\bf u} \right |^q
  \rangle \sim {\bf r}^{\xi(q)}.
\end{equation}

When $\xi(q)$ is a linear function of $q$ we are in the monofractal
formalism. On the other hand, $\xi(q)$ is observed experimentally to
be nonlinear in $q$ which means a multifractal behaviour.
The nonlinearity of ${\xi(q)}$ as a function of $q$ is thus
interpreted as a consequence of intermittency, it has a geometric
origin and the link between statistics and geometry is as
follows. Let us denote by ${\cal F}_{\bf h}$ the set of points
${\bf x}$ whose velocity increments, measured in the unitary direction
${\bf u}$ as before, behave as ${\bf r}^{\bf h}$ for a certain ${\bf
  h}$ i.e.
\begin{equation}
\label{definitionfh}
{\cal F}_{\bf h} = \{ {\bf x} ~\mbox{s.t.} ~\left | \left( {\bf v}({\bf x} + {\bf r}{\bf u}) - {\bf v}({\bf x} ) \right) \cdot \displaystyle {\bf u} \right | \sim {\bf r}^{\bf h} ~~\mbox{when} ~~{\bf r} \rightarrow 
0\}. 
\end{equation}
These sets ${\cal F}_{\bf h}$ for small or even negative values of
${\bf h}$ feature a complex geometrical organization which is closely
related to the nonlinear behavior of $\xi(q)$. To quantify how the
geometry of the sets ${\cal F}_{\bf h}$ is related to the turbulent
behavior of the velocity field, one can introduce a notion of
dimension for these sets, called the Hausdorff fractal
dimension \citep{falconer1997}, which is a positive real number
describing the way they fill in the space. We will come back to this
point in Sect.~\ref{ss} but let us denote for the moment $D({\bf h})$ the
Hausdorff fractal dimension of the sets ${\cal F}_{\bf h}$:
\begin{equation}
\label{Hausdorff}
D({\bf h}) = \dim_H ({\cal F}_{\bf h}).
\end{equation}
The link between geometrical complexity and statistical behavior is then given by the relation:
 \begin{equation}
 \label{turbulencelegendretransform}
 \xi(q) = \underset{{\bf h} > 0}{\operatorname{inf}} (q {\bf h} + d - D({\bf h})),
 \end{equation}
 where $d$ is the dimension of embedding space of the experiment ($d=2$ in 2D turbulence, $d=3$ in 3D turbulence),
 i.e. $\xi(q)$ appears as a Legendre transform. The mapping 
\begin{equation}
\label{defsingularityspectrum}
{\bf h} \mapsto D({\bf h})
\end{equation}
is called the singularity spectrum of the velocity field. In the
monofractal formalism, ${\xi(q)}$ is a linear function of $q$ and the
mapping~(\ref{defsingularityspectrum}) reduces to a single point in
the graph of $D({\bf h})$. In the multifractal
formalism eq.~(\ref{defsingularityspectrum}) defines a distribution of
fractal dimensions as a function of ${\bf h}$. In general $\displaystyle {\bf h} = \frac{\mbox{d}\xi}{\mbox{d}q}(q)$.

The notion applies of course to any signal $s({\bf x})$ or random
process $(X_t)_{t \geq 0}$ with values in $\mathbb{R}^d$ displaying a
multiscale behaviour i.e. a nonlinear scaling of moments $  \langle \| X(t) \|^q \rangle = c(q)t^{\xi (q)} $ for a nonlinear mapping $q \mapsto \xi(q)$, or to even more general multifractal 
random processes \citep{GRAHOVAC2020109735}. As
we will see in Sect.~\ref{ss}, it also applies to measures. The
singularity spectrum thus contains a lot of information on the
statistics of turbulence but its computation is a fundamental problem
of the multifractal formalism.

The intermittency of the velocity field, which is responsible of the
observed scaling behavior and the partition of energy at different
scales, has been the subject of a description in terms of random
multiplicative cascades proposed by the Russian school 
\citep{Kolmogorov1941a,Kolmogorov1962,Yaglom1966,Novikov1990,Novikov1994}. Denoting $(\delta v)_{ {\bf r}}$ the
longitudinal increments $\left | \left( {\bf v}({\bf x} + {\bf r}{\bf
  u}) - {\bf v}({\bf x} ) \right) \cdot \displaystyle {\bf u} \right
|$, it was proposed to model the cascade through the probability
distribution for $(\delta v)_{ {\bf r}}$
\citep{refId0,PhysRevLett.80.708}
\begin{equation}
 \label{scalelaw}
{\cal P}_{(\delta v)_{ {\bf r}_1}}= \displaystyle  \int_{\mathbb{R}} G_{{\bf r}_1 {\bf r}_2} (u) e^{-u} {\cal P}_{e^{-u}(\delta v)_{ {\bf r}_2}}\, \mbox{d} u
\end{equation} 
which expresses the change in scales of the probability distribution
depending on a kernel measure $G_{{\bf r}_1 {\bf r}_2} $ (${\bf r}_1 <
{\bf r}_2$). For instance if $G_{{\bf r}_1 {\bf r}_2}$ were a Dirac
measure, then eq.~\ref{scalelaw} implies that ${\cal P}_{(\delta v)_{
    {\bf r}_1}} $ would have the same shape as ${\cal P}_{(\delta v)_{
    {\bf r}_2}} $ within a scaling factor in the velocity
amplitudes. Equation~\ref{scalelaw} is satisfied if the kernel $G$
satisfies the convolution law (for each monotonic sequence
of scales ${\bf r}_1 < {\bf r}_2 < \cdots < {\bf r}_n$)
\begin{equation}
\label{decomplaw}
G_{{\bf r}_1 {\bf r}_n} = G_{{\bf r}_1 {\bf r}_2} * \cdots * G_{{\bf
    r}_{n-1} {\bf r}_n}
\end{equation}
which implies that the laws ${\cal P}_{(\delta v)_{ {\bf r}}}$ are
indefinitely divisible. In turn, this can be physically interpreted in
the form of the multiplicative cascade.
 
\subsection{$\log$-normal and $\log$-Poisson processes}
 \label{subsec:lognormallogpoisson}
 
\begin{figure*}[h]
  \centering
  \includegraphics[width=0.395\textwidth]{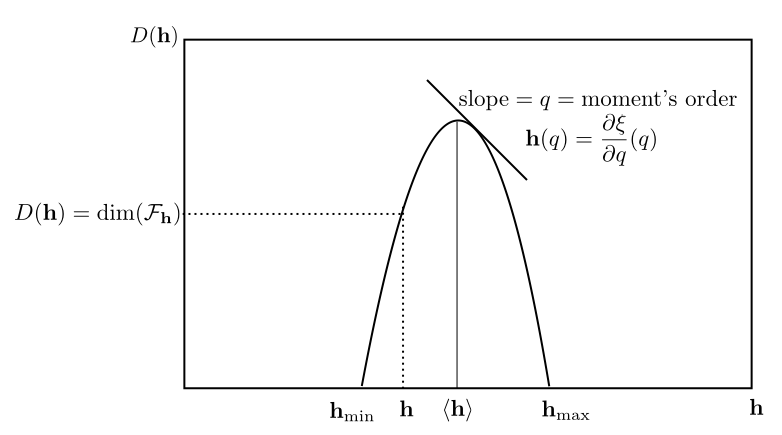}  \includegraphics[width=0.295\textwidth]{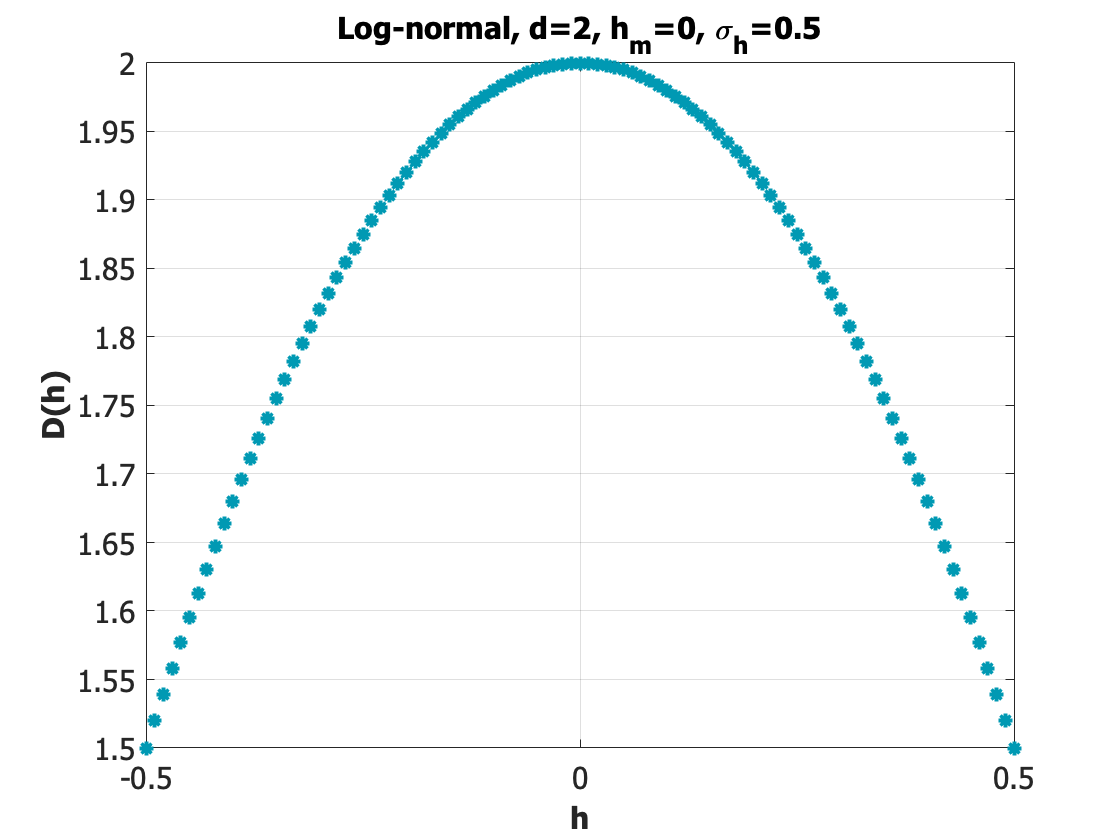}  \includegraphics[width=0.295\textwidth]
{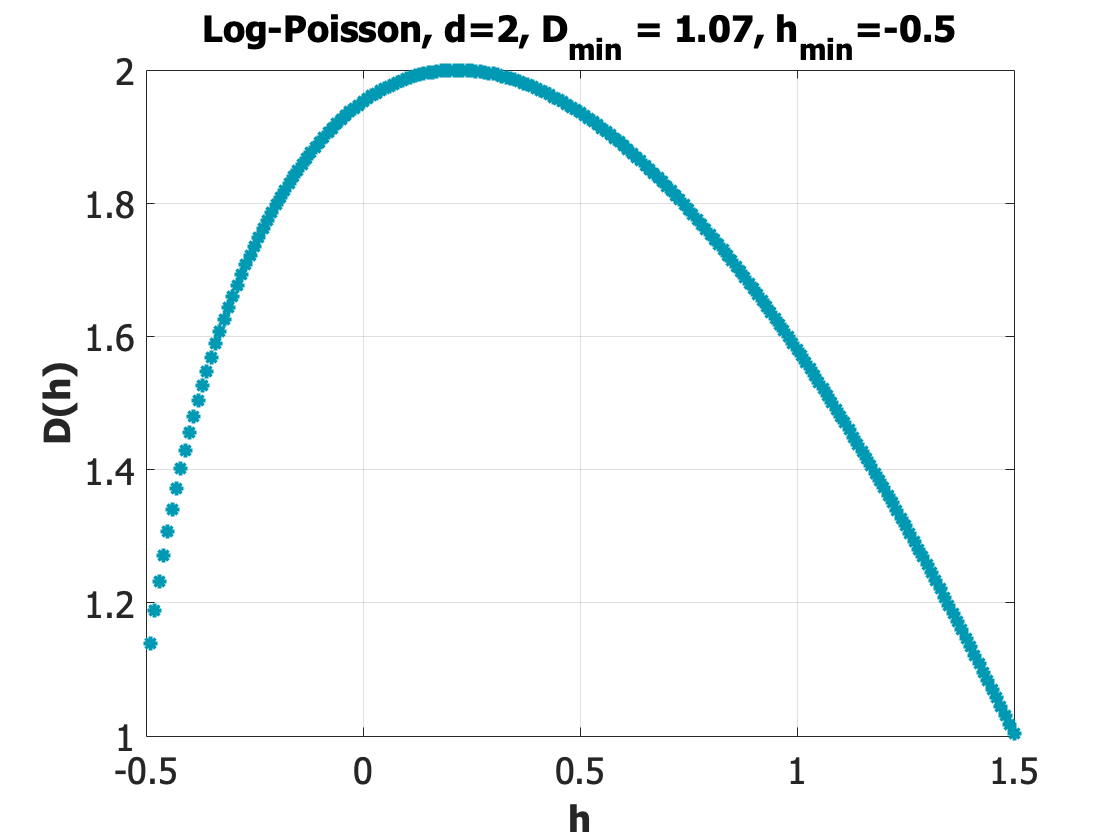}
  \caption{Left: general form of the singularity spectrum $D({\bf
      h})$, Legendre transform of $\xi(q)$. See text for a definition
    of $D({\bf h})$, $\xi(q)$ and ${\cal F}_{\bf h}$. Middle and
    right: singularity spectrum of a $\log$-normal process with $d=2$,
    ${\bf h}_m=0$ and $\sigma_{\bf h} =0.5$ (middle panel) and of a
    $\log$-Poisson process with $d=2$, $D_{\min} = 1.07$ and ${\bf
      h}_{\min} = -0.5$ (right panel). \label{genericss}}
\end{figure*}  
There are only few cases for which a singularity spectrum can be
computed exactly. We give here two examples, well-known for their
physical interpretation, that will also play a key role in this study:
the $\log$-normal and the $\log$-Poisson processes.
 
The energy cascading model for intermittency in a turbulent medium
mentioned previously implies a multiplicative form for the energy
dissipation rate $\dot{\varepsilon}({\bf x})$~\citep{VazquezSemadeni1994}. Indeed, an initial volume of space
is divided into 8 equal cubes of size half the size of the initial
volume; the energy dissipation rate is then multiplied by 8 random
variables, equally distributed. When this scheme is recursively
repeated, the field $\dot{\varepsilon}({\bf x})$ at a given scale does
have a multiplicative form. Its logarithm is then a sum of random
variables and, under the hypotheses of the central limit theorem, the
energy dissipation rate can be approximated by a $\log$-normal
process. In that case, the space intermittency comes from the
existence of a large number of regions with nearly equal dissipation
rates. The $\log$-normal model was introduced by N. Kolmogorov \citep{Kolmogorov1962}.

There are, however, different intermittency models which depend on
various assumptions about the energy dissipation
rate~\citep{Hopkins2013}. In~\citet{Gledzer1996} an energy cascade model of intermittency
is introduced, involving rare localized regions of both large and weak
energy dissipation areas, in the spirit of~\citep{PhysRevLett.72.336},
leading to $\log$-Poisson statistics. It is a model with dissipatively
active and passive localized regions, allowing the existence of "holes
of dissipation"; $\log$-Poisson processes are particularly interesting
in this context for their potential ability to describe distribution
of filaments, as we will see in this work.

The singularity spectrum ${\bf h} \mapsto D({\bf h})$ can be computed in a closed form in the case of $\log$-normal and $\log$-Poisson processes. Precisely:
\begin{itemize}
  \item A $\log$-normal process in $\mathbb{R}^d$ has a parabolic singularity spectrum 
  \begin{equation} 
  \label{lognormalspectrum}
  D({\bf h}) = \displaystyle  d - \frac{1}{2} \left ( \frac{{\bf h} - {\bf h}_m}{\sigma_{{\bf h}}} \right )^2
  \end{equation} 
  with $\xi(q) = {\bf h}_m q - \displaystyle \frac{1}{2} \sigma_{ {\bf
      h}}q^2$, $ {\bf h}_m$: mean singularity, $\sigma_{
    {\bf h}}$: singularity dispersion.
  \item A $\log$-Poisson process in $\mathbb{R}^d$ is  a translationally invariant indefinitely divisible process with a singularity spectrum given by
  \begin{equation}
  \label{logpoissonspectrum}
  D({\bf h}) = D_{\min}  + (d -D_{\min}) \omega({\bf h})(1 - \log \omega({\bf h}))
   \end{equation}
   with $\omega({\bf h}) = \displaystyle \frac{{\bf h} - {\bf
       h}_{\min}}{(d - D_{\min})(-\log \beta)}$, ${\bf h}_{\min}$:  
   minimum value of singularities, $D_{\min}$: associated dimension of
   set ${\cal F}_{{\bf h}_{\min}}$ and $\beta$ is the dissipation parameter
   ($0 < \beta < 1$): $\beta = \displaystyle 1 + \frac{{\bf
       h}_{\min}}{d - D_{\min}}$. Accordingly, $\xi(q) = {\bf
     h}_{\min}q + (d - D_{\min})(1 - \beta^q)$.
\end{itemize}

In general, for a wide class of processes, the singularity spectrum
$D({\bf h})$, being the Legendre transform of $\xi(q)$, has a concave
shape, shown the left panel of Fig.~\ref{genericss}. The same
figure displays the singularity spectra of typical $\log$-normal
processes (middle panel), and $\log$-Poisson processes (right panel).  

A $\log$-normal singularity spectrum, being parabolic, is symmetric, while a $\log$-Poisson process has a non-symmetric singularity spectrum. The lack of symmetry is a strong 
indication of very different dynamic properties compared to those described by a $\log$-normal model. In this work we focus on the $\log$-normal and $\log$-Poisson processes because they are associated to geometrical models. It must be 
noted however, that there are other models among them the $\log$-$\alpha$-stable which also has also a non-symmetric spectrum \citep{schmitt-huang-2016}.
\subsection{Existing computational approaches}
\label{existingappraoches}

In this subsection we summarize previous works done in astronomy which
made use of the multifractal approach. We recapitulate previous
introductions given in \citet{Elia2018} and
\citet{Khalil2006,McAteer_2007,Salem_2009,Kestener_2010,Macek_2014,Robitaille2020},
and combine those with more fundamental presentations given in
\citet{Arneodo1995,Turiel2008} that describe general signal processing
approaches.

Equation~(\ref{definitionfh}) defines the scaling exponents $\bf{h}$ from the true turbulent velocity field ${\bf v}({\bf x})$, which is not directly accessible either in astronomy, or in many other geophysical sciences. 
The basic information from which numerical computations are performed
is one or more observational maps, which constitute an original signal
denoted $s$ in the following. Consequently, in signal processing, the multifractal formalism consists in defining and computing singularity exponents {\bf h}({\bf x})  and other characteristics like the singularity spectrum from the data in $s$.  This is done by considering, from the data in $s$, general positive measures, denoted $\mu$ instead of the unknown velocity field ${\bf v}({\bf x})$. These can be for instance 
probability measures built out of the signal, or some kind of additive
quantity defined in the domain of the signal $s$. 

If $\mu$ is a positive measure, and if ${\cal B}({\bf x},{\bf r})$ is a ball centered at
point ${\bf x}$ and of radius ${\bf r}$ in the signal's domain, we can
evaluate the measure of that ball, $ \mu({\cal B}({\bf x},{\bf r}))$
and, since the measure is supposed to be positive, one can consider
the limit
\begin{equation}
\label{defse}
{\bf h}({\bf x}) = \displaystyle \lim_{{\bf r} \rightarrow 0}
\displaystyle \frac{\log \mu({\cal B}({\bf x},{\bf r}))}{\log ({\bf r}
  )}
\end{equation}
called the singularity exponent ${\bf h}({\bf x})$ of $\mu$ at ${\bf
  x}$ \citep{Arneodo1995,Venugopal2006b,Turiel2008}. The singularity
exponent ${\bf h}({\bf x})$ appears as a power law exponent in the
limiting measures of the balls:
\begin{equation}
\label{singularitybehavior}
 \mu({\cal B}({\bf x},{\bf r})) \sim {\bf r} ^{{\bf h}({\bf x}) }
\end{equation}
when ${\bf r} \rightarrow 0$. A singularity exponent ${\bf h}({\bf x})$ encodes a limiting scaling information at every point ${\bf x}$. It is a pointwise generalization of other scaling 
parameters often used by astrophysicists and defined globally such as the power law of a power spectrum or the $\Delta$-variance \citep{Ossenkopf2008}. If the singularity exponents ${\bf
  h}({\bf x})$ can be computed at every point ${\bf x}$ in the
signal's domain, then we can generalize equation~\ref{definitionfh}
and define the sets
\begin{equation}
\label{definitionfh2}
{\cal F}_{\bf h} = \{ ~{\bf x} ~| ~{\bf h}({\bf x}) = {\bf h} ~\}.
\end{equation} 

Like in equation~\ref{Hausdorff}, the complex organization of the sets
${\cal F}_{\bf h}$ is measured by their fractal dimension. This
implies that if we cover the support of the measure $\mu$ with balls
of size ${\bf r}$, the "histogram" $N_{\bf h}(\bf r)$, or number of
such balls that scale as $ {\bf r} ^{{\bf h}}$ for a given ${\bf h}$
behaves like
\begin{equation}
\label{histofh}
N_{\bf h}({\bf r}) \sim {\bf r}^{-D({\bf h})}
\end{equation}
(see \citep{Arneodo1995}). Hence, the singularity spectrum ${\bf h}
\mapsto D({\bf h})$ is a distribution which represents the limiting
behavior of the histograms $N_{\bf h}({\bf r})$ when ${\bf r}
\rightarrow 0$; it is one of the fundamental tools to study complex
and turbulent signals. Obviously, the main problem in the analysis of
observational maps in astronomy is to be able to compute the
singularity spectrum for the whole map or parts of it.

The most direct approach, which consists in estimating the slope
of $\log \mu({\cal B}({\bf x},{\bf r})$ vs. $\log {\bf r}$ for
various ${\bf r}$ ("box-counting method"), is known to be inefficient
and can potentially lead to errors
\citep{Arneodo1995,Chappell2001}. In \citep{Elia2018} the authors make
use of the generalized fractal dimensions $D_q$: the support of the
signal's domain is covered with boxes ${\cal B}({\bf x}_i,{\bf r})$ of
size ${\bf r}$ ($1 \leq i \leq N({\bf r})$), and $\mu_i({\bf r})$ is
defined to be the proportion of signal values inside a ball ${\cal
  B}({\bf x}_i,{\bf r})$. In this case the measure $\mu$ reduces 
to consider the signal as a probability distribution. Then a
partition function is defined:
\begin{equation}
\label{partitionone}
Z(q,{\bf r}) = \displaystyle \sum_{i=1}^{N({\bf r})}\mu_i({\bf r})^q.
\end{equation}
In the limit ${\bf r} \rightarrow 0$ one has, just like in equation~\ref{scaling-turbulence}:
\begin{equation}
\label{scalingpartition}
Z(q,{\bf r}) \sim {\bf r}^{\tau (q)}
\end{equation}
and the generalized dimensions are defined as 
\begin{equation}
\label{generalizeddimension}
D_q = \displaystyle \frac{\tau (q)}{q-1}.
\end{equation}
Some of the quantities $D_q$ can be interpreted: $D_0$ is the box
dimension of the support of $\mu$, i.e., the dimension of the signal's
support as defined in \citet{Elia2018}, $D_1$ is the
"information dimension", and for $q \geq 2$, $D_q$ encodes the scaling
of correlation integrals. In \citet{Arneodo1995} it is shown that
the relation between $\tau(q) = (q-1)D_q$ and the singularity
spectrum $D({\bf h})$ is as follows:
\begin{equation}
\label{legendredq}
\tau (q) = \underset{{\bf h}}{\operatorname{min}}  ( q{\bf h} -D({\bf h}) )
\end{equation}
i.e., a Legendre transform. This leads to an interpretation of the
multifractal formalism with thermodynamics: $q$ is identified with a
Boltzmann inverse temperature and the multifractal formalism allows
the study of the self-similarity phases of the measure $\mu$. The
limit ${\bf r} \rightarrow 0$ is the thermodynamic limit at infinite
volume ($V = \log \displaystyle \frac{1}{\bf r}$) and ${\bf h}_i =
\displaystyle \frac{ \log \mu_i}{\log {\bf r}} = \frac{- \log
  \mu_i}{\log \left (\frac{1}{\bf r} \right )}$ corresponds to $E_i$,
the energy per unit volume of miscrostate $i$. The partition
function $Z(q,{\bf r}) $ is rewritten in its usual form:
\begin{equation}
\label{energypartition}
Z(q,{\bf r})  = \displaystyle \sum_i e^{-qE_i}.
\end{equation}
In this context, the singularity spectrum ${\bf h} \mapsto D({\bf h})$
is the entropy per unit volume, so that the computation of the
singularity spectrum is equivalent to the computation of the entropy
per internal energy in a large multi-body system. This is one {\it
  microcanonical} formulation of the multifractal formalism, which is
correct only in the thermodynamic limit, because it corresponds to
microcanonical ensembles of thermodynamics. It can be implemented
numerically through the box-counting or the histogram methods. This is
the approach used in \citet{Chhabra1989,Chappell2001,Elia2018}, and
also in \citet{Movahed2006} with detrending in the context of solar
data. This approach however, gives poor results due to severe finite
size effects \citep{Arneodo1995}. Very importantly, an essential
criterion for verifying the quality of the results obtained with this
or any other multifractal approach is to test the results on several
rescaled versions of the same signal. The singularity spectrum, for
example, must be the same for different scales by the very definition
of scale invariance. This is not the case with the box-counting
implementation, as we will demonstrate in Sect.~\ref{h-loglog}.

To cope with the difficulties of the microcanonical approach, {\it canonical} formulations have been introduced in the literature. They
consist in evaluating the ${\bf h}$ values and $D({\bf h})$ as
averages among many realizations, i.e., in a {\it canonical}
ensemble. This approach supposes the availability of many realizations
of a same system. The canonical approach to multifractal formalism is
presently the most used when grand ensembles of signal's realizations
are available. The most advanced canonical numerical implementations
include: the Wavelet Transform Maximum Modulus (WTMM) technique (see
\citep{Arneodo1995,Venugopal2006b,Khalil2006} for a detailed exposition), the
cumulant analysis method, which we use in this work for analyzing $\log$-correlations
\citep{Venugopal2006b}, the wavelet leaders technique
\citep{SERRANO20092793}, and the multifractal detrended fluctuation
analysis (MDFA) \citep{KANTELHARDT200287}.

As an example, the WTMM methodology applied on a signal $s({\bf x})$
from an observational map (so that ${\bf x}$ refers to the spatial
coordinates in the signal), makes use of an analyzing wavelet $\psi$
with sufficient vanishing moments to filter out spurious long-range
correlations, and the wavelet-projected signal ${\cal T}_{\psi}({\bf
  x},{\bf r})$ evaluated at position ${\bf x}$ and scale ${\bf r}$, a
suitably chosen $q$th order partition function $Z(q,{\bf r})$ built out
of ${\cal T}_{\psi}$  so that $Z(q,{\bf r}) \sim {\bf r}^{\tau (q)}$ when ${\bf r} \rightarrow
0$ (see \citet{Venugopal2006b} for computationally effective choices of partition functions). The singularity spectrum is recovered as $D({\bf h}) =
\underset{q}{\operatorname{min}} ( q{\bf h} - \tau(q) )$. This
supposes the correct evaluation of $\tau(q)$ through $\log$-$\log$
regression with large amounts of data coming from ensembles of
realizations of the signal.

In the study of the ISM, in most of the cases grand ensembles of
observational maps of the same cloud are not available; this is the
reason why \citep{Elia2018} favors a microcanonical computational approach over the canonical
formulations. Using box-counting or histogram methods, however, do not produce
satisfactory evaluations of the singularity spectrum. In this work, we
develop an alternative and much more reliable  implementation of the
microcanonical approach, introduced  in the next section.

\subsection{Singularity spectrum from "microstates"}
 \label{ss}

One major goal of the microcanonical multifractal formalism is to
evaluate precisely the quantities ${\bf h}({\bf x})$ for a well
defined measure $\mu$ and to derive the singularity spectrum ${\bf h}
\mapsto D({\bf h})$ in order to obtain information on the statistics
of a complex and turbulent signal.  We present in this section and
Appendix A the

determination of the singularity spectrum and the singularity
exponents in a microcanonical formulation based on the theory of
predictability in complex systems \citep{Turiel2008}, which overcomes
the drawbacks of the box-counting and histogram methods.

\begin{figure}[h]
  \centering
  \includegraphics[width=0.45\textwidth]{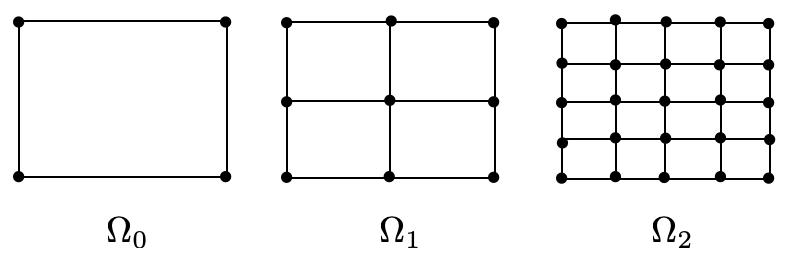} 
\caption{Increasing sequence of lattice nets  $\Omega_0 \subset  \Omega_1  \subset \cdots \subset  \Omega_n \subset \cdots$  inside the unit square.\label{net}}
\end{figure}

In the case of observational maps in astronomy, but also in the
analysis of general turbulence data, scaling is valid only within a
certain inertial range of scales (Sect.~\ref{multifractal})
and we have to define a measure from a finite set of discrete
acquisitions values in the map. Hence, we follow the exposition of the {\it multifractal ansatz} in the
physics of disordered systems, which allows for a rigorous presentation
of the multifractal formalism in the case of discrete signals
\citep{FYODOROV20104229,Fyodorov2012}. First, we have to define a
measure from the data of an observational map. In previous works
\citet{Chhabra1989,Chappell2001,Elia2018} the measure is often defined
from the the signal $s$ itself. Following \citet{Turiel2008}, it turns
out that better accuracy is obtained when starting from discrete
gradient information. Consequently we consider an inertial range in
the image domain $[ {\bf r} _1, {\bf r} _2]$ which we identify for
convenience with the unit square $[0,1]^2$.  We define a collection of
refined lattice points as follows: for each $n \in \mathbb{N}$, let us
consider the set $\Omega_n$ of points whose coordinates are integer
multiples of $2^{-n}$ inside the unit square. We get an increasing
sequence $ \Omega_0 \subset \Omega_1 \subset \cdots \subset \Omega_n
\subset \cdots$ of lattice points as shown in Fig.~\ref{net}.
 
\begin{figure}[h]
  \centering
  \includegraphics[width=0.35\textwidth]{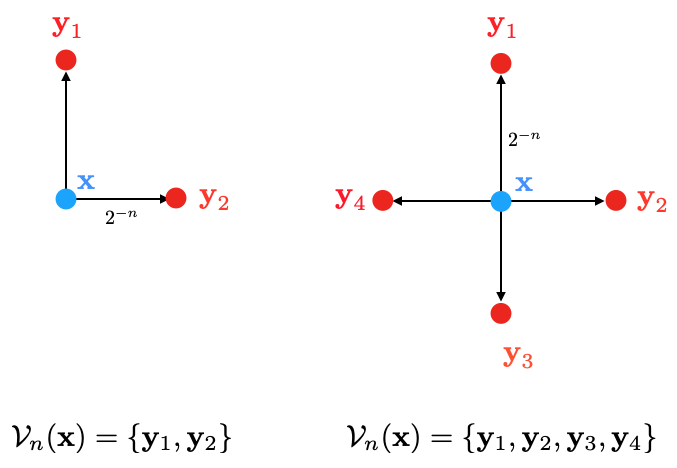} 
\caption{Discrete neighborhood sets  ${\cal V}_n({\bf x})$ at scale $2^{-n}$ around point ${\bf x}$.\label{voisin}}
\end{figure}

For each ${\bf x } \in \Omega_n$ we define a finite set of neighboring points ${\cal V}_n({\bf x })$ as shown in Fig.~\ref{voisin}. 
 If ${\bf x } \in \Omega_n$, we have the information of the signal
 data $s({\bf x})$ at point ${\bf x }$ together with the discrete
 gradient's norms determined by the differences $D({\bf x},{\bf y}) =
 \displaystyle \frac{| s({\bf x}) - s({\bf y}) |}{\| {\bf x} - {\bf
     y}\|}$, with ${\bf y} \in {\cal V}_n({\bf x})$ of vicinity points
 around ${\bf x}$ at scale $2^{-n}$. We define a measure $\mu_n$ at
 that scale $2^{-n}$ from the discrete gradient information data. The
 exact definitions are given in Appendix~\ref{definitionmeasure}, and
 $\mu_n$ is seen as a "gradient measure" associated to the
 observational map data at scale $2^{-n}$, which means that the
 $\mu_n$-measure of a ball is the sum of discrete gradient's norms
 information inside that ball.

The application of the multifractal formalism to the signal $s$ and
its associated measure $\mu_n$ is valid only if the measure $\mu_n$
satisfies the {\it scaling hypothesis} shown in
equation~\ref{scalinghypothesis}. In canonical implementations of the
multifractal formalism, checking the scaling is a necessary first step
before applying the numerical tools, and is generally done by
$\log$-regression performed on the chosen partition functions. In
existing microcanonical implementations such as
in \citet{Elia2018}, the scaling is shown only for output
simulations of fractional Brownian motions realizations. In
Sect.~\ref{inertialrange} we will make use of the 2D structure
function methodology \citep{Renosh2015}) to check the scaling of
analyzed data.

The key idea in a computation of the singularity exponents in a
microcanonical setting based on predictability is to relate the
quantities ${\bf h}({\bf x})$ to the signal's reconstruction. In physical
signals, the set of possible values taken by ${\bf h}({\bf x})$ is
bounded, and the lowest singularity exponents' value ${\bf h}_{\infty}
= \min \{ {\bf h}({\bf x}) \}$ corresponds to the sharpest transitions
in a signal according to equation~\ref{singularitybehavior}. The
points ${\bf x}$ such that ${\bf h}({\bf x}) = {\bf h}_{\infty}$ form
a particular subset ${\cal F}_{\infty}$ in the signal's domain, of
fractal nature, encoding the strongest transitions: in a 2D signal,
${\cal F}_{\infty}$ encodes the strongest edges, while the other sets
${\cal F}_{\bf h}$ correspond to more smoother edges and transitions
as ${\bf h}$ increases. Under the assumption that ${\cal F}_{\infty}$
coincides with the set of most unpredictable points (in the sense of
complex systems theory), it can be shown that the computation of ${\bf
  h}({\bf x})$ at scale $2^{-n}$ involves only immediate neighboring
points around ${\bf x}$. When these intuitive considerations are
expressed rigorously into a more mathematical manner, as shown in
Appendix~\ref{h-upm}, we arrive to a computation of the singularity
exponents ${\bf h}({\bf x})$ using a correlation measure denoted
${\cal H}$ in Appendix~\ref{h-upm} which can be evaluated locally
around any point ${\bf x}$ in the signal domain's, leading to
equations~\ref{hupm} and~\ref{defH} which we use as our fundamental
methodology to compute the singularity exponents in a microcanonical
formulation.

\section{Experimental comparison of microcanonical methodologies: scale invariance}
\label{h-loglog}
In this section, we test the validity of the scaling hypothesis  eq.~\ref{scalinghypothesis} by comparing the singularity spectra computed in a
microcanonical formulation of the multifractal formalism using the two
approaches mentioned previously: first, an enhanced version of the
counting-box method presented in Sect.~\ref{existingappraoches} called the {\it gradient modulus wavelet projection method} detailed in \citep{Turiel2006} and
second, the method explained in Sect.~\ref{ss}. The experiment is
performed on the Musca {\sl Herschel} 250 $\mu$m observational map
described in more detail in Sect.~\ref{musca}.

Scale invariance eq.~\ref{scalinghypothesis} implies that singularity spectra computed from two scaled versions of an original
observational map must coincide. Consequently, we will perform the
following experiment: we take the Musca map, generate out of it two
downscaled versions of the map, compute the singularity spectrum with
either method on the two downscaled versions of the map (see below),
and check if the resulting spectra coincide.

\begin{figure}[h]
  \centering
  \includegraphics[width=0.20\textwidth]{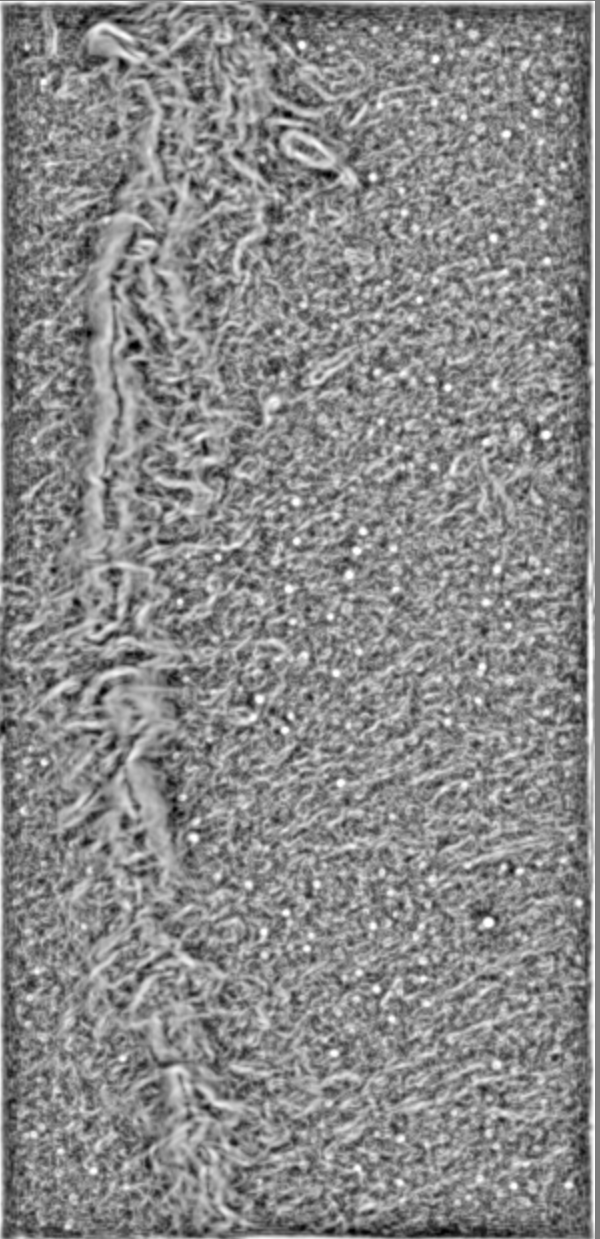}~~ \includegraphics[width=0.197\textwidth]{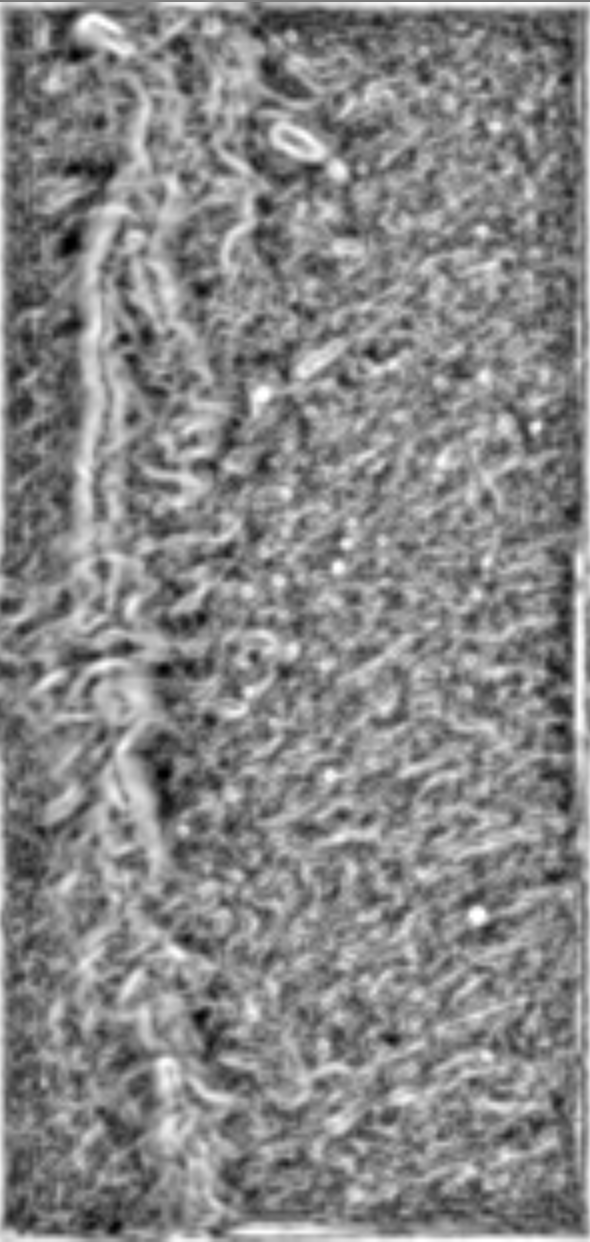} \\
    \includegraphics[width=0.48\textwidth]{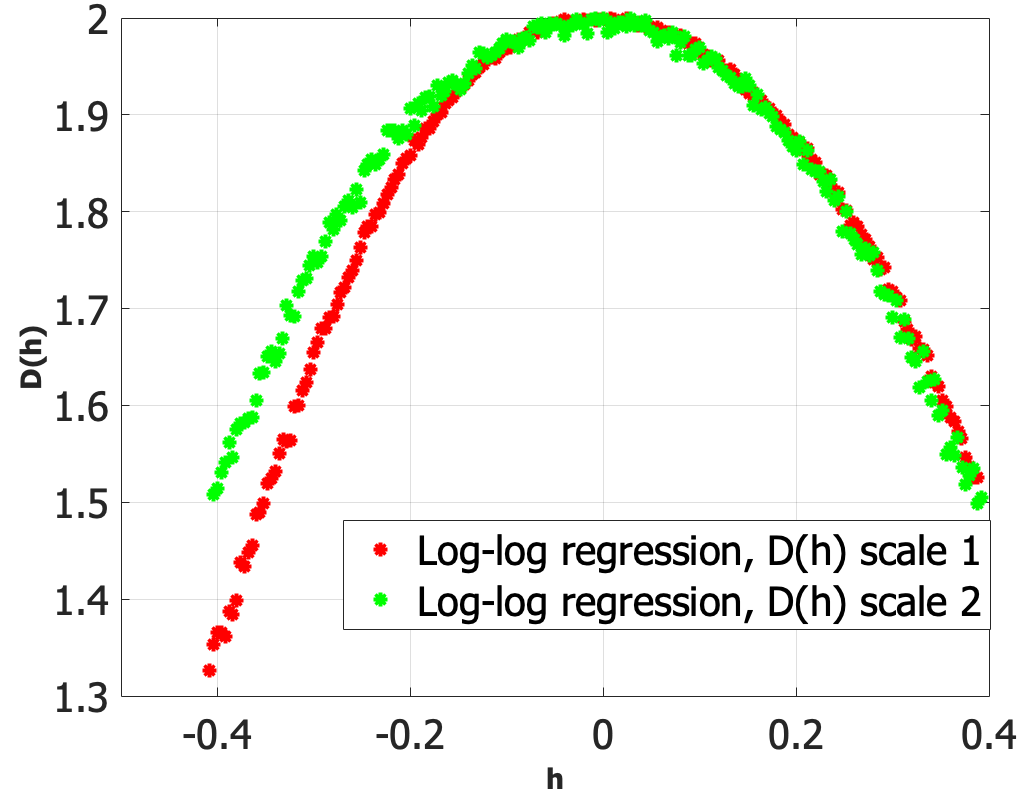}
\caption{Top row, from left to right: visualization of the singularity
  exponents computed through $\log$-regression using a Lorentz wavelet
  of order 3 on the Musca 250 $\mu$m {\sl Herschel} flux map presented
  in Sect.~\ref{musca}. Algorithm used to compute the singularity exponents: gradient modulus wavelet projection method. The two images correspond to two consecutive
  downscaled wavelet projections of the original signal using a
  discrete wavelet projection. Bottom: the resulting mappings ${\bf h}
  \mapsto D({\bf h})$ using the singularity exponents computed on the
  two consecutive scales (red: half scale of the original Musca
  250$\mu$m {\sl Herschel} flux map, green: 1/4 scale of original
  Musca 250$\mu$m {\sl Herschel} flux map). We observe that, inside
  the domain corresponding to ${\bf h}
  \leq 0$, the two mappings do not coincide. Consequently, the
  $\log$-regression method for computing the singularity exponents
  provide poor result in this case. \label{loglog}}
\end{figure}
\begin{figure}[h]
  \centering
  \includegraphics[width=0.20\textwidth]{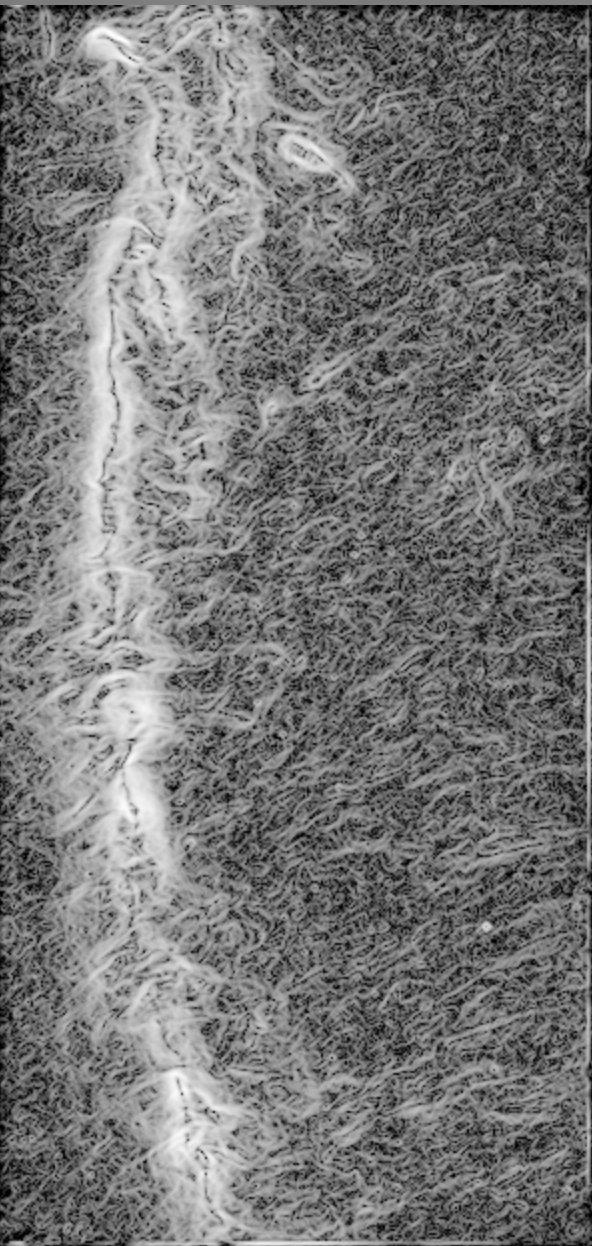}~~ \includegraphics[width=0.197\textwidth]{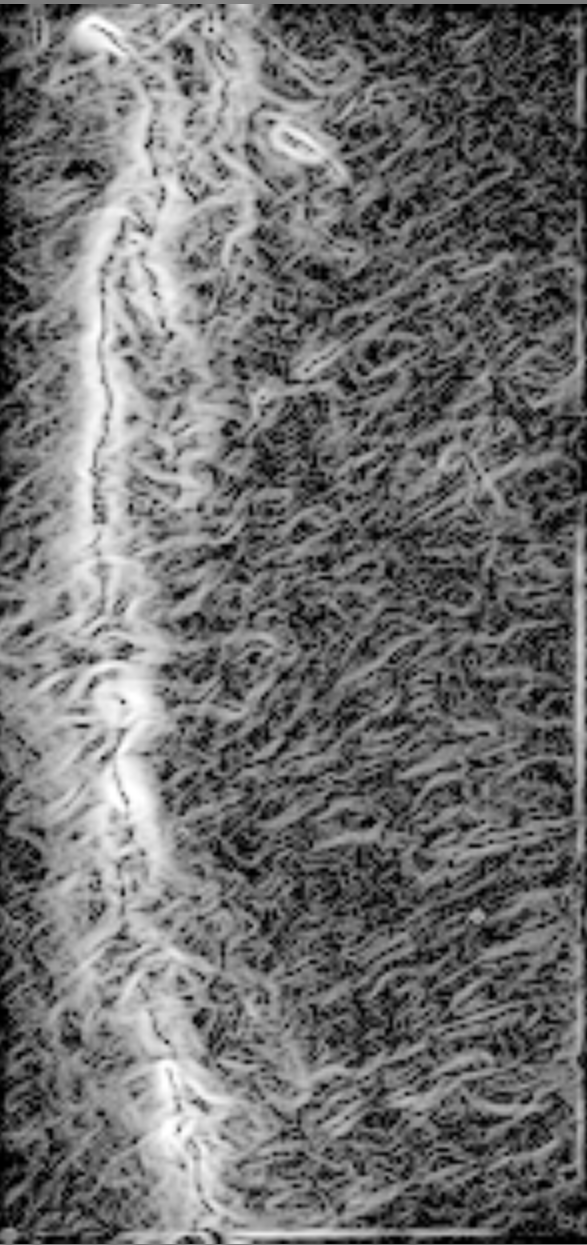} \\
    \includegraphics[width=0.47\textwidth]{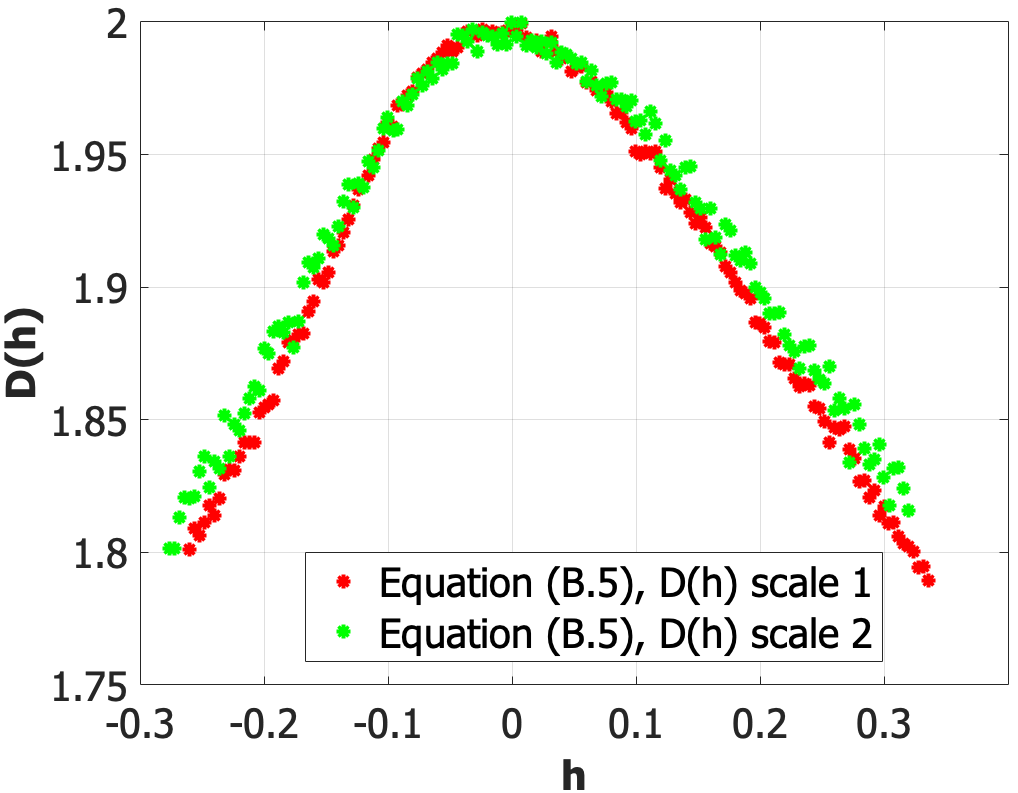}
\caption{Top row, from left to right: visualisation of the singularity
  exponents computed with the local correlation measure described in
  equations~\ref{defH} and~\ref{hupm} on the Musca 250 $\mu$m {\sl
    Herschel} flux map presented in Sect.~\ref{musca}. The two
  images correspond to two consecutive downscaled wavelet projections
  of the original signal using a discrete wavelet projection. Bottom:
  the resulting mappings ${\bf h} \mapsto D({\bf h})$ using the
  singularity exponents computed on the two consecutive scales (red:
  half scale of the original Musca {\sl Herschel} map, green: 1/4 scale of
  original Musca {\sl Herschel} map). We observe that the two mappings
  coincide much better than with the previously exponents computed
  from $\log$-regression and displayed in
  Fig.~\ref{loglog}. Consequently, the local correlation measure
  algorithm exhibits the scaling in a much better
  way. \label{upmmeasure}}
\end{figure}

\subsection{Gradient modulus wavelet projection method}
The algorithm is described in detail in \citep{Turiel2006}. It is an enhanced counting-box method which allows for the computation of the singularity exponents  ${\bf h}({\bf x})$ and the singularity spectrum  ${\bf h}
\mapsto D({\bf h})$ through  a $\log$-regression performed, not directly on 
the scaling hypothesis  eq.~\ref{scalinghypothesis}  but rather over wavelet
projections of the measure for better computational accuracy. This
approach is allowed because if a measure $\mu$ scales with singularity
exponents ${\bf h}({\bf x})$, as in equation~\ref{scalinghypothesis},
wavelet projections of $\mu$ scale with the same exponents ${\bf
  h}({\bf x})$ as long as the analyzing wavelet has $n$ vanishing
moments with $ n > {\bf h}({\bf x})$
\citep{Venugopal2006b,Turiel2008}: if ${\bf r} > 0$ is a scale, $\mu$
a measure on $\mathbb{R}^2$, $\psi$ a real wavelet, and $\lambda_{\bf
  r}$ the measure $\displaystyle \frac{1}{\bf r}\psi \left (
\frac{-{\bf x}}{\bf r}\right ) \, \mbox{d}{\bf x}$, the wavelet
projection of $\mu$ at scale ${\bf r}$ is another measure noted ${\cal
  T}_{\psi}(\mu, {\bf r})$ which is the convolution of the measures
$\mu$ and $\lambda_{\bf r}$:
\begin{equation}
\label{measureproj}
{\cal T}_{\psi}(\mu, {\bf r})= \lambda_{\bf r} * \mu.
\end{equation}
If $\mu$ possesses a density, then ${\cal T}_{\psi}(\mu, {\bf r})$ has
a density given by the usual continuous wavelet transform of the
original density with mother wavelet $\psi$. The measure
$\mu_n$ considered here is the one defined by eq.~\ref{mu-n} with the
neighboring set ${\cal V}_{n}({\bf x})$, corresponding to the left
part of Fig.~\ref{voisin} (2 neighboring points). To compute the
singularity exponents, we take as an analyzing wavelet a
$\beta$-Lorentzian defined by
\begin{equation}
\label{lorentz}
L_{\beta}({\bf x}) = \displaystyle \frac{1}{(1 + \| {\bf x} \|^2)^{\beta}}
\end{equation}
with $\beta = 3$, with its support scaled to adapt to the signal
Nyquist frequency. The $\log$-regression is performed over 30 scales.

To check the scaling of the measure we downscale the original signal
on two consecutive scales less than the original signal maximal
resolution using a standard discrete wavelet transform defined by the
reverse bi-orthogonal projection of order 4.4. Then, for each of these
two downscaled wavelet projections, we compute the singularity
exponents through $\log$-regression with the Lorenztian wavelet, as
explained previously, to get two scalar fields of singularity exponents
${\bf h}({\bf x})$ at the two consecutive resolutions. From each of
these scalar fields, we compute a mapping ${\bf h} \mapsto D({\bf h})$
using eq.~\ref{historesolinf}: if the measure is scale invariant
and if the singularity exponents are correctly computed, then the two
mappings corresponding to each resolution should be coincident. We
show the result of the $\log$-regression in Fig.~\ref{loglog}.  We
observe in the figure that, inside the most informative part of the
mapping, corresponding to ${\bf h} \leq 0$, the two mappings do not
coincide. Consequently, the $\log$-regression method for computing the
singularity exponents provide poor result in this case and the
quantities computed through $\log$-regression do not show the scaling
of the measure.

\subsection{Local correlation measure of section~\ref{ss}.}

In Fig.~\ref{upmmeasure}, we reiterate the previous experiment on the
Musca 250 $\mu$m {\sl Herschel} map, this time with the singularity
exponents computed using eqs. ~\ref{defH} and~\ref{hupm}.  The
resulting maps ${\bf h} \mapsto D({\bf h})$ reveal much more
satisfactorily the scaling behavior of the measure because they
coincide very well for the two consecutive scales over the range of
singular values. This agreement is much stronger than the one of the
exponents previously computed from $\log$-regression and displayed in
Fig.~\ref{loglog}. Consequently, the local correlation measure
algorithm exhibits the scaling in a much better way. Note also that
the resulting graphs differ notably from the ones computed using
$\log$-regression and shown in the bottom of Fig.~\ref{loglog}: they
are much less "parabolic". It becomes obvious that the
$\log$-regression method (gradient modulus wavelet projection algorithm) tends to produce singularity spectra of the
$\log$-normal type in that case.
 \begin{figure*}[h]
 \centering
 \includegraphics[width=1.0\textwidth]{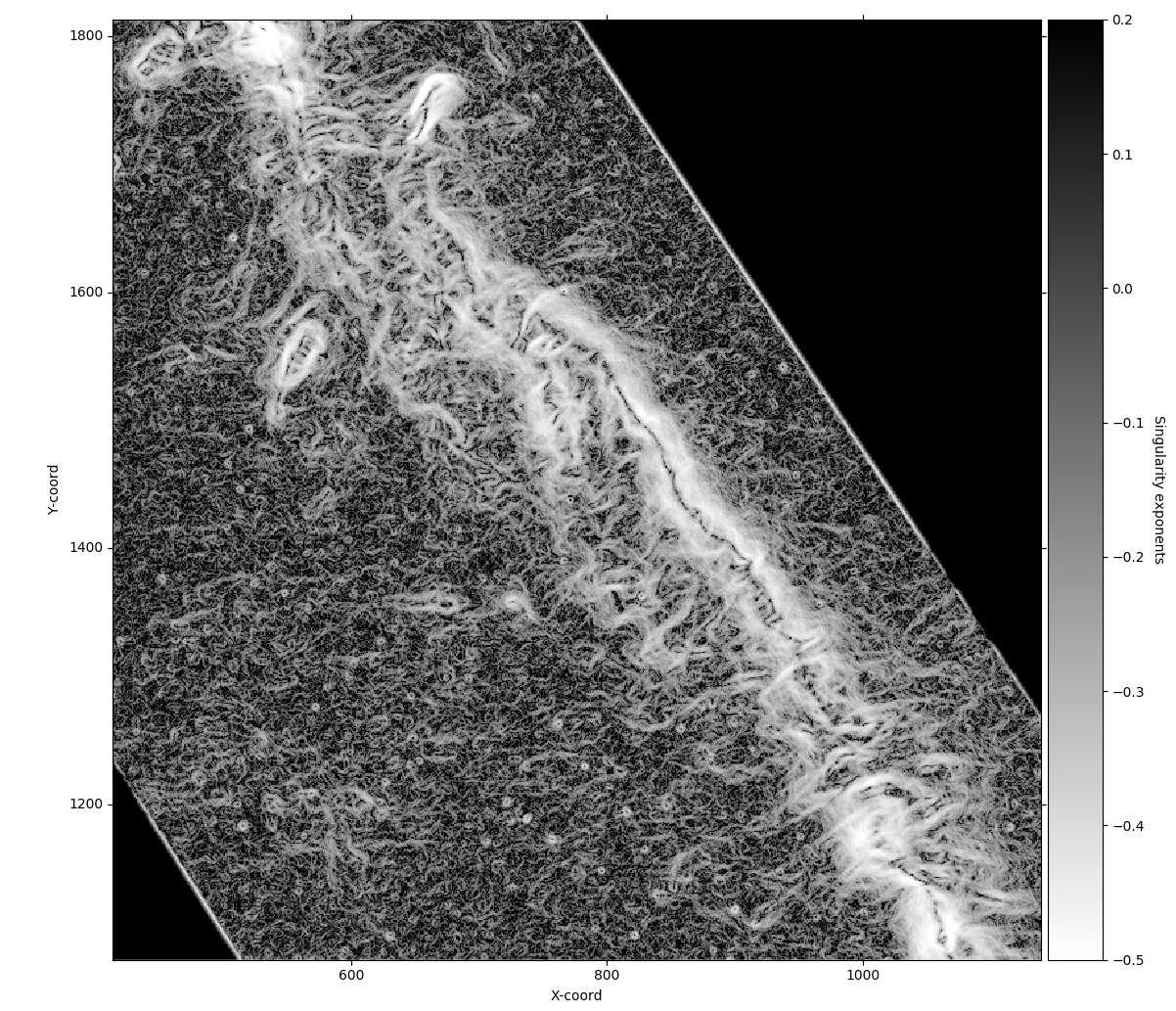} 
\caption{Map of the singularity exponents computed on filtered Musca
  250 $\mu$m data.  The image is a zoom of the singularity map over the central part of the Musca  observation map, to
  better show how the singularity exponents reflect the complex
  distribution of filamentary coherent structures.\label{expsmusca}}
  \end{figure*}
  
\section{Application to the Musca observation map. Sparse filtering. Determination of the inertial range.}
\label{musca-appli}
\subsection{Map of the singularity exponents}
\label{sing-ex-musca}

In Sect.~\ref{multifractal} we introduced the multifractal formalism
through the analysis of intermittency in the velocity field as this is
the usual presentation in the turbulence literature. In this study,
the analysis is done on thermal continuum emission using the discrete
gradient measures introduced in Sect.~\ref{ss}.
In Fig.~\ref{expsmusca} we display the singularity exponents ${\bf
  h}({\bf x})$ on the {\sl Herschel} 250 $\mu$m map, computed using
the local correlation measure defined by eqs.~\ref{defH} and
\ref{hupm}.  The computation has been done on a filtered version of
the map, described in Sect.~\ref{eaf}, to eliminate some background
noise. As we will see in the following, background noise elimination
is performed not only for better visualization, but also to compute
more precise singularity spectra.  The wedge used in the figure is
such that most negative singularity exponents are brighter. We can see
how the singularity exponents trace the transitions in the signal,
particularly along the filamentary structures, which are very well
rendered by negative values of ${\bf h}({\bf x})$. Note also the thin
dark region along the crest of the main filament: it follows a
generatrix-like curve of a cylindrical structure, along which the flux
appears constant. We observe the same phenomenon in MHD simulations,
see middle picture of Fig.~\ref{simusdib}. In the case of the Musca
cloud, however, a close inspection of the crest shows that this thin
dark region is filled with small elongated structures perpendicular to the crest,
making the picture even more complex. In addition, there are small
circular features distributed all over the map that are most likely
background galaxies that were not eliminated.

\subsection{Sparse filtering of an observational map}
\label{eaf}
\begin{figure*}[h]
\includegraphics[width=0.424\textwidth]{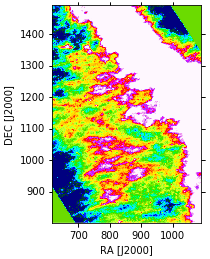}\includegraphics[width=0.495\textwidth]{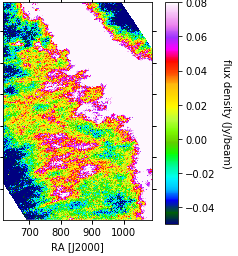} 
\caption{Visualisation of the filtering (computed with $\lambda =
  0.7$) zoomed in the central part of the Musca 250 $\mu$m map. The
  level was chosen to emphasize low values. The left panel shows the
  filtered data and the right one the unfiltered data.  The positive
  effect of the filtering on the flux density data becomes obvious.
  The visualisation done on the singularity exponents, displayed in
  Fig.~\ref{filtredsemusca}, demonstrates the importance of the
  filtering on filamentary structures, which are of low fractal
  dimension.\label{filtredsemuscaflux}}
\end{figure*}
\begin{figure*}[h]
\includegraphics[width=0.45\textwidth]{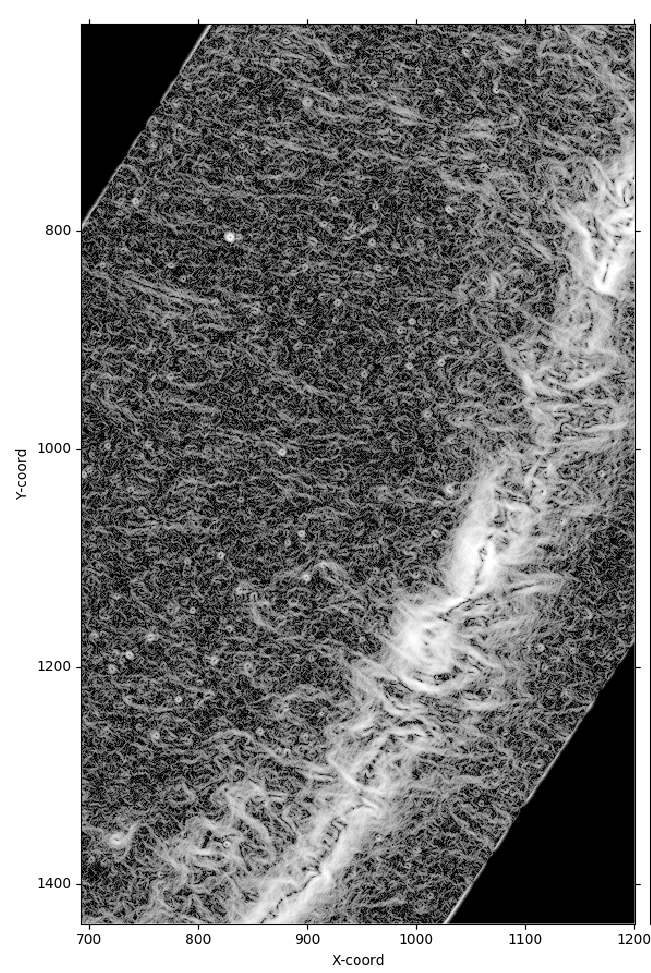}\includegraphics[width=0.515\textwidth]{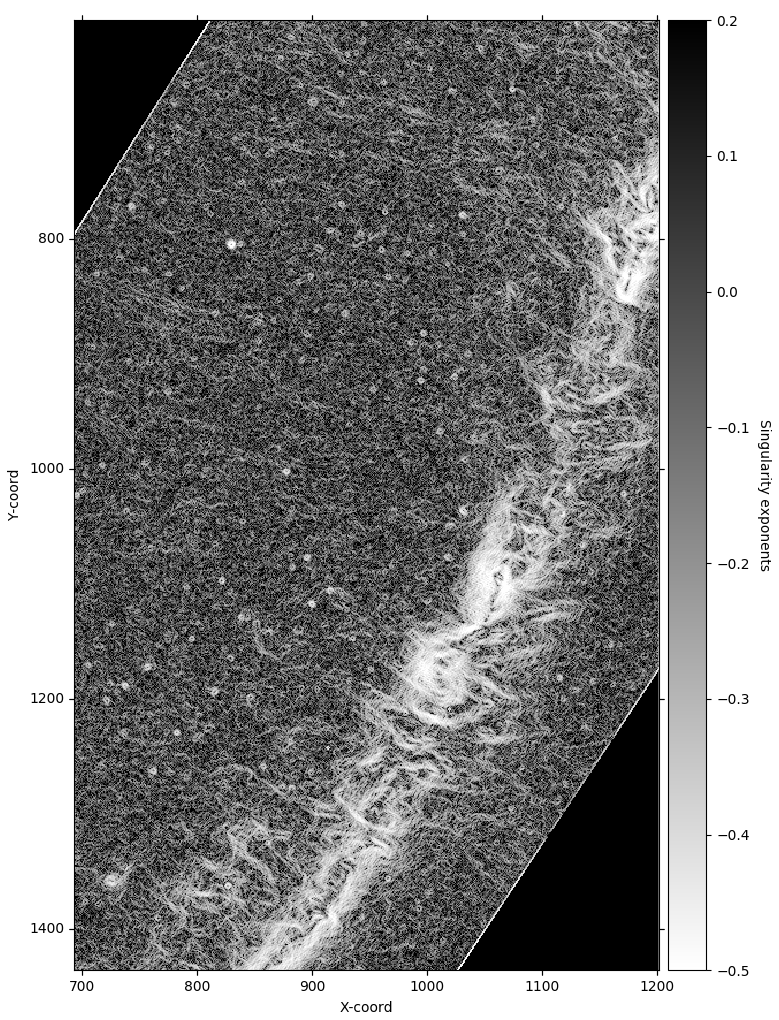} 
\caption{Illustration of background noise elimination and enhancement of the
  filamentary structures in the Musca 250 $\mu$m map by edge-aware nonlinear filtering while
  promoting a signal's gradients sparsity with $L^1$ norm. The 
  images show the singularity exponents of a region in Musca 
  that is rich in filamentary structures.
  The right panel displays the singularity exponents derived from the non-filtered original
  map, while the left panel shows the singularity exponents after filtering the map
  with $\lambda = 0.7$ (eq.~(1). The same greyscale wedge is used for both 
  images.
    \label{filtredsemusca}}
\end{figure*}

\begin{figure}[h]
\centering
\includegraphics[width=0.47\textwidth]{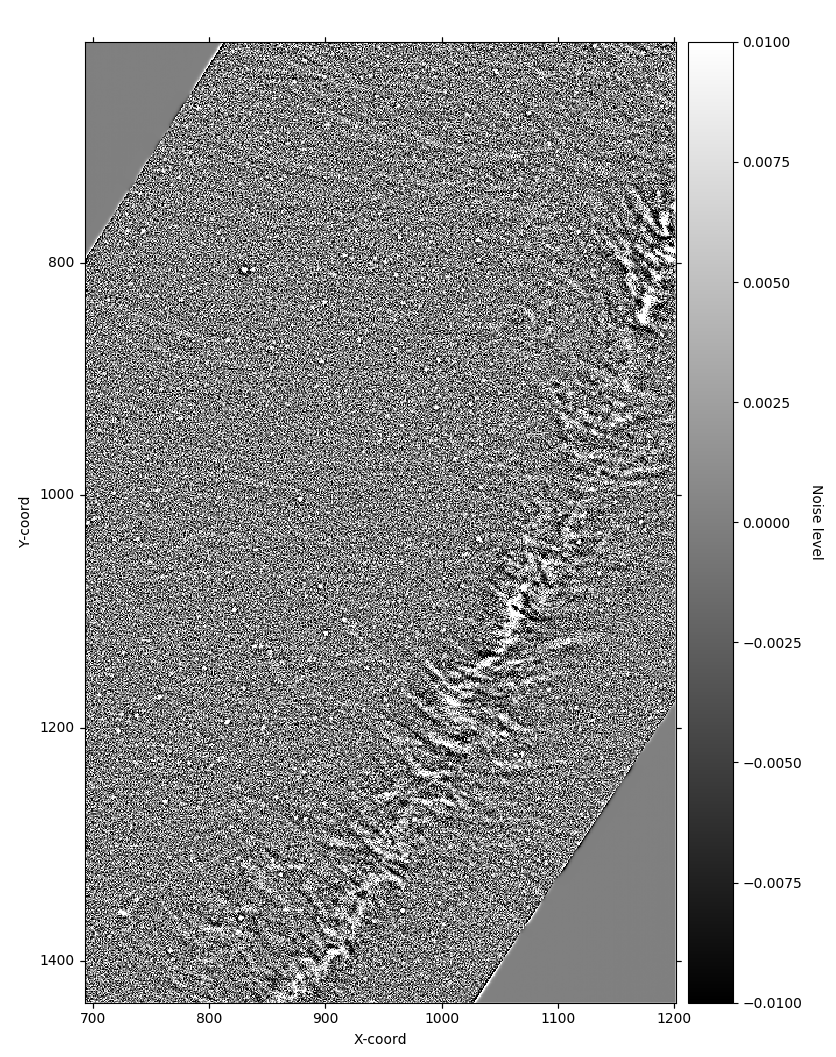}
\caption{Visualisation of the noise eliminated by the edge-aware noise
  filtering: the data displayed is $s - s_f$ in the same area as shown
  in figure~\ref{filtredsemusca}. \label{noisemusca}}
\end{figure}

\begin{figure}[h]
  \centering
  \includegraphics[width=0.48\textwidth]{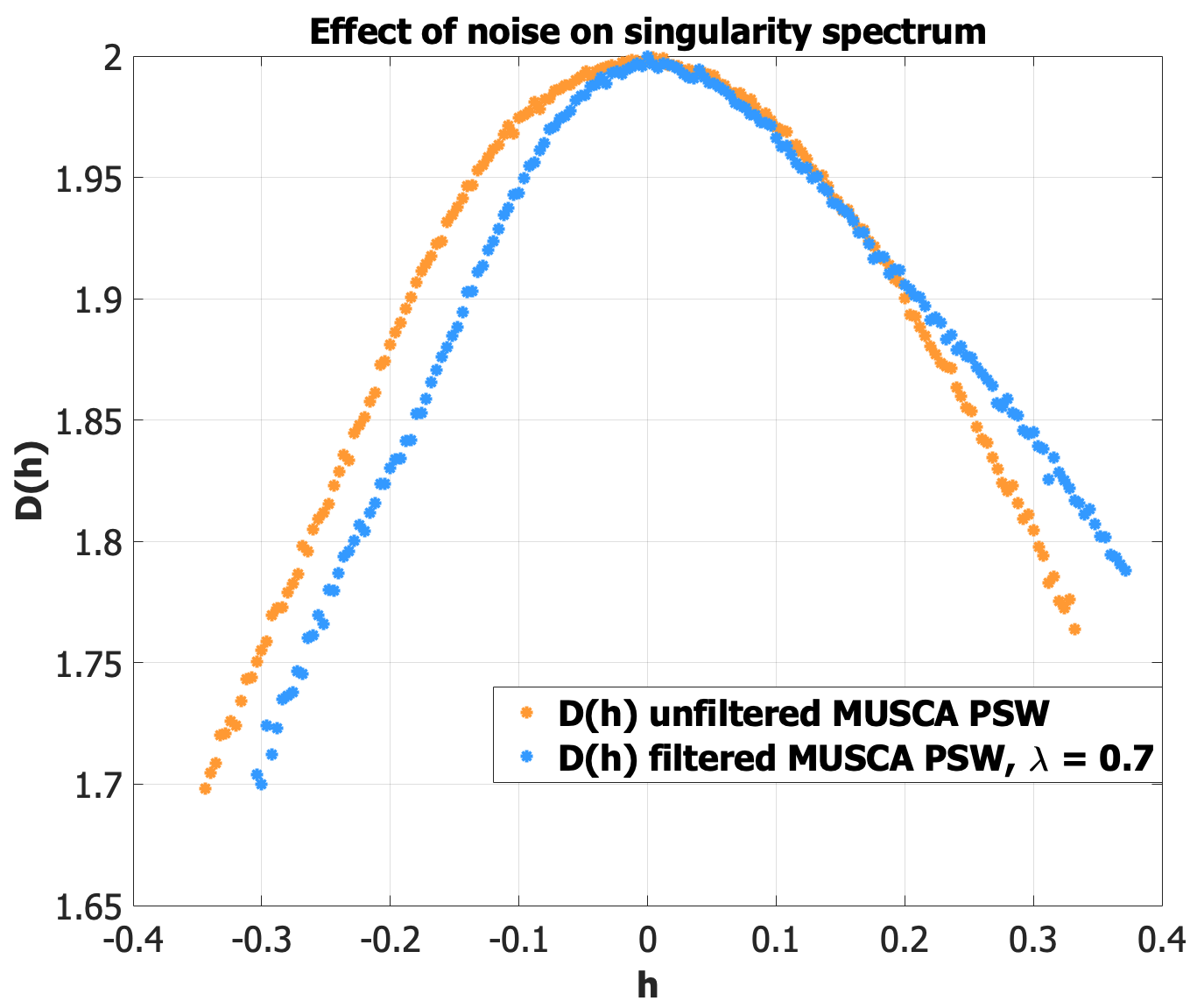}
  \caption{Singularity spectrum of the Musca 250 $\mu$m {\sl Herschel} map,
    computed with and without edge-aware filtering.  The orange curve shows the
    singularity spectrum computed on raw non-filtered data while the
    blue curve displays the singularity spectrum computed on
    edge-aware filtered data using the filtering introduced in
    Sect.~\ref{eaf} with parameter $\lambda = 0.7$. The horizontal
    axis gives the ${\bf h}$-values of the scaling exponents, the
    vertical axis the fractal dimension $D({\bf h}) = \dim ({\cal
      F}_{{\bf h}})$ (see Sect.~\ref{ss}). First, it becomes obvious
    that the part of the spectrum corresponding to ${\bf h} \leq 0$,
    i.e. the strongest transitions, is not accurately computed on the
    nonfiltered data, the presence of the background noise leads to
    an overestimation of the multifractal spectrum. Second, the slope
    of the right part of the graph is poorly evaluated in the
    presence of noise. As a result, the singularity spectrum computed
    on the filtered observation map is less symetrical w.r.t. the
    vertical axis ${\bf h} = 0$. \label{muscafiltering}}
\end{figure}

The Musca {\sl Herschel} 250 $\mu$m flux map 
contains point-like sources, which are mostly galaxies,
and the cosmic infrared background\footnote{The CIB in the
  far-infrared consists mostly of Galactic cirrus emission and
  zodiacal emission, i.e. thermal emission of dust in the Solar
  system.} (CIB) and the cosmic microwave background
\citep{Padoan2001a,Robitaille2019}, which have isotropic low amplitude
values very close to those of low scale filamentary structures. As a
consequence, eliminating this 'noise' is of primary importance in the
multifractal analysis of the Musca gas.
One possible method, presented in
\citet{Robitaille2019}, makes use of a filtering algorithm aiming at
separating (and reconstructing) the large space-filling monofractal
content of an image from the coherent structures which have a
multifractal nature. This technique applies a threshold on the
probability distribution function (PDF) of wavelet coefficients at
different spatial scales and successfully identifies both the
monofractal component signature of the noise and the turbulent
component of the ISM. However, we do not employ this method here 
because thresholding at low scales a wavelet decomposition blurs
irremediably the filamentary structures after
reconstruction. Consequently, a multifractal analysis applied on the
reconstruction can have substantial impact on the singularity
spectrum. Instead, one needs to keep any coherent structure at low
scales intact, while eliminating noise. For that reason, it is preferable
to use an edge-aware noise filtering approach~\citep{badri:tel-01239958} for the multifractal
analysis of astronomical data which we present in the following.

We compute a filtered image $s_f({\bf x})$
such that the result $s_f$ remains close to the original data $s$ while having sparse gradients, i.e., we
want to eliminate the noise but keep coherent
structures. \citet{Candes2008} show that the $L^1$ norm favours
sparsity in a much better way than the $L^2$ norm while remaining
convex, which greatly facilitates the numerical implementation and
guarantees the existence of a global minimum. Consequently, the image
$s_f$ is computed as the solution of the optimization problem:
\begin{equation}
\label{sparsemin}
\underset{s_f}{\operatorname{argmin}} ~~~ \| s - s_f \|_1 + \lambda \| \nabla s_f \|_1 
\end{equation}
with 
\begin{equation}
\label{norml1}
\| \nabla s \|_1 = \displaystyle \int_{\mathbb{R}^2}\displaystyle \left ( \left | \frac{\partial s}{\partial x_1 } \right |  +   \left | \frac{\partial s}{\partial x_2}   \right | \right )  \mbox{d}{x_1} \,
\mbox{d}{x_2} 
\end{equation}
and $\lambda > 0 $ is a tuning parameter. The first term $ \|  s
-  s_f \|_1$ guarantees that the filtered image is very similar
to the original ("data fitting term"), while the second term $ \lambda \,\| \nabla s_f
\|_1$, when minimized, promotes sparsity of the filtered
gradients. Hence, the minimisation of the two terms eliminates noise,
while keeping gradient information.  To solve the minimization problem
defined by eq.~(\ref{sparsemin}), we make use of the Half Quadratic Splitting
resolution method \citep{392335,6909751}. We show the result of sparse
filtering first on the original flux density data in
Fig.~\ref{filtredsemuscaflux}, where it appears as standard noise
reduction. However, turning to the visualisation on the singularity
exponents, as shown in Fig.~\ref{filtredsemusca}, reveals the power of
this sparse filtering with respect to the filamentary structures, as
they are clearly enhanced with respect to the unfiltered
version. We also observe that although the filtering does not eliminate much of background
 galaxies, it clearly eliminates sufficiently cosmic infrared background (CIB) and cosmic microwave background to significantly
 expose filamentary structures that were hidden by Gaussian noise in the original data. Figure~\ref{noisemusca} shows the image $s - s_f$, i.e., the
resulting noise eliminated by the edge-aware filtering.
Figure~\ref{muscafiltering} displays the results of applying filtering
for the singularity spectrum on the Musca map, with and without
background noise elimination.  Note how the background noise leads to
an overestimation of the low dimensional filamentary structures, and
how the gaussianity of background noise makes the nonfiltered spectrum
more parabolic than the filtered one. Consequently,  noise elimination is a necessary preprocessing step in a multifractal analysis of astronomical data for a better computation of the singularity spectrum.

Nevertheless, a question can be raised: does the edge-aware filtering defined by
eq.~(\ref{sparsemin}) introduce spurious new gradients, initially not 
present in the original data? The answer is no, as can be seen
from the numerical implementation of the minimization algorithm, which
is performed in two steps:
\begin{enumerate}
\item Low level gradients are set to 0, and gradients whose norm is
  greater than a given threshold are kept. This operation does not
  introduce new gradients.
\item In a second step a new image is generated whose gradient is, at
  each point, a weighted sum of the original gradient image and the
  image obtained at the precedent step. This operation does not
  introduce new gradients either.
\end{enumerate}
\begin{figure}[h]
  \centering
   \includegraphics[width=0.48\textwidth]{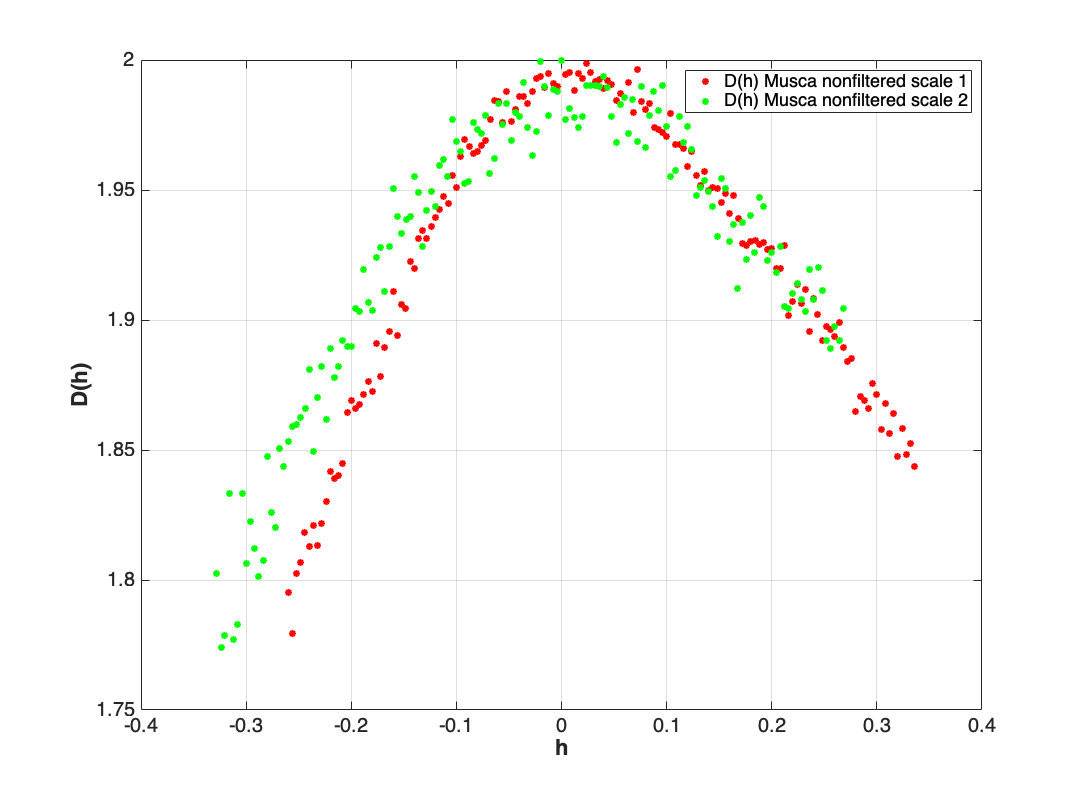}  
\caption{The mappings ${\bf h} \mapsto D({\bf h} )$ computed for two
  consecutive scales of the unfiltered Musca 250 $\mu$m {\sl Herschel}
  observational map using a reverse bi-orthogonal discrete wavelet
  transform of order 4.4.  We see from the graphs that the presence of
  the noise alters the scaling of the measure.\label{scaling}}
\end{figure}
\begin{figure}[h]
  \centering
  \includegraphics[width=0.48\textwidth]{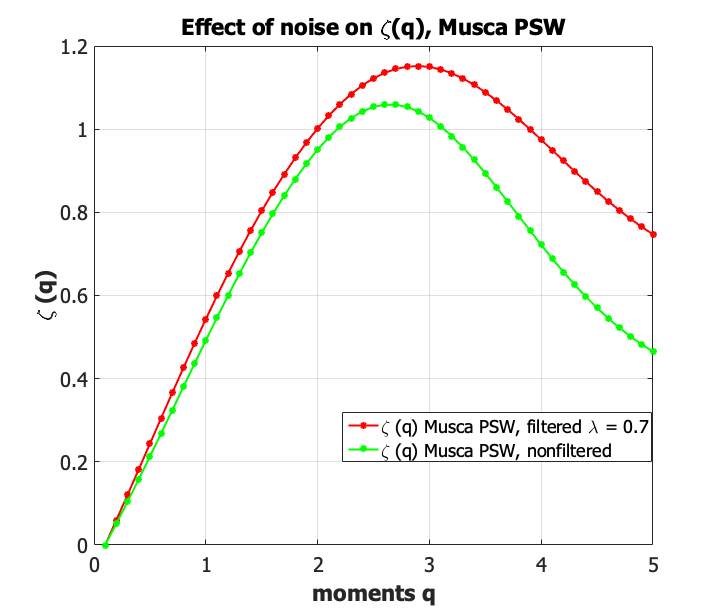} 
\caption{Plot of $\zeta(q)$ for the Musca 250 $\mu$m {\sl Herschel}
  observational map with filtering ($\lambda = 0.7$, curve in red) and
  without filtering (curve in green); the mapping $q \mapsto \zeta(q)$
  is defined by the structure function method (equation~\ref{eqnSSF}).
  We see on this graph that the background noise does affect the
  statistics of scaling. \label{zetas}}
\end{figure}

In Sects.~\ref{h-loglog} and Appendix~\ref{h-upm}, we checked the
scaling of the gradient measure by computening the singularity spectra
at two consecutive scales in a wavelet projection of the signal.
We observed from that experiment the scaling of the measure when the
singularity spectra are derived from the computation of the
singularity exponents using the method presented in Appendix~\ref{h-upm}.
These experiments were performed on the filtered Musca map with
$\lambda = 0.7$. We show in Fig.~\ref{scaling} the result of the
same test using the nonfiltered Musca map. The two scaled versions are generated,
like previously, using a reverse bi-orthogonal discrete wavelet
transform of order 4.4. We see from the graphs that the presence of
noise alters the scaling of the measure. This also advocates for the
use of filtered observational maps.

Figure~\ref{zetas} displays the result of the computation of the
function $q \mapsto \zeta(q)$ defined in the structure function method
(see Sect.~\ref{inertialrange}, eq.~\ref{eqnSSF}) on the
unfiltered and filtered Musca map. The two graphs are
clearly distinct which adds to non negligible effects of the
background noise on the scaling properties of the observational map.

\begin{figure}[h]
  \centering
  \includegraphics[width=0.49\textwidth]{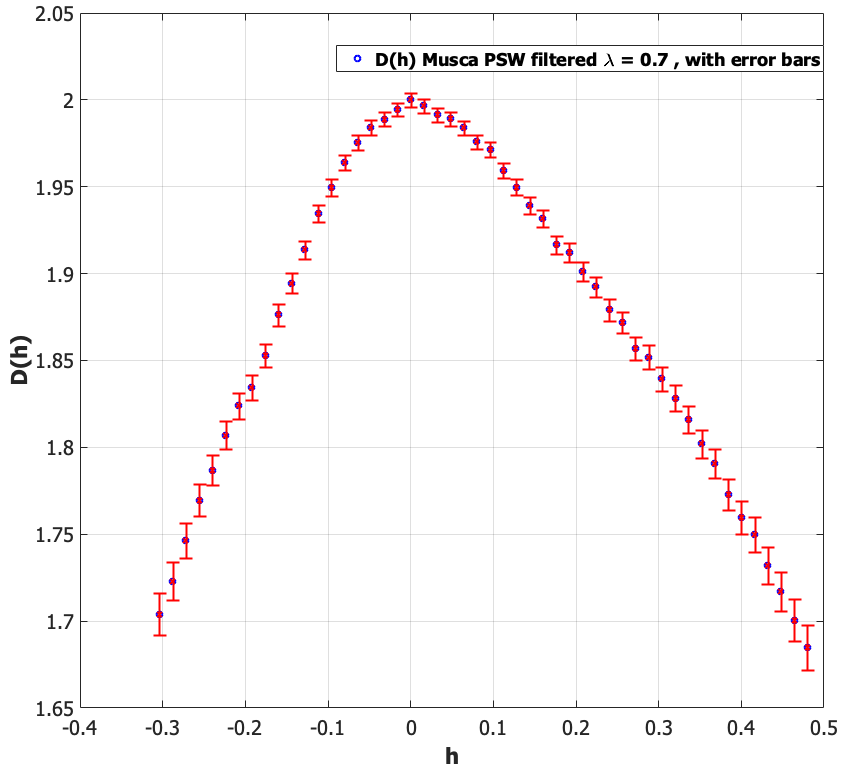} 
\caption{Singularity spectrum of the Musca {\sl Herschel} 250 $\mu$m
  map with error bars as defined by
  eq.~\ref{errspectre}; the observational map is edge-aware filtered with a 
  tuning parameter $\lambda = 0.7$.\label{errorbars}}
\end{figure}

Since the singularity spectrum is estimated from the histogram of the
singularity exponents (see eq.~\ref{historesolinf}), it is possible to
estimate the error bars in the spectrum. If one discretizes the
histogram of singularity exponents with a large number of bins, the
probability $p_{\alpha}$ that a singularity exponent belongs to bin
$B_{\alpha}$ can be estimated by $N_{\alpha}/N$, where $N$ is the
total (large) number of realizations of the singularity exponents, and
$N_{\alpha}$ is the number of samples falling in the bin indexed by
$\alpha$. Then, if $N$ is large enough, $N_{\alpha}$ can be estimated
by a Gaussian, and one can assign an error bar to the measurement by
setting a 99\% confidence interval. A computation using this
hypothesis leads to the following values of the error
bars \citep{Turiel2006}: 
\begin{equation}
\label{errspectre}
\Delta D({\bf h}) = \displaystyle \frac{3}{\log {\bf r}_1}\cdot \frac{1}{\sqrt{N_{\bf h}}}
\end{equation}
with $N_{\bf h}$ the number of events in the bin associated to
singularity exponent ${\bf h}$, and ${\bf r}_1$ as the minimum
resolution of the inertial range. Figure~\ref{errorbars} 
displays the singularity spectrum with its error bars of the
edge-aware filtered Musca 250 $\mu$m map with a 
parameter value $\lambda = 0.7$.

 \subsection{Inertial range and the 2D structure function method}
 \label{inertialrange}
  \begin{figure*}[h]
  \centering
  \includegraphics[width=0.30\textwidth]{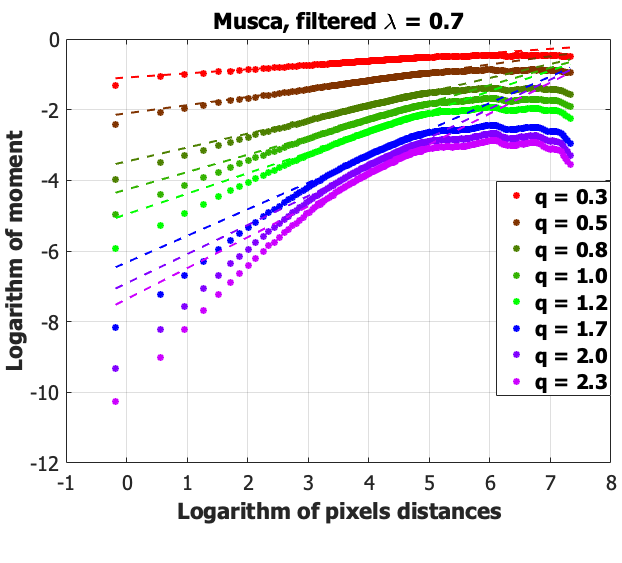}\,\includegraphics[width=0.313\textwidth]{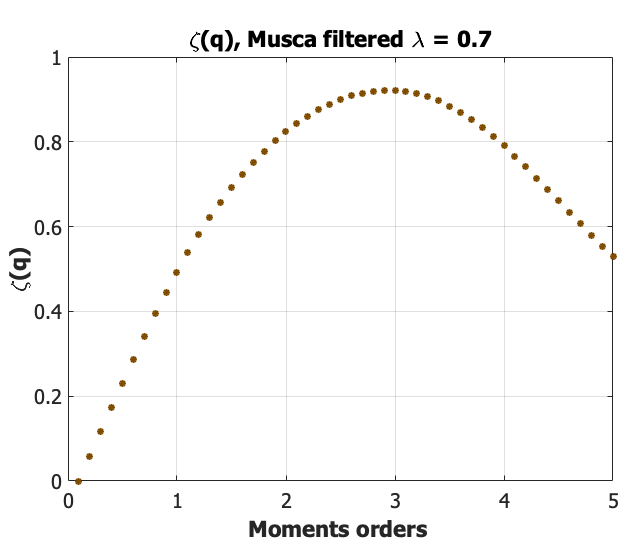}\,\includegraphics[width=0.358\textwidth]{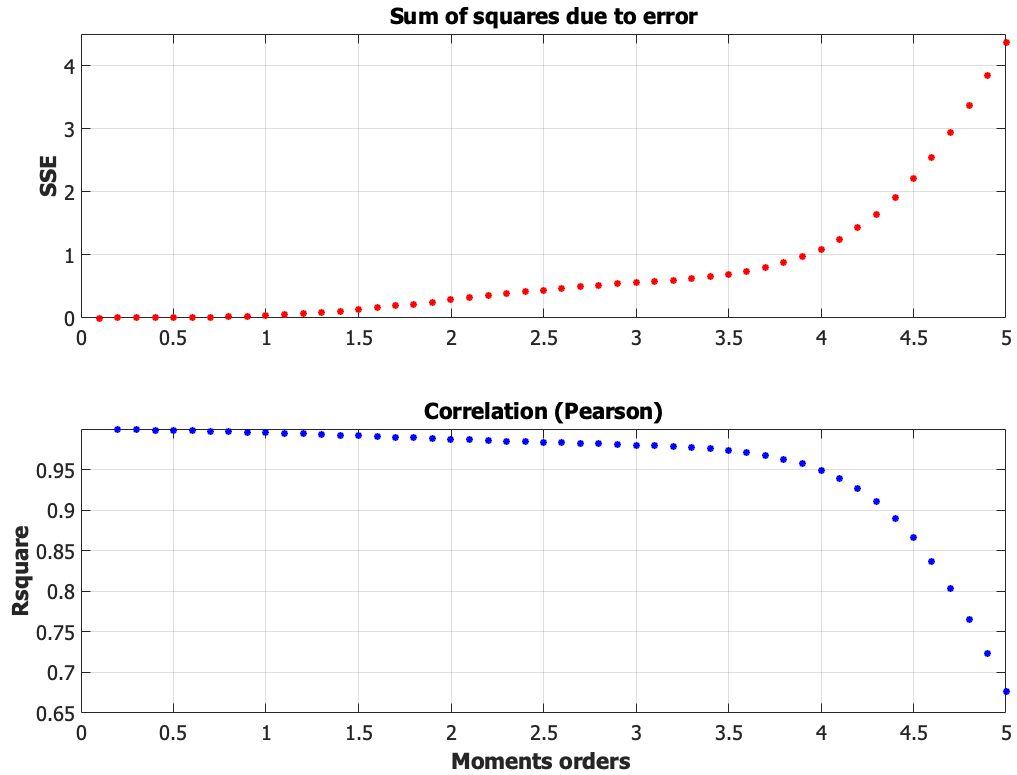} 
\caption{Result of the 2D structure function method for the
  determination of an inertial range of spatial scales. Computations are done
  on the Musca 250 $\mu$m map filtered data with $\lambda = 0.7$
  (eq.~\ref{sparsemin}). $5 \times
  10^8$ couples of points $({\bf x}_1, {\bf x}_2)$ chosen uniformly in
  the spatial domain; moments $q$ are between 0.1 and 5 with increment
  step 0.1 (only some of the moments are shown for clarity).  The left
  panel shows the graph of a $\log - \log$ plot (eq.~\ref{eqnSSF})
  with colors indicating some moment values $q$.  The image also shows the graphs of linear
  regression fits (dotted lines), performed in an inertial range
  covering the interval $[13, 160]$ pixels, corresponding to the range
  of distance $[0.053, 0.65]~\mbox{pc}$. The middle panel displays the
  resulting map $q \mapsto \zeta(q)$. The two panels on the right show
  the quality of linear regression: sse (sum of squares due to error),
  and the $r$ correlation coefficient. Hence the quality of the fit
  decreases with the moment order $q$.\label{inertial}}
\end{figure*}
  \begin{figure*}[h]
  \centering
  \includegraphics[width=0.245\textwidth]{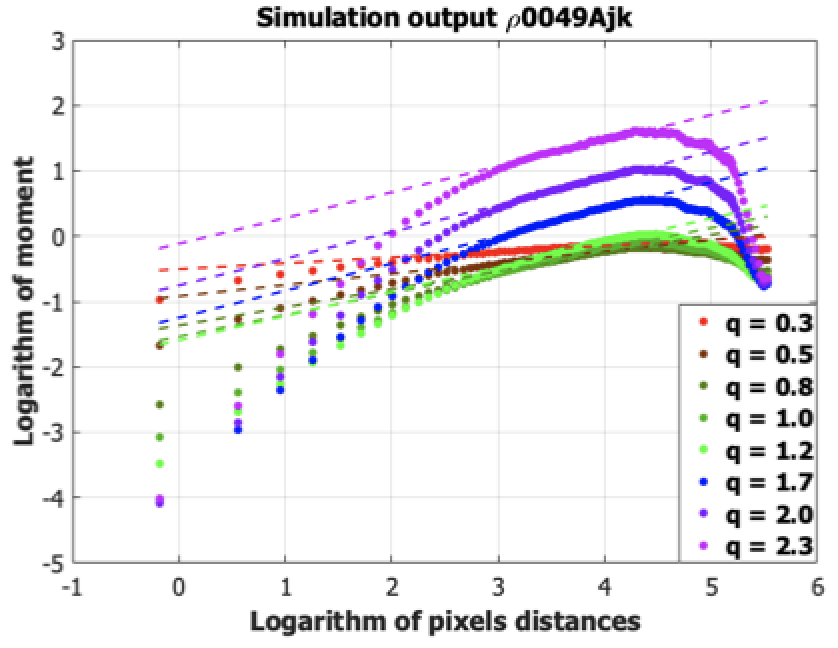}  \includegraphics[width=0.245\textwidth]{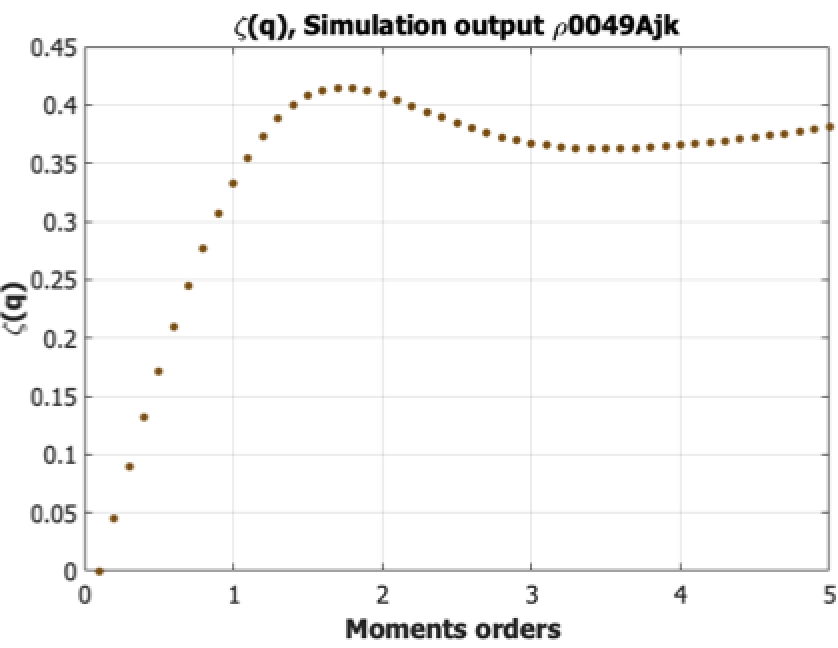} \includegraphics[width=0.245\textwidth]{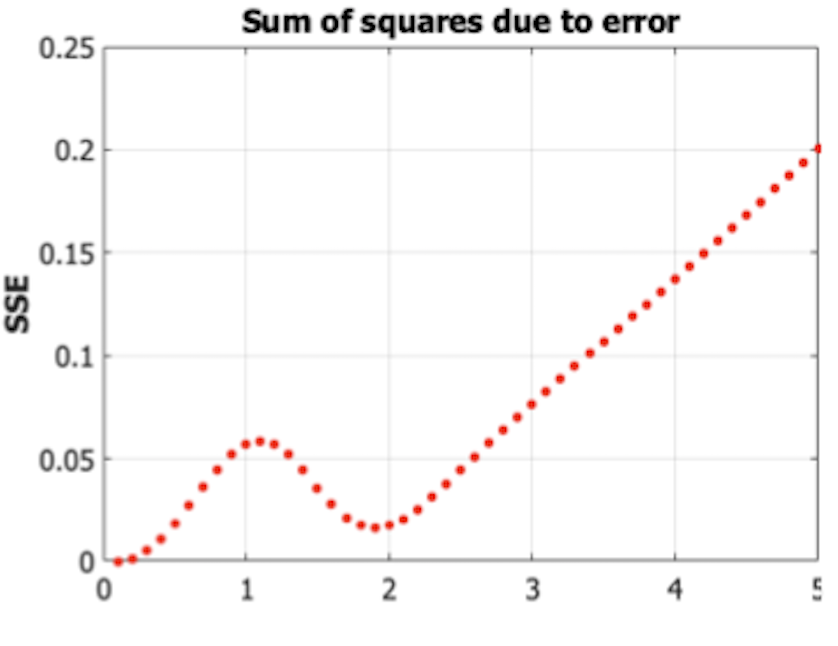}  
\includegraphics[width=0.245\textwidth]{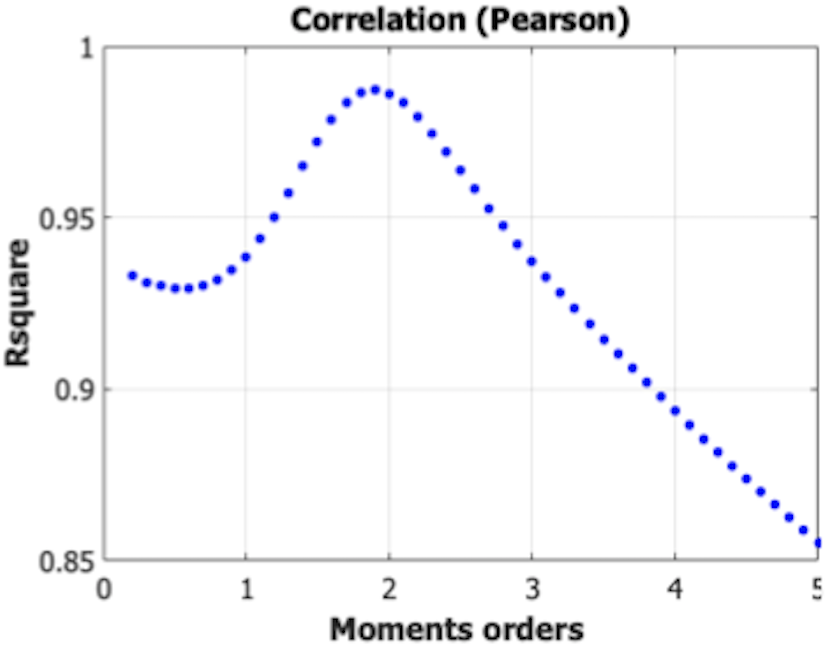} 
\caption{Result of the 2D structure function method, applied over one
  of the MHD simulation outputs presented in Sect.~\ref{simu}, for the
  determination of an inertial range of spatial scales. $5 \times 10^8$ couples of points $({\bf x}_1, {\bf x}_2)$ chosen uniformly in the spatial domain; the moments $q$ have values
  between 0.1 and 5 with increment step 0.1 (only few of the moments
  are shown for clarity).  The first left panel shows the graphs of
  the $\log - \log$ plot (eq.~\ref{eqnSSF}, with colors indicating
  some moment values $q$).  Like in Fig.~\ref{inertial}, we also draw
  the graphs of linear regression fits (dotted lines) performed in an
  inertial range covering the interval of distance $[0.18,
    1.5]~\mbox{pc}$. The second panel displays the map $q \mapsto
  \zeta(q)$. The third and fourth panels show the quality of linear
  regression: sse (sum of squares due to error), and the $r$
  correlation coefficient.\label{inertial2}}
\end{figure*}

In a canonical approach to multifractality, the determination of
scaling laws is achieved by statistical analysis of the moments of
wavelets projections of the signal over a large range of scales, and a
grand ensemble of realizations \citep{Venugopal2006b}.  As mentioned before, verifying the existence of scaling laws through
the computation on grand ensembles of realizations 
cannot be achieved with our single Musca image. 

We thus have to use a spatial approach, based on 2D structure functions,
introduced in \citet{Renosh2015} to check the existence of a significant inertial range.  If ${\bf x}_1$ and ${\bf x}_2$ are
points in the 2D signal domain, the existence of scaling laws for a
certain range of spatial distances is verified when
\begin{equation}
\label{eqnSSF}
\langle | s({\bf x}_1) - s({\bf x}_2) |^q \rangle \sim \| {\bf x}_1 - {\bf x}_2 \|^{\zeta (q)}.
\end{equation}
Consequently, the spatial moments $\langle | s({\bf x}_1) - s({\bf
  x}_2) |^q \rangle$ are $\log - \log$ plotted against the distances
$\| {\bf x}_1 - {\bf x}_2 \|$ for a very large ensemble of couples
$({\bf x}_1, {\bf x}_2)$.  We carried out experiments on both original
and edge-aware filtered data for $5\times 10^8$ couples of points
$({\bf x}_1, {\bf x}_2)$, chosen randomly in the Musca map. We also
conduct the experiment on MHD simulation outputs presented in
Sect.~\ref{simu}. In Fig.~\ref{inertial} we show the result of the
structure function method for the Musca 250 $\mu$m map
filtered data with $\lambda = 0.7$ (eq.~\ref{sparsemin}). From left to right,
the first image (top left) is the graph of a $\log - \log$ plot with colors indicating some
moment values $q$. The image also shows the graphs of linear
regression fits (dotted lines) performed in an inertial range covering
the interval $[13, 160]$ pixels, corresponding to the range of
distance $[0.053, 0.65]~\mbox{pc}$. Middle image displays the resulting map $q \mapsto \zeta(q)$.  The two images on
the right show the quality of linear regression, evaluated both with
sse (sum of squares due to error), and the classical Pearson
correlation coefficient $r$. 

We note that the derived inertial range is probably affected by the beam size ($\sim$3 pixels, 0.012 pc) for the lower boundary (0.05 pc), and by the sizes of the rectangular map (1786 pixels in the $x$ direction  i.e. 7.144 pc, 2135 pixels in the $y$ direction i.e. 8.54 pc) for the upper boundary (0.65 pc).

\section{MHD simulation data}
\label{simu}
The simulations used in this work are presented in
\citep{Dib_2007,Dib2008}.  Here, we recall their basic features. The ideal MHD
equations are solved on a uniform 3D cubic grid using a total
variation diminishing scheme (TVD), which is a second-order-accurate
upwind scheme \citep{Kim_1999}. The boundary conditions used in the
three directions are periodic. The Poisson equation is solved to
account for the self-gravity of the gas using a standard Fourier
algorithm. In order to achieve second-order accuracy in time, an
updated step of the momentum density due to the gravitational force is
implemented, as in \citep{Truelove_1998}.

\begin{figure*}[h]
  \centering
\includegraphics[width=0.35\textwidth]{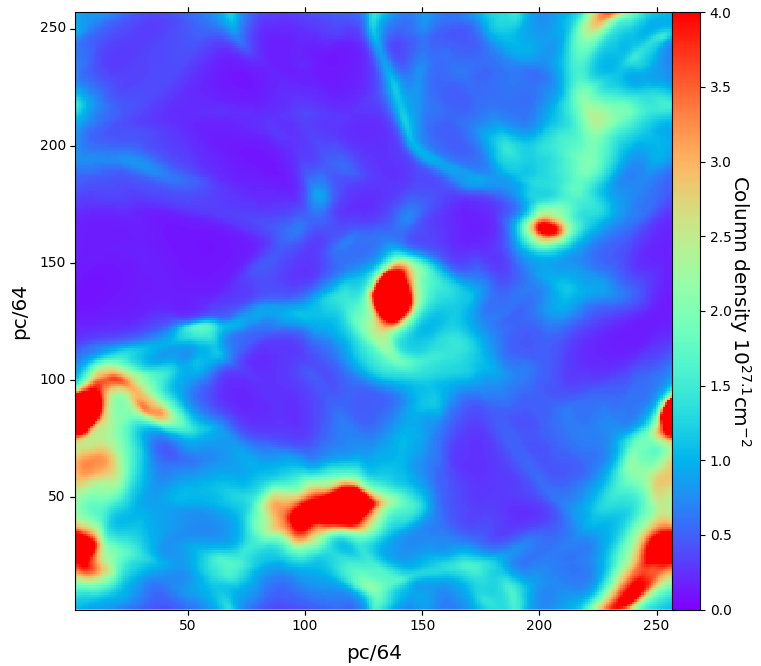}  ~~
\includegraphics[width=0.306\textwidth]{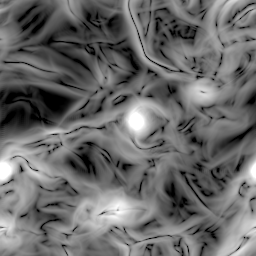}     ~~
 \includegraphics[width=0.296\textwidth]{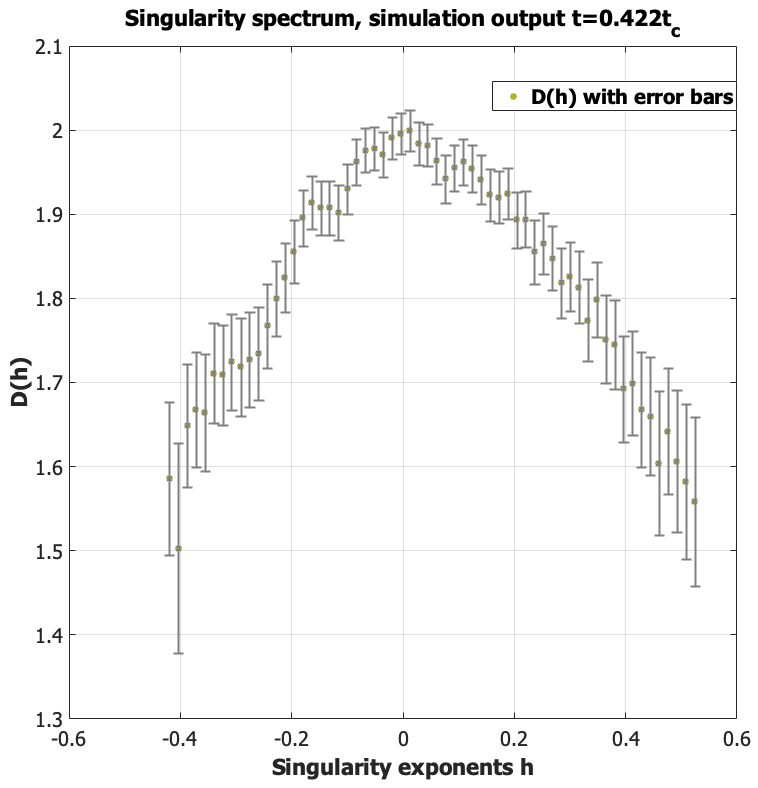} 
\caption{Left panel: one snapshot of column density output of
  the MHD simulation data used in this work, integrated along one axis
  of the cube. The output shown here corresponds to the hydrodynamical
  case with no magnetic field at time $t=0.422 \,t_{c}$. Middle panel: 
  map of the singularity exponents. Right panel: singularity spectrum with
  its error bars computed using eq.~(\ref{errspectre}).
 \label{simusdib}}
\end{figure*}

Following the method described in \citep{Stone_1998}, turbulence is
continuously driven in the simulation box and the kinetic energy input
rate is adjusted such as to maintain a constant specified ${\it rms}$
sonic Mach number $M_{s}=10$. Kinetic energy is injected at large
scales, in the wave number range $k=1-2$. When converted into physical
units, the models correspond to a box size of 4 pc and an average number
density of $500$ cm$^{-3}$. The corresponding column density is thus $\sim
5\,10^{21}$ cm$^{-2}$, which is similar to that of many molecular
clouds except for those associated with massive star formation. The 
temperature is 11.4 K, the sound speed $0.2$ km s$^{-1}$, and the 
initial {\it rms} velocity is 2 km s$^{-1}$ (i.e., therefore the
initial sonic Mach number is $M_{s}=10$). The four simulations vary by
the strength of the initial magnetic field ranging from a magnetically
subcritical cloud model to a non-magnetic cloud. The strength of the
initial magnetic field in the box for the subcritical, moderately
supercritical, and strongly supercritical magnetized models are
$B_{0}=45.8, 14.5$, and $4.6$ $\mu$G, respectively. This corresponds
to $\beta$ plasma and mass-to-magnetic flux values of the box for
these runs of, $\beta=0.01, 0.1$ and $1$, and $\mu_{box}=0.9, 2.8$,
and $8.8$, respectively. The simulations start with a uniform density
field, and are evolved for half a sound crossing time (the sound
crossing time is $t_{s}=20$ Myrs), equivalent to 5 turbulent crossing
times (the turbulent crossing time is $t_{c}=2$ Myrs), before
self-gravity is turned on. This is a common practice in such
simulations and a necessary step in order to allow for the full
development of the turbulent cascade. Left image of Fig.~\ref{simusdib} displays the integrated column density maps of
one snapshot corresponding to the hydrodynamical case with no magnetic
field at a time $t=0.422 \,t_{c}$ Myrs after gravity is turned on.

In Fig. ~\ref{inertial2} we show the result of the determination of inertial range with the 2D structure function method (Sect.~\ref{inertialrange})  applied over one MHD simulation output. We see considerable differences compared to the Musca 250 $\mu$m observational map results, notably in terms of quality of the
linear regression. The range of distances chosen for performing the
linear regression fit is $[0.18, 1.5]~\mbox{pc}$. Other outputs in the
MHD simulation show similar plots. Hence, the scaling properties of MHD simulation outputs are different compared to real observation maps.

\begin{figure}[h]
  \centering
  \includegraphics[width=0.47\textwidth]{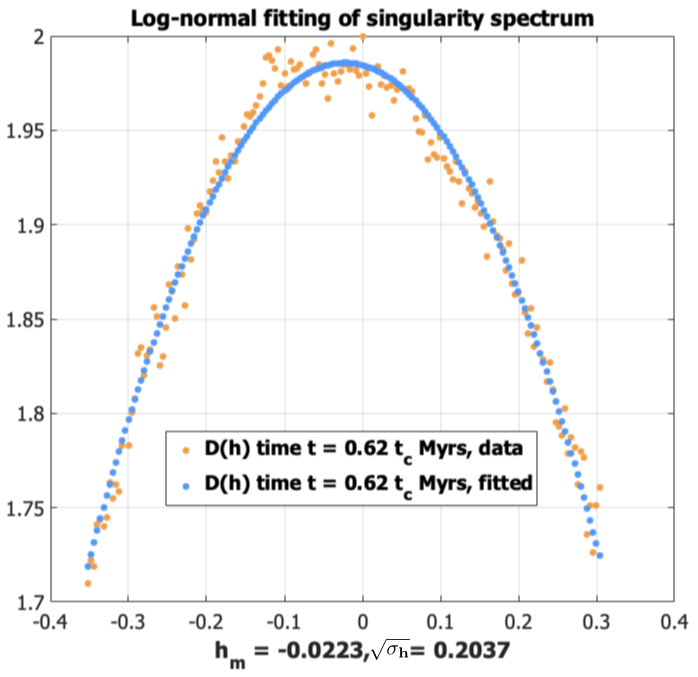}
\caption{Log-normal fit of the singularity spectrum for the data
  corresponding to time $t=0.62 t_{c}$ Myrs after gravity is turned
  on. In orange, the singularity spectrum computed in the
  microcanonical framework. In blue, a fitted quadratic $\log$-normal
  singularity spectrum. The fit is very good, which indicates a good
  approximation by a $\log$-normal process. \label{parabolicspectrum}}
\end{figure}

\begin{figure}[h]
  \centering
  \includegraphics[width=0.4\textwidth]{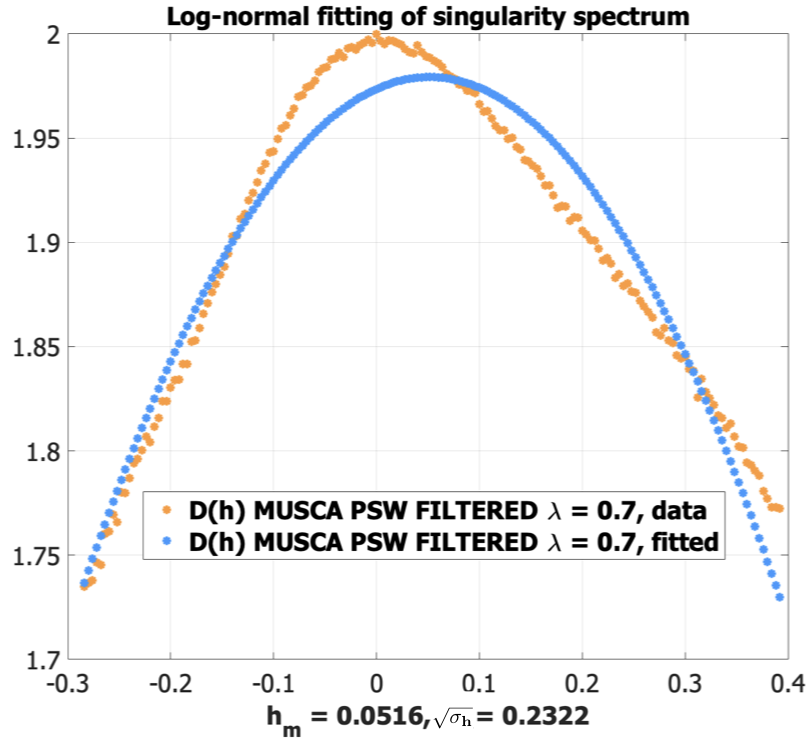} \\
   \includegraphics[width=0.4\textwidth]{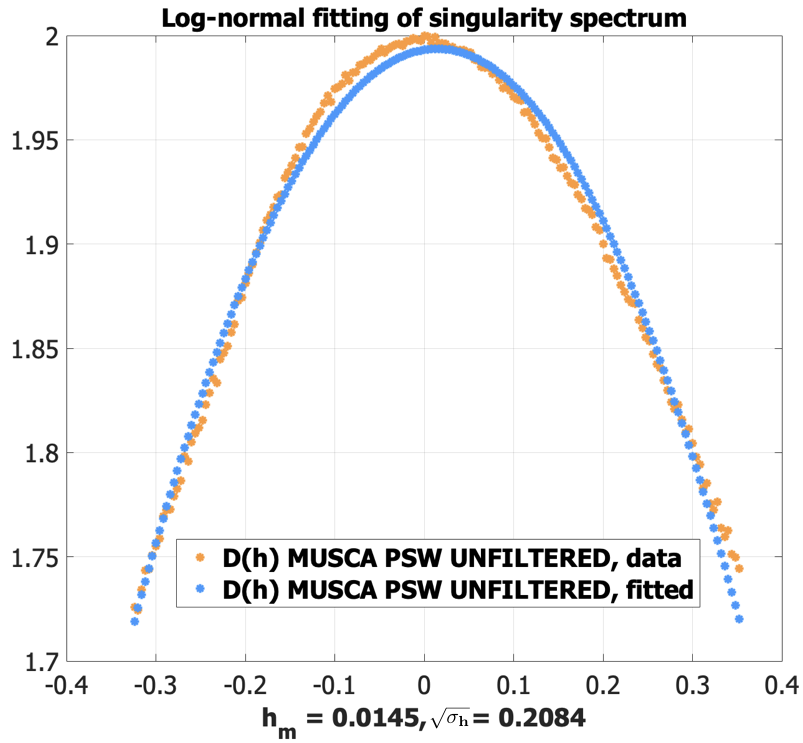}
\caption{These two figures show the results of a $\log$-normal fit for
  the singularity spectrum of the full Musca 250 $\mu$m map (also
  called 'PSW'), edge-aware filtered (top image, $\lambda = 0.7$) and
  unfiltered (bottom image). In each image, the singularity spectrum
  ${\bf h} \mapsto D({\bf h})$ is shown in orange, and the
  $\log$-normal parabola in blue. The values obtained for the
  coefficients of the fit (equation~\ref{lognormalspectrum}) are given
  in each case.  The differences in the two fits underline the
  importance of the edge-aware filtering for multifractal analysis of
  astronomical observational maps. It appears that the background
  noise "$\log$-normalizes" the data, while the filtered version,
  which better takes into account fine filamentary structures,
  deviates from $\log$-normality.  Note in particular, in the filtered
  version, the change in the orange curve around ${\bf h} =
  -0.2$. This is an indication that different types of turbulent
  processes are present in the Musca cloud, in particular those
  processes inside the main filament and the surrounding
  cloud.\label{muscafitDh}}
 \end{figure}

In the middle and right panels of Fig.~\ref{simusdib} we
also show the singularity exponents and the singularity spectrum as
described in Sect.~\ref{ss} .  Note that the error bars display more
uncertainty than those of the Musca 250 $\mu$m map displayed in
Fig.~\ref{errorbars}: this larger uncertainty arises from the low 
number of samples (the image data is of size ($256 \times 256$).  In
Fig.~\ref{parabolicspectrum}, we display both the singularity spectrum
(curve in orange) and the computation of a fitted $\log$-normal
spectrum (in blue) $D({\bf h}) = \displaystyle 2 - \frac{1}{2} \left (
\frac{{\bf h} - {\bf h}_m}{\sigma_{{\bf h}}} \right )^2$
(eq.~\ref{lognormalspectrum}) for the data corresponding to time
$t=0.62 \,t_{c}$ Myrs after gravity is turned on. The fitting is
performed using the Quasi-Newton minimization algorithm implementation
available in Matlab. We get here $ {\bf h}_m = -0.0223$, $\sigma_{{\bf
    h}} = 0.2037$.  Hence, in this case $D({\bf h}) = 2 - \displaystyle
\frac{({\bf h}+ 0.0223)^2}{0.0829}$. The fit is very good, which
indicates a good approximation by a $\log$-normal process. This good
fit was observed for all our simulation outputs. In comparison,
looking at Figs.~\ref{errorbars} and~\ref{muscafitDh} we conclude that
the singularity spectrum of the Musca 250 $\mu$m map does not
fit so well with a $\log$-normal process. We will discuss this point 
in the next sections.

 
\section{Multifractal analysis of data} \label{results}
\subsection{Detection of a multiplicative cascade}  \label{result-b}
\begin{figure*}[h]
  \centering
  \includegraphics[width=0.41\textwidth]{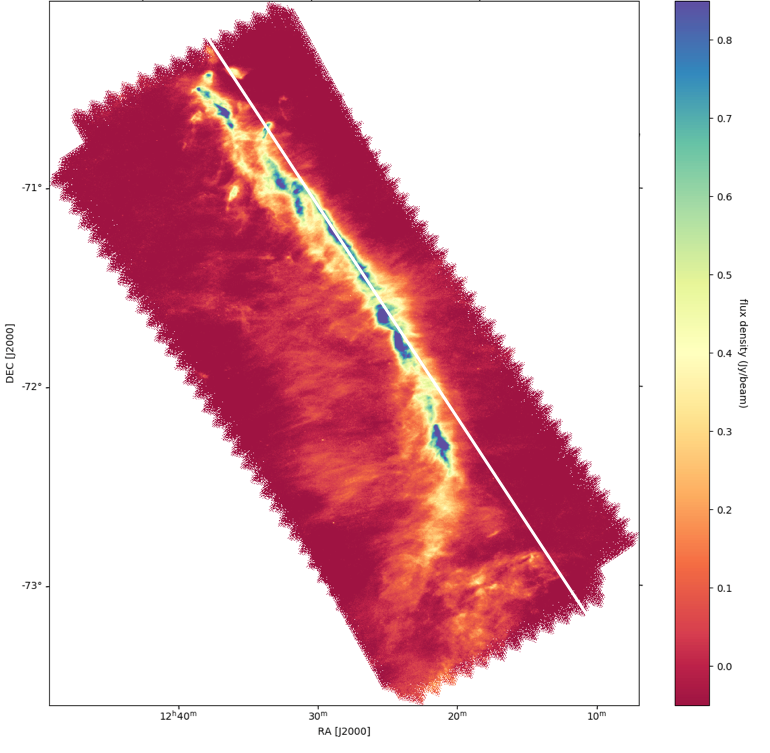}  \includegraphics[width=0.58\textwidth]{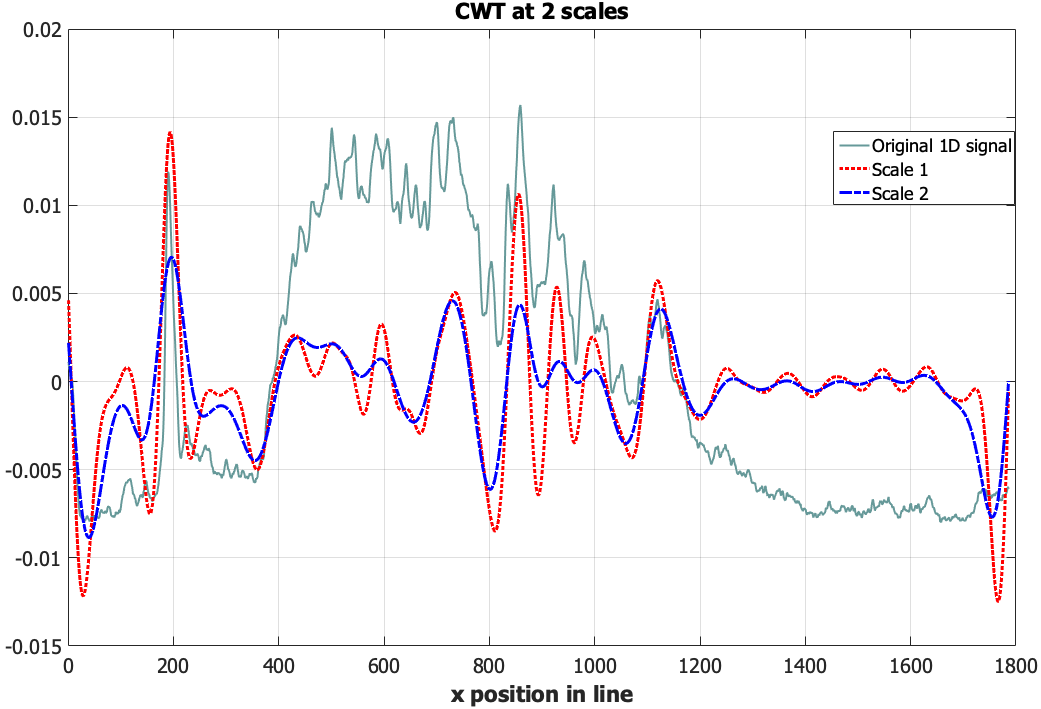}
  \caption{Left panel: Musca 250 $\mu$m map in which the white column
    defines a 1D signal. Right panel: original 1D signal (centered and
    rescaled) and CWT projections of the selected 1D signal with the
    Mexican hat wavelet respectively at scales 5 (red) and 60
    (blue).\label{projcol888}}
\end{figure*}

To investigate the existence of a multiplicative cascade in the Musca
map, we use the cumulant approach described in
Appendix~\ref{cumulant}. The cumulant analysis is performed over 1D signals which are extracted from columns of a 2D observational map  as shown in the
left panel of Fig.~\ref{projcol888}. In this figure we display, as an example,  a 1D signal that comes from a particular column in the Musca observation map; that  column crosses low and high flux regimes, i.e. regions outside and
inside the main filament. We perform the cumulant analysis
using the mexican hat as analyzing wavelet:
\begin{equation}
\label{defmh}
\psi({\bf x} ) = \displaystyle \frac{2}{\pi^{1/4} \sqrt{3 \sigma}}  \left ( \frac{{\bf x^2}}{\sigma^2} - 1 \right ) e^{\frac{-{\bf x}^2}{2 \sigma^2}}
\end{equation}
(second derivative of a gaussian) with a deviation $\sigma = 3.2$
chosen for the data. If $f({\bf x})$ is a 1D signal, the continuous
wavelet projection (CWT) of $f$ at scale ${\bf r} > 0$ is
\begin{equation}
\label{defcwt}
{\cal T}_{\psi}(f)({\bf x},{\bf r}) = \displaystyle \frac{1}{{\bf r}}\int_{\mathbb{R}} f({\bf u})\psi \left ( \displaystyle \frac{{\bf x} - {\bf u}}{{\bf r}} \right ) \, \mbox{d}{\bf u}. 
\end{equation}
Note that we use $L^1$ normalisation instead of $L^2$: to study
correlations, the conservation of energy is not necessary, while $L^1$
normalisation better adapts to strong variations in the
signal \citep{Venugopal2006b}. We use the {\sl Herschel} 250
$\mu$m Musca dataset filtered with $\lambda = 0.7$
(eq.~\ref{sparsemin} in Sect.~\ref{eaf}). Figure~\ref{projcol888} 
shows a selected column in the Musca map and the associated 1D signal
$f({\bf x})$ with two of its wavelet projections ${\cal
  T}_{\psi}(f)({\bf x},{\bf r})$ at two respective scales ${\bf r}_1$
and ${\bf r}_2$ with ${\bf r}_2 < {\bf r}_1$ (graphs in red and blue
respectively).
\begin{figure*}[h]
\centering
\includegraphics[width=0.5\textwidth]{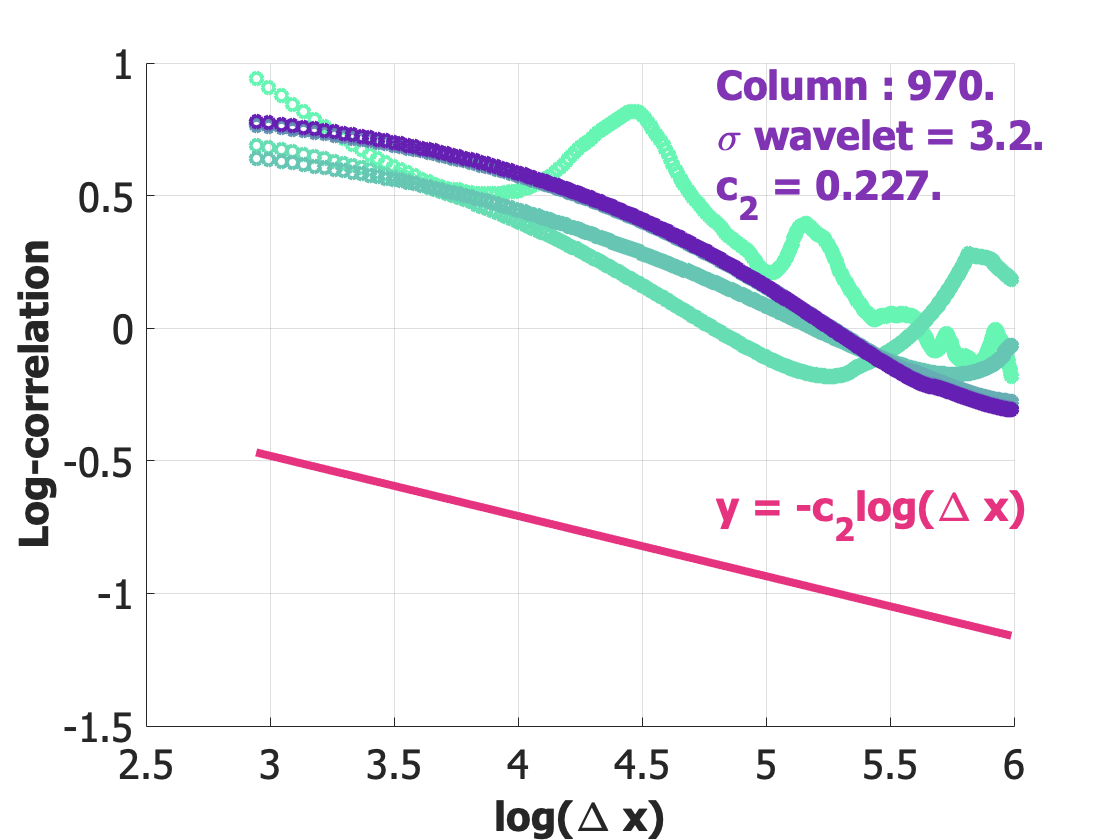}~\includegraphics[width=0.5\textwidth]{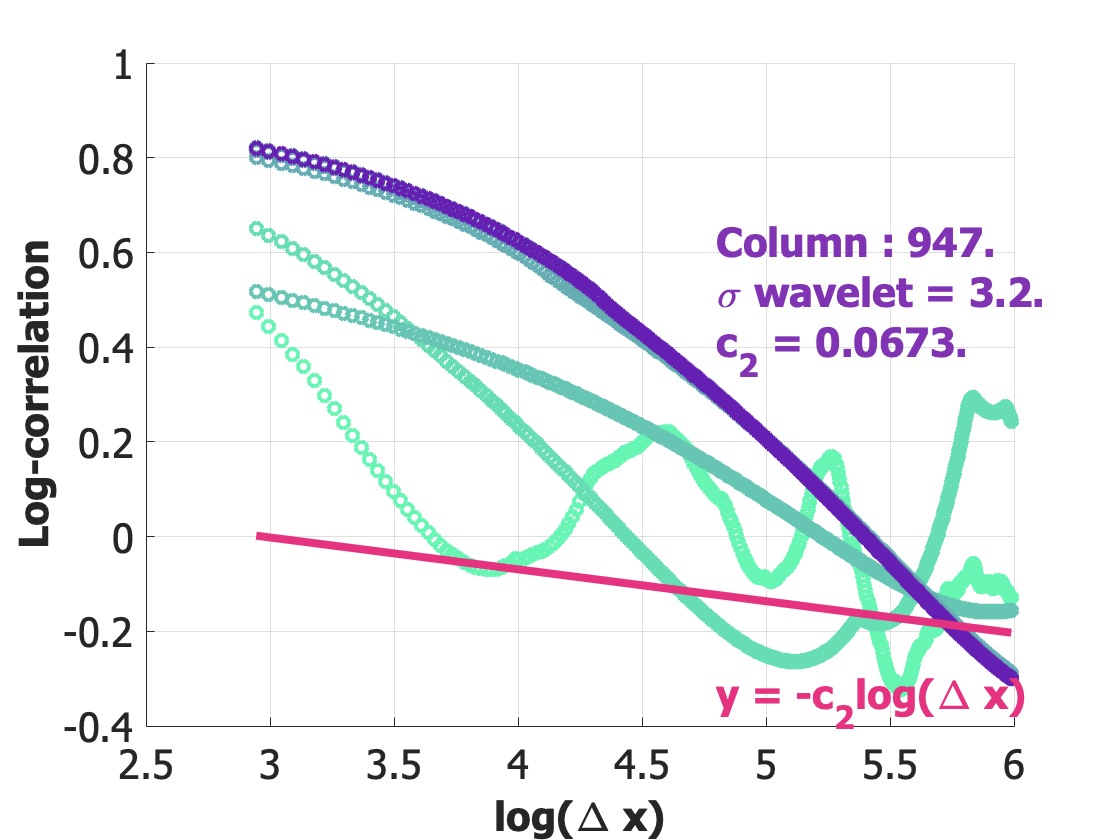}
\caption{Plots of the $\log$-correlations ${\cal C}({\bf r},\Delta
  {\bf x})$ vs $\log (\Delta {\bf x})$ for two 1D signals extracted
  from the Musca 250 $\mu$m map according to the scheme
  shown in Fig.~\ref{projcol888}. The analyzing wavelet is given by
  equation~\ref{defcwt} with $\sigma = 3.2$. We use 500 scales from
  ${\bf r} = 1$ to ${\bf r} = 500$ and 400 values for $\Delta x$
  between 1 and 400. The chosen values for the scales are enough to
  highlight more than 2 nodes in a cascade, if such a cascade
  exists. Each image shows the graphs of the log-correlations for few
  scales (curves from green to purple corresponding to increasing
  scales), and the line $y = -c_2\log (\Delta x)$ with the slope $c_2$
  computed through linear regression, according to
  eq.~\ref{calcul-cumulants2}. The linear regression used to compute
  $c_2$ is performed for scales between 10 and 200. The left panel
  shows the $\log$-correlations corresponding to column 970 in the
  Musca observational map. We see that the graphs of ${\cal C}({\bf
    r},\Delta {\bf x})$ for different scales follow the slope given by
  the line $y = -c_2\log (\Delta x)$. The right panel, corresponding to column 947, displays a
  different behavior: we observe long-range correlations, but the
  behavior seen in a multiplicative cascade is not
  present.\label{logcorr}}
\end{figure*}

\begin{figure*}[h]
  \centering
  \includegraphics[width=0.49\textwidth]{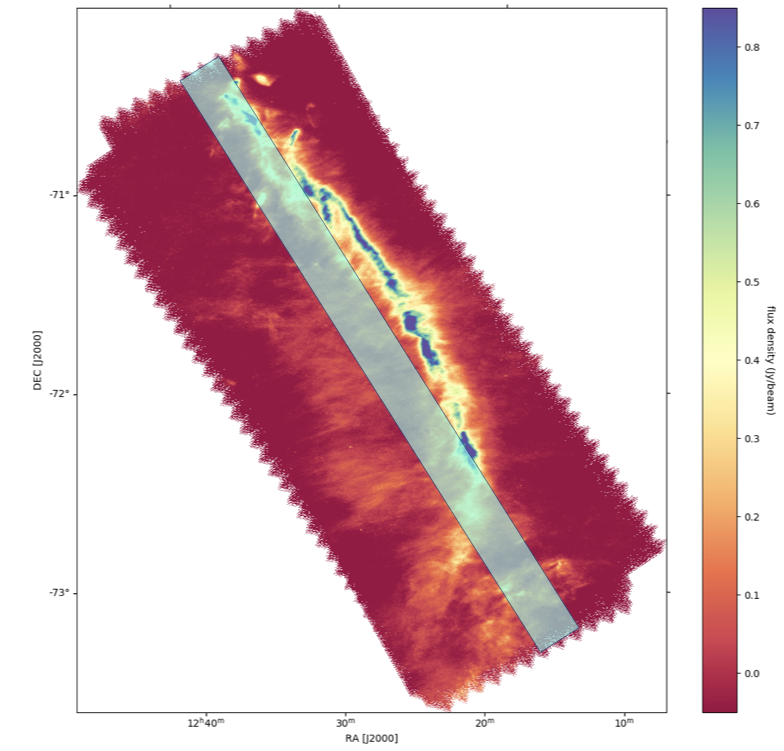} \includegraphics[width=0.48\textwidth]{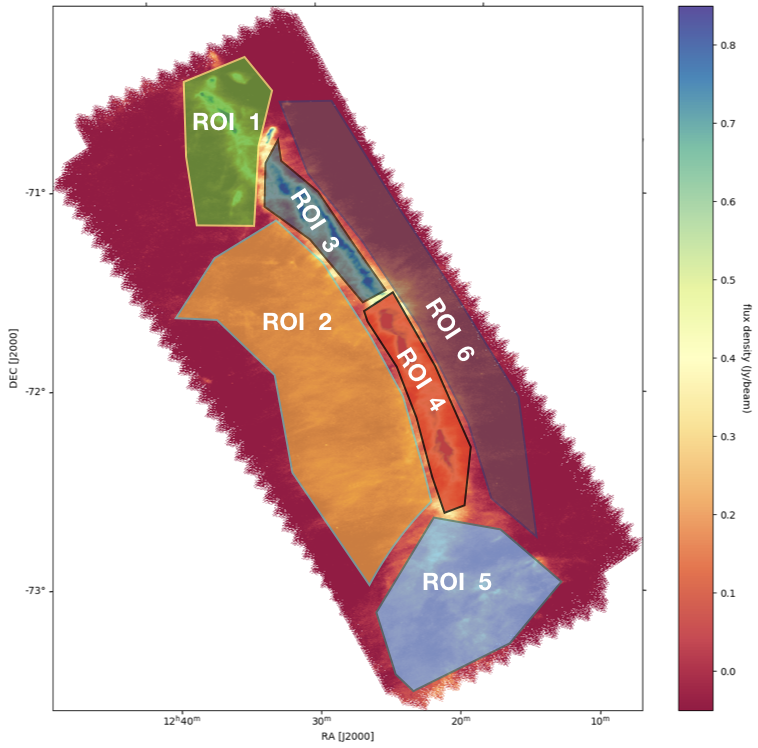}
\caption{Left: in the rectangular area shown in the picture, the
  statistics of the $\log$-correlations are different than in other
  parts of Musca). Right: definition of 6 ROIs (regions of interest)
  inside Musca ISM.\label{stat-corr-pb}}
\end{figure*}

We apply this cumulant analysis on each 1D signal (columns) by computing
the $\log$-correlations ${\cal C}({\bf r},\Delta {\bf x})$
(eq.~\ref{deflogcorr}) and plot them against $\log (\Delta {\bf
  x})$. As explained in Appendix~\ref{cumulant}, we look for
fingerprints of long-range correlations and the existence of a
multiplicative cascade according to ${\cal C}({\bf r}, \Delta {\bf x})
\sim \log \Delta {\bf x}$ and eq.~\ref{verif-cascade}. We use 500
scales from ${\bf r} = 1$ to ${\bf r} = 500$ and 400 values for
$\Delta x$ between 1 and 400. The slope $c_2$ is computed through
linear regression, according to eq.~\ref{calcul-cumulants2}. The
linear regression used to compute $c_2$ is performed for scales
between 10 and 200. Our experiment shows that long-range correlations
are found for all 1D columns, and that a multiplicative cascade does
exist in the Musca cloud. The left panel in Fig.~\ref{logcorr} displays an
example, corresponding to column 970 in the Musca map, of a typical
graph obtained in Musca: the $\log$-correlations corresponding to
different scales decrease in accordance with the equation of the line $y
= -c_2\log (\Delta x)$ for the computed $c_2$ value, which is an
indication for the presence of a multiplicative cascade in the
column. However, there exists an interval of columns, corresponding to
the rectangular area shown in Fig.~\ref{stat-corr-pb} (left), for
which the curves behave as shown in the right panel of
Fig.~\ref{logcorr}. This does not imply the absence of a
multiplicative cascade, but there is an area, inside this region,
which brings perturbations to the statistics; this is also an
indication that different processes are at work inside the Musca
cloud.
\begin{figure*}[h]
\centering
\includegraphics[width=0.5\textwidth]{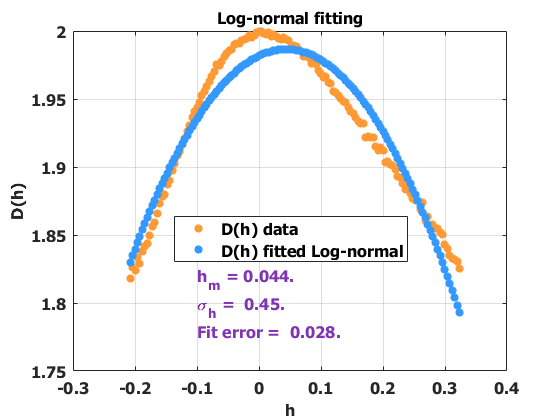}\includegraphics[width=0.5\textwidth]{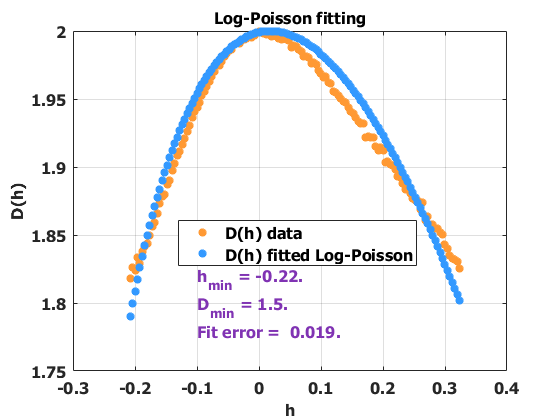}
\caption{Left: the singularity spectrum of the filtered ($\lambda =
  0.7$) Musca 250 $\mu$m observational map (in orange) is fitted
  against a $\log$-normal parabolic spectrum (blue) defined by
  equation~(\ref{lognormalspectrum}). The fit is performed using the
  Quasi-Newton minimization algorithm implementation available in
  Matlab. Right: the singularity spectrum of the of the filtered
  ($\lambda = 0.7$) Musca 250 $\mu$m observational map (in orange) is
  fitted against a $\log$-Poisson singularity spectrum (blue) defined
  by equation~(\ref{logpoissonspectrum}). Same minimization
  algorithm. Both images display the values found by minimization
  algorithm and the error
  fit.\label{fitlognormallogppoissonmuscafiltered}}
\end{figure*}
\begin{figure*}[h]
\centering
\includegraphics[width=0.35\textwidth]{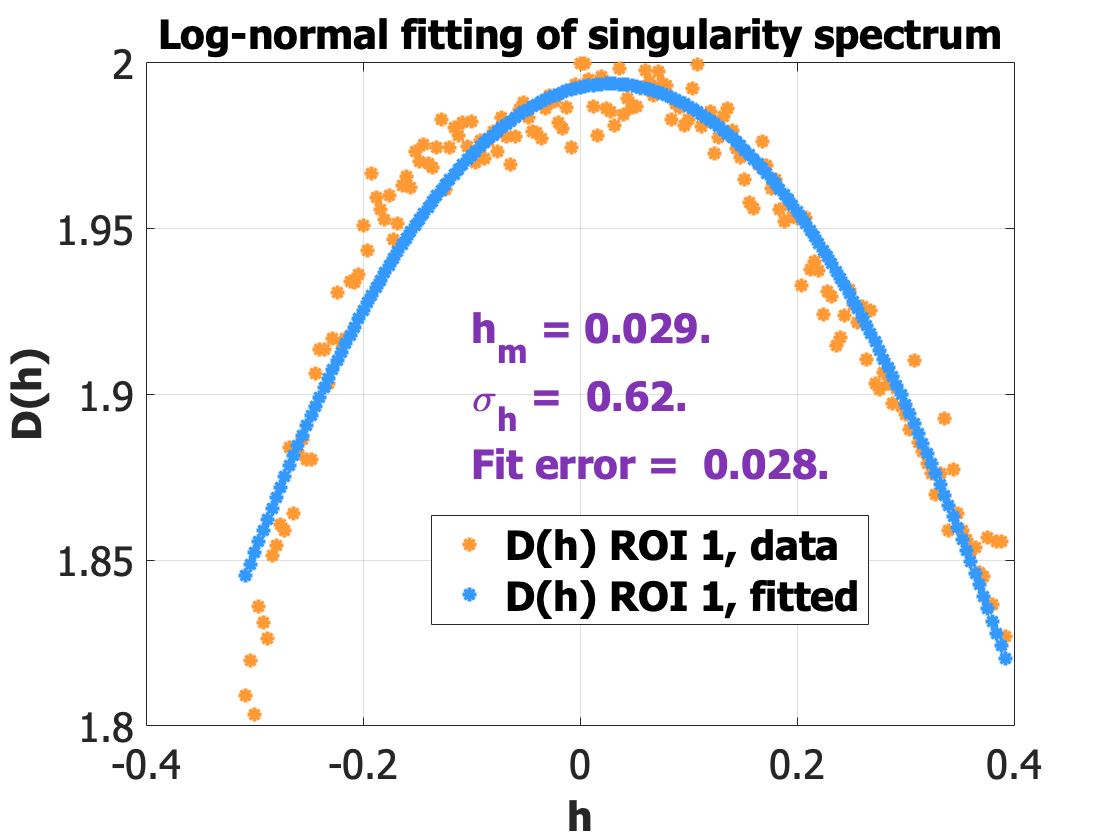}~\includegraphics[width=0.35\textwidth]{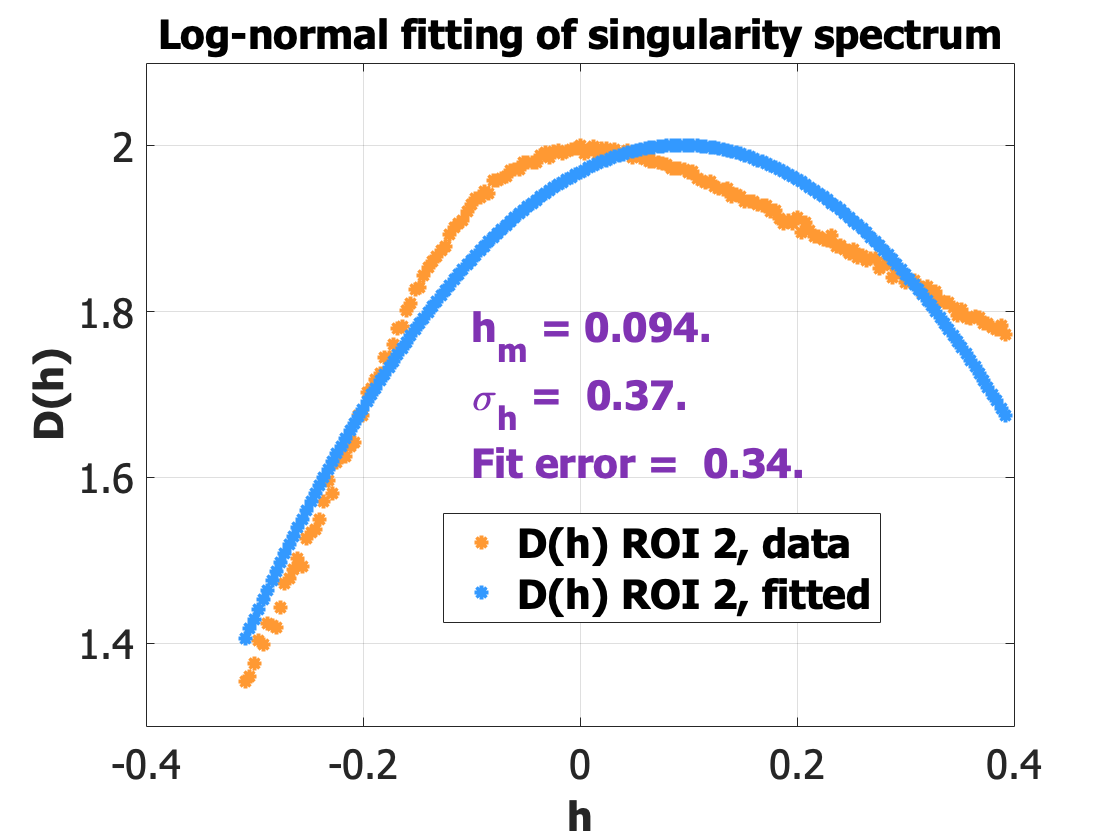}~\includegraphics[width=0.35\textwidth]
{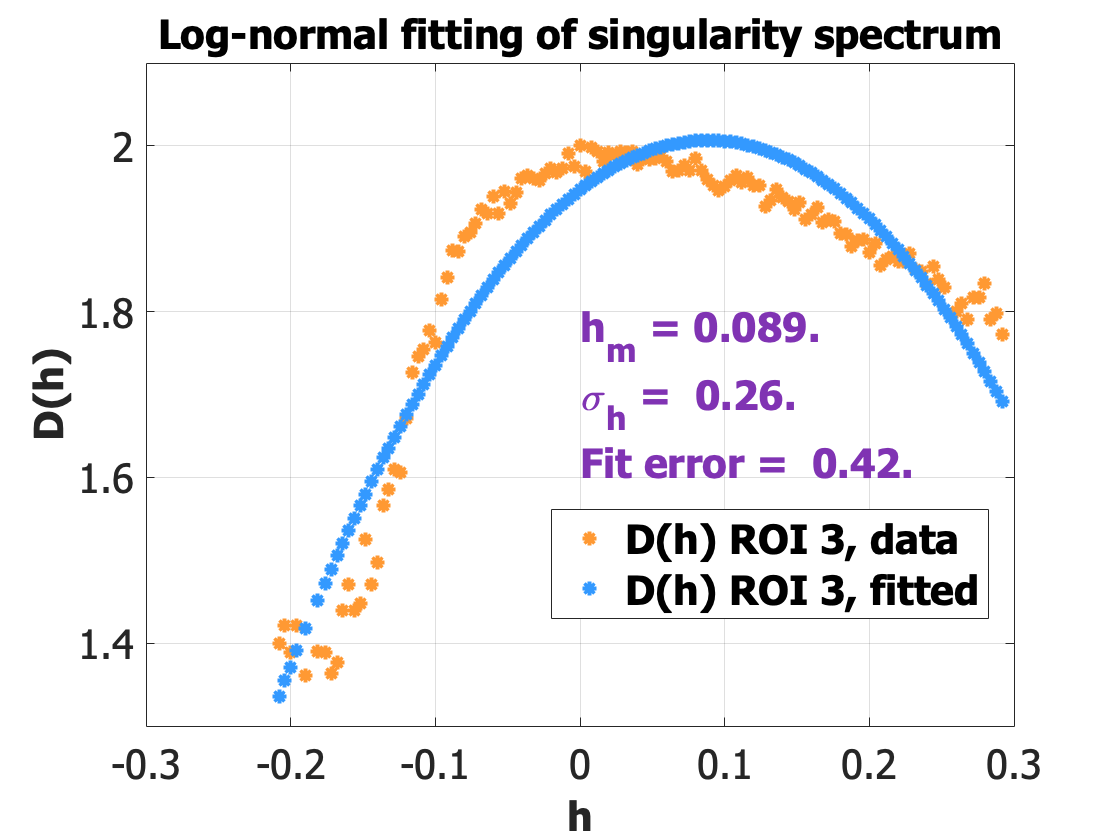} \\
\includegraphics[width=0.35\textwidth]{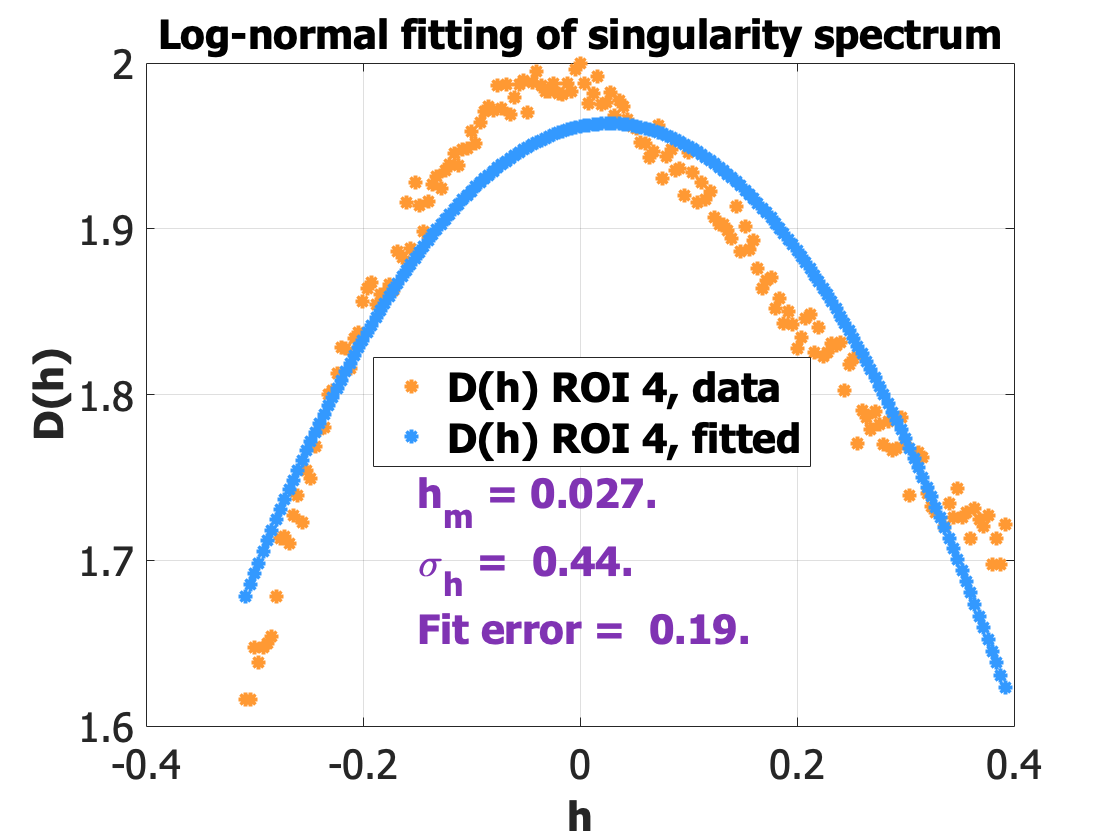}~\includegraphics[width=0.35\textwidth]{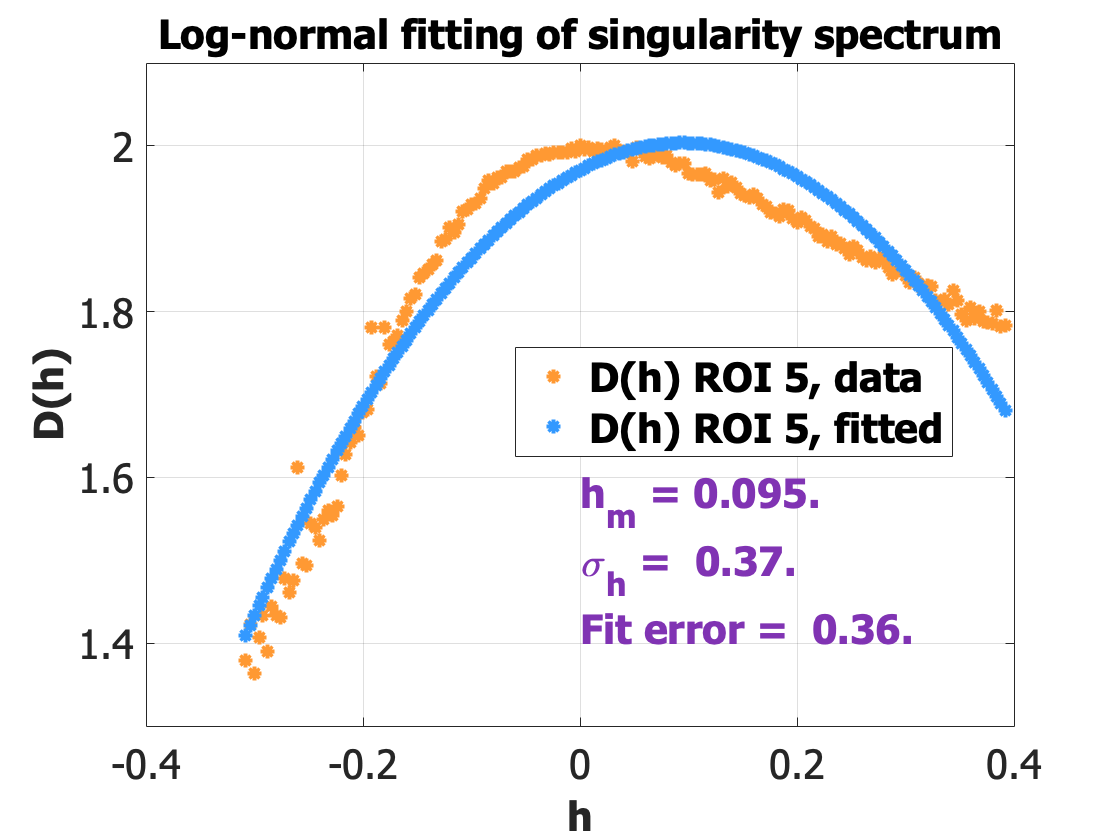}~\includegraphics[width=0.35\textwidth]
{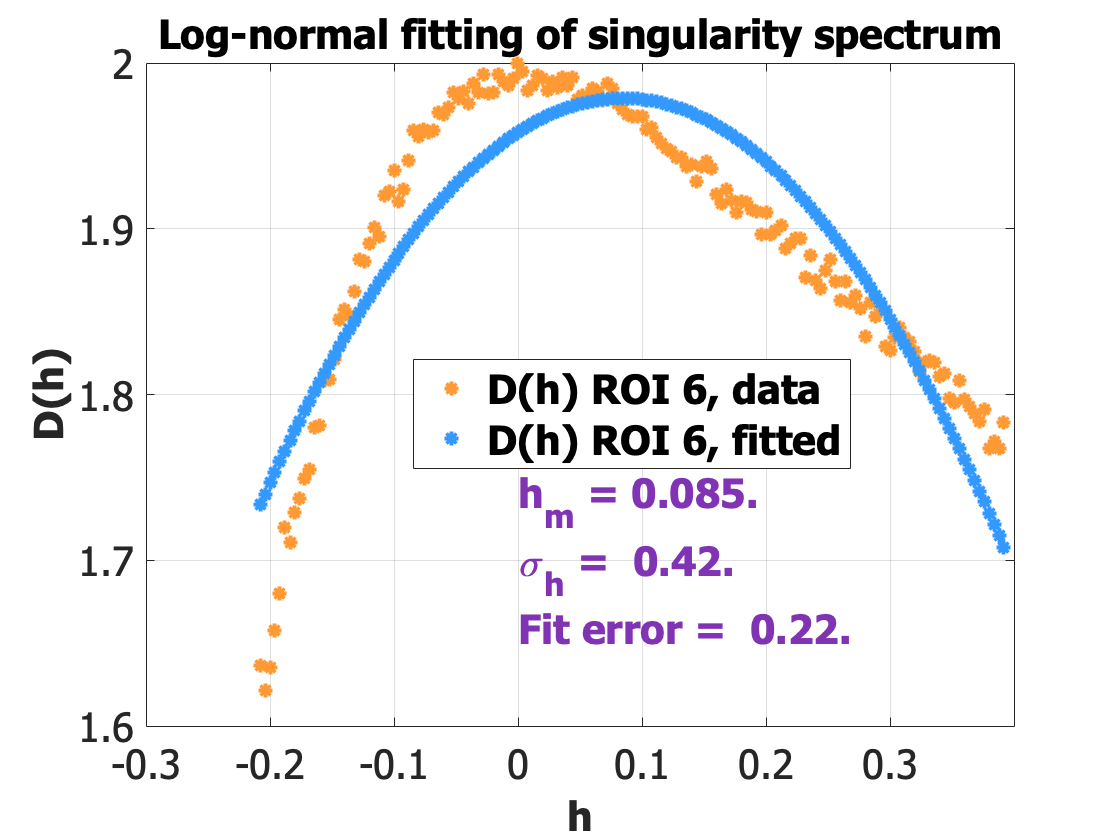}
\caption{The images display, for each region of interest in the Musca
  cloud delimited in Fig.~\ref{stat-corr-pb}: the singularity
  spectrum ${\bf h} \mapsto D({\bf h})$ of a ROI in orange color, a
  $\log$-normal fitted approximation singularity spectrum in blue
  (equation~\ref{lognormalspectrum}), the obtained numerical values
  ${\bf h}_m$ and $\sigma_{{\bf h}}$, the error of the fit. The
  various values of ${\bf h}_m$ and $\sigma_{{\bf h}}$ for each ROI
  are also given in table~\ref{tablerois}\label{spectraroisfitted}}
\end{figure*}

\subsection{Musca spectrum: multifractality, $\log$-normality} \label{result-a}

Our final singularity spectrum obtained for the whole Musca region is displayed in Fig.~\ref{errorbars}. In contrast to previously obtained singularity spectra for the ISM~(e.g. \citet{Khalil2006,Elia2018}), this improved spectrum appears not very symmetrical with a clearly steep, almost linear, descend in the ${\bf h} \leq 0$ side (left) and a more round, gaussian like shape for the ${\bf h} \geq 0$ side (right). We note that this clear asymmetric shape has been enhanced and well revealed thanks to the noise filtering (see Fig.~\ref{muscafiltering}). Despite this clear asymmetry, we fitted in Figure~\ref{fitlognormallogppoissonmuscafiltered}-left the obtained spectrum with a parabolic curve which would represent a log-normal behavior of the turbulent fluctuations. We find, ${\bf h}_m =
0.0206$, $\sigma_{{\bf h}} = 0.4564$. The part of the singularity
spectrum corresponding to ${\bf h} \leq 0$ shows a clear deviation between
the spectrum computed in the microcanonical formalism (orange) and the
parabolic (blue), which seems to strongly point to some non $\log$-normal process in the turbulent flow of Musca.
We note that there is actually more resemblance with a
$\log$-Poisson spectrum as shown by the fit displayed in Fig.~\ref{fitlognormallogppoissonmuscafiltered}-right. A $\log$-Poisson spectrum is indeed asymmetric with a steep descend in the negative ${\bf h}$ side (see Fig.~\ref{genericss}).
These fit results can be compared with the bottom panel of
Fig.~\ref{muscafitDh}\footnote{The slight difference in ${\bf h}_m$
  values between top image of Fig.~\ref{muscafitDh} and left image of
  Fig.~\ref{fitlognormallogppoissonmuscafiltered} comes from small
  differences in initial ${\bf h}$ value intervals chosen: $[-0.3,
    0.4]$ in Fig.~\ref{muscafitDh} and $[-0.3, 0.33]$ in
  Fig.~\ref{fitlognormallogppoissonmuscafiltered}.}, which corresponds
to the non-filtered case. Given that the noise filtering enhances mostly the contrast of filamentary structures (which were otherwise merged in the noise data), it suggests that the asymmetry in the singularity spectrum and the $\log$-Poisson behavior are related to 1D structures, i.e. filaments as expected for the most singular structures in a turbulent cascade with intermittency (see~\citep{She1990226} and Sect.~\ref{subsec:lognormallogpoisson} below for a discussion). 
In the next section, we will see that some regions of Musca are "more $\log$-normal" than others.

Furthermore, in addition to the possible proxy of the whole spectrum by a $\log$-Poisson curve, by a close inspection of Fig.~\ref{errorbars}, 
we note around ${\bf h } = -0.2$ a slight non-convex area in
the graph. This observation can be made only on the edge-aware
filtered signal, because the background noise "smoothes" the spectrum
as seen in the orange graph of
Fig.~\ref{muscafiltering}. Consequently, the singularity spectrum in
this case seems to present two local extrema. This might be an indication that at
least two different processes can be present in the signal.

\begin{table}[h!]
\centering
\begin{tabular}{||c c c c c c||} 
 \hline
   ROI  &  ${\bf h}_m$ & $\sigma_{{\bf h}}$ & Error & $[{\bf h}_1,{\bf h}_2]$ & Std dev.\\ [0.5ex] 
 \hline\hline

   ROI1 & 0.029& 0.62 & 0.028 &  $[-0.3,0.4]$ & 0.057\\
   \hline
   ROI2 & 0.094& 0.37 & 0.34 &  $[-0.3,0.4]$ & 0.16 \\
   \hline
	 ROI3 &  0.089 & 0.26 & 0.42 &  $[-0.2,0.3]$ & 0.18 \\
   \hline
	 ROI4 &  0.027 & 0.44 & 0.19 &  $[-0.3,0.4]$ & 0.1 \\
   \hline
	 ROI5 &  0.095 & 0.37 & 0.36 &  $[-0.3,0.4]$ & 0.16 \\
   \hline
	 ROI6 &  0.085 & 0.42 & 0.22 &  $[-0.2,0.4]$ & 0.087\\
   \hline
	\end{tabular}
\caption{Parameters of fitted $\log$-normal process for each ROI:
  column 1: ROI, column 2: ${\bf h}_m = \langle {\bf h} \rangle$
  average of singular values, column 3: $\sigma_{{\bf h}}$ singularity
  dispersion (see equation~\ref{lognormalspectrum}), column 4: fit
  error = $\| D({\bf h}) - D_{\mbox{logn-fit}}({\bf h}) \|_2^2$,
  column 5: interval $[{\bf h}_1,{\bf h}_2]$ chosen according to error
  bars of singular values in which the statistics ${\bf h}_m$,
  $\sigma_{{\bf h}}$, the standard deviation and the fit error are
  computed, column 6: standard deviation of each ROI singularity
  spectrum within the specified interval of ${\bf h}$ values shown in
  column 5.\label{tablerois}}
\end{table}

\subsection{Distinct statistical properties observed in the data}
\label{rois}

To go one step further, we now compare singularity spectra computed
over different regions of interest (ROI) to reveal potential different
turbulent behaviors inside the cloud. Figure~\ref{stat-corr-pb} shows
the geometric location of 6 ROIs inside the Musca cloud that were
selected based on different physical properties.  ROI1 is the northern
end of the filament where a single protostellar object is
located. ROI2 represents the eastern cloud area with prominent
striations (weak filamentary structures perpendicular to the main crest of Musca filament). ROI3 and ROI4 are the highest density crest regions. We define two ROIs on the crest with ROI4 being the southern part with already signs of fragmentation while in ROI3 the northern part, the crest is still very much homogeneous. ROI5
is the southern end of the filament which is less organized in
filaments or striations. ROI6 is the western part of the embedding cloud with less
prominent striations. For each ROI, we compute: the singularity
spectrum and the values ${\bf h}_m$ and $\sigma_{{\bf h}}$ of a fitted
$\log$-normal spectrum $D_{\mbox{logn-fit}}({\bf h}) = \displaystyle 2
- \frac{1}{2} \left ( \frac{{\bf h} - {\bf h}_m}{\sigma_{{\bf h}}}
\right )^2$. The quality of the fit is estimated by the $L^2$ error
$\| D({\bf h}) - D_{\mbox{logn-fit}}({\bf h})\|_2^2$. The best defined
curves are the ones with the highest number of pixels, i.e. ROI2 and
ROI5. Figure~\ref{spectraroisfitted} and Table~\ref{tablerois} show
the results of the computations. The obtained graphs and values
confirm that the processes inside each ROI can display strong
deviation from a $\log$-normal process.
First, it becomes obvious that we can distinguish two classes of
curves, the ones that are not too far from a $\log$-normal fit, ROI1
and - to a lesser extent - ROI4 (with respective fit errors 0.028 and 0.19), and the remaining ones that deviate
significantly from $\log$-normality. ROI1 and ROI4 have the largest
singularity dispersions $\sigma_{{\bf h}}$ of a fitted $\log$-normal
with $\sigma_{{\bf h}}$ = 0.62 and 0.44, respectively. Moreover, ROI1 and ROI4 feature the most symmetric singularity spectra among the ROIs. The other ROI's
all show a steeper slope at negative {\bf h} values than at positive
ones and ROI2 features the worst $\log$-normal fit among the 6 ROIs; this
result that can be put in relation with, for one part, the large density of striations seen in ROI2, and for another part with Fig.~\ref{stat-corr-pb} and
the discussion in Sect.~\ref{result-b}: it is likely that the part of
rectangle shown in Fig.~\ref{stat-corr-pb} that intersects ROI2 is
responsible for deviation also observed in the $\log$-correlations
${\cal C}({\bf r},\Delta {\bf x})$.  For all ROI's except of ROI1 and
ROI4, the singularity spectrum is more of the $\log$-Poisson type than
$\log$-normal. This finding also applies to the singularity spectrum
of the whole (edge-aware filtered) Musca observational map as can be
seen in Figs.~\ref{errorbars},~\ref{muscafitDh}, and~\ref{fitlognormallogppoissonmuscafiltered}. 

 \subsection{Application on data from simulations}
 \label{result-3}
 \begin{figure}[h]
\centering
\includegraphics[width=0.5\textwidth]{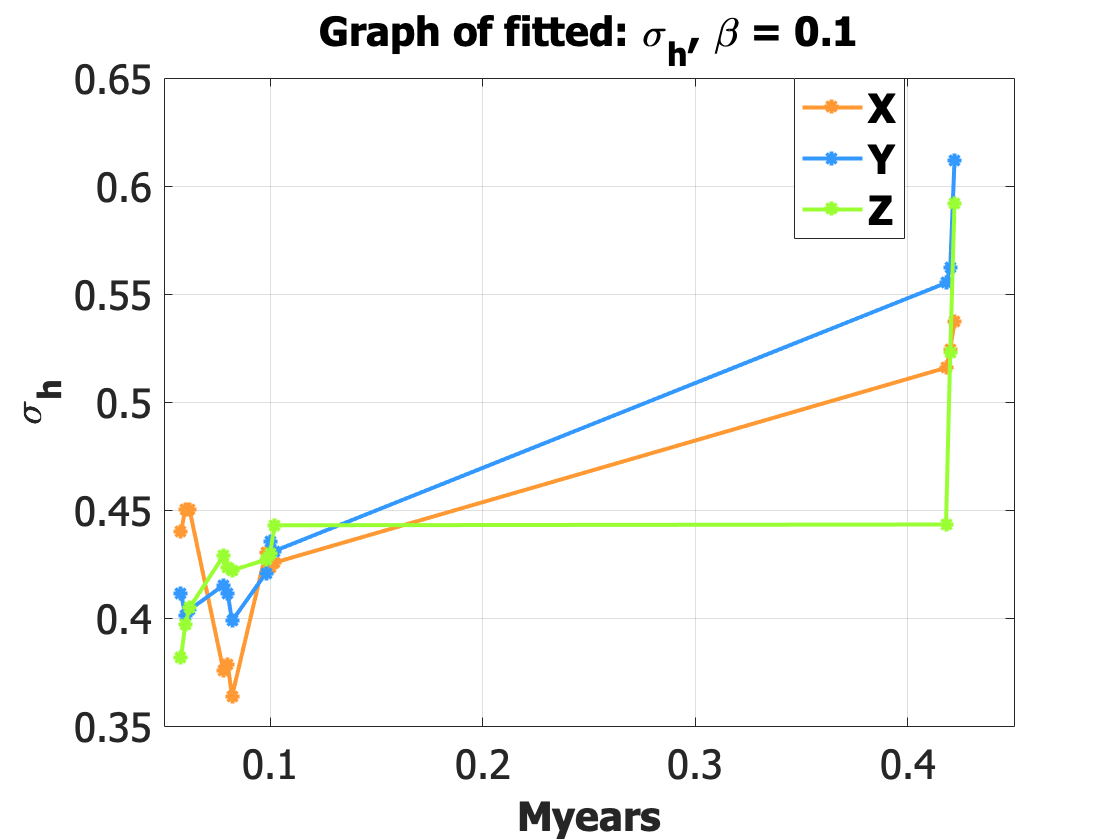}
\caption{Variation in time of the fitted $\sigma_{\bf h}$ of
  $\log$-normal singularity spectrum
  (equation~\ref{lognormalspectrum}) for the simulation outputs
  described in Sect.~\ref{simu}. Value of the magnetic pressure
  $\beta = 0.1$,\label{variationsigmasdib}}
\end{figure}

Figure~\ref{parabolicspectrum} shows a typical result of a
$\log$-normal fit applied on one projection of the MHD simulation data
presented in Sect.~\ref{simu} (by projection we mean here in one of
the three spatial directions -X, Y or Z- since the data is in the form
of a 3D cubic volume): the fit is very good, in all directions, and
this is true for all available simulation outputs. Consequently, we
can say that the MHD simulation outputs are $\log$-normal processes,
and it is then possible to see the variation in time of the width of
the $\log$-normal fitted singularity spectrum (the coefficient
$\sigma_{\bf h}$ of
eq.~\ref{lognormalspectrum}). Figure~\ref{variationsigmasdib} displays
the graph of the dispersion of the fitted $\log$-normal process, or
$\sigma_{\bf h}$ of eq.~\ref{lognormalspectrum} as a function of time,
in MYr, for the value $\beta = 0.1$. The dispersion becomes wider
as gravitational effects become much more pronounced (more bound and
collapsing structures), as was already stated in \citet{Elia2018},
where the authors note that when gravity is present one gets a broader
spectrum, and this is also the case when turbulence is driven with
compressive modes. Compressive driving generates larger density
contrasts in the simulations volume and in the resulting 2D maps. This
effect mimics the presence of gravity. Accordingly, in the
$\log$-normal approximation, the dispersion $\sigma_{\bf h}$ can be
used as an indication of the gravitational effects.

\section{Discussion of results: Musca turbulent dynamics and implication on star formation processes}
 \label{discussion}
 
 \begin{figure*}[h]
\centering
\includegraphics[width=0.323\textwidth]{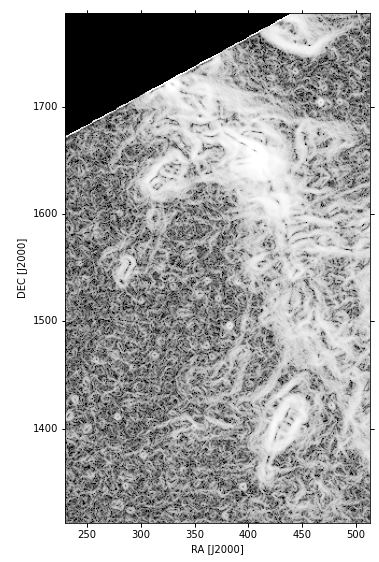}~\includegraphics[width=0.65\textwidth]{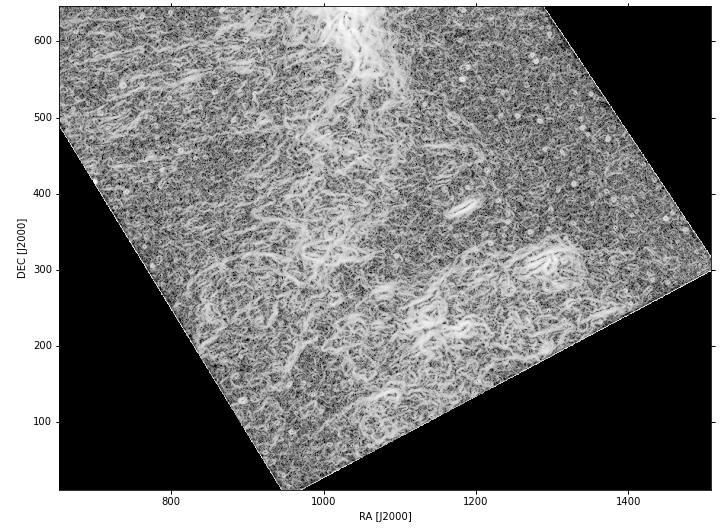}
\caption{Distribution of the singularity exponents in ROI1 (left) and in the southern region of Musca ROI5 (right). Both regions show complex linear structures, but of different sizes and organisation. \label{se-rois-filamentary}}
\end{figure*}

As discussed for instance in \citet{Bonne2020b}, several physical
processes are presumably at work to explain the formation of the Musca
filament in particular, as well as the global behavior of the ISM in
this molecular cloud. We have to consider the turbulence injected from
large scales, self-gravity presumably acting at small scales, magnetic
field pressures and tensions at all scales, and possibly some local
effects such as shocks. Our long term goal is to seek for statistical
signatures of these different processes thanks to the multifractal
analysis, introduced in this paper, and profiting from the
unprecedented depth and richness of {\sl Herschel} maps. Facing the
numerous difficulties in deriving such clean statistical probes, in
investigating the possible different signature of processes and the
inherently complex situations of the real data, the present article
should be seen as a first step towards this long term ambitious
goal. The following discussion thus only provides hints for new
directions which will need confirmation and further interpretations thanks to extensive studies
on simulations and on a large number of star-forming regions.

The fact that we clearly found a multiplicative cascade with a
significant and large inertial range from at least 0.05 to 0.65$\,$pc
is an important step to support the turbulence interpretation of the
dust emission from {\sl Herschel} towards a weak column density cloud surrounding the low-mass Musca filament.
This result points to a dominance of turbulent motions, and confirms
that this turbulent behavior originates from at least pc
scales. 

With the help of this new microcanonical approach which does not require multiple realizations (single map), we could then derive for the first time
a precise enough singularity spectrum of the observed flow to characterize its turbulent
behavior. We verified that we have a good
invariance of the singularity spectrum to the scale which was not
obtained with box counting method for instance. This
is highly important as we believe this is the missing step to be able
to derive the true, relatively precise singularity spectra of our
datasets (real data and simulations) and to quantify some differences
between regions and between observations and simulations to progress on the understanding of the physics leading to cloud and filament formation.
 
The obtained singularity spectrum confirms that the ISM towards Musca and its surroundings 
has a clear multifractal structure. It implies the existence of intermittency
and non-gaussian behavior which are believed to be critical
to explain the formation of the densest and star forming
structures. Moreover, while the so far derived
singularity spectra were often found to have a $\log$--normal behavior in previous works, we here clearly
obtain a global singularity spectrum which strongly deviates from
$\log$-normality. Deviation from $\log$-normality here is apparent as the singularity spectra are clearly non-symmetric. Among the many multifractal processes having a non-symmetric spectrum,  we discussed here the simplest case which is the $\log$-Poisson behavior which has been extensively discussed in the literature on turbulence. According to
\cite{Gledzer1996} for instance a $\log$-Poisson behavior is expected for an
energy cascading model of intermittency involving rare localized
regions of both large and/or weak energy dissipation (dynamical
intermittency) while a $\log$-normal behavior is obtained for
intermittency arising from widespread regions with nearly equal
dissipation rates (space intermittency). The existence of such geometric models for the $\log$-Poisson spectrum is interesting in astronomy, although further analysis must be done in order to relate the observed statistics with a particular type of process. Indeed, in all the ROIs chosen in the Musca map, and even in the ROI1, we can see filamentary and linear structures of various sizes and orientations on the map of singularity exponents, Consequently, it could be the distribution of these structures which affects the $\log$-normality of a singularity spectrum. See Fig.~\ref{se-rois-filamentary}. Interestingly enough according to~\citet{She1990226} the most intermittent structures in incompressible purely hydrodynamical turbulence have to be filamentary to be stable. In the case of compressible and magnetised turbulence, we can only expect that these filaments are more stable.

We therefore propose that the
ISM associated with the Musca filament region shows a turbulent
behavior which is better reproduced by a {\it dynamical intermittency}
than by a {\it space intermittency} with localized enhanced
dissipation locations. From the global view discussed for instance in
\cite{Bonne2020b} and from the realistic view that most intermittent regions have to be filamentary~\citep{She1990226,PhysRevLett.72.336}  we could then speculate that these localized
enhanced dissipations are associated with the formation of the Musca
filament and with the striations in the surroundings, and could be related to the efficient guiding and focusing
by the magnetic field (reducing the space dissipation) towards a
region of strong accumulation of matter where dissipation could be
enhanced in accretion events. Local (accretion) shocks, providing strong local turbulence dissipation
may explain and complement our understanding of the particular
statistics ($\log$-Poisson type) of the large singularities of the map
(large ${\bf h}$ values, i.e. the locations of large local gradients
as imaged in white in the singularity map of
Fig.~\ref{expsmusca}). The fact that we find the most non-symmetric (hence, possibly $\log$-Poisson)
singularity spectra towards the ROIs associated with the filaments of
Musca reinforces this interpretation that the dynamical intermittency
is associated with filament formation.

Interestingly enough, using the same microcanonical analysis we could
not identify in simulated data a similar $\log$-Poisson
behavior. From simulated datasets we always obtain $\log$-normal
spectra. This behavior could perhaps reflect the fact that these
simulations are run with a continuous injection of well behaved
turbulence at large scales. The fact that self gravity is included
does not seem to affect the singularity spectra. We clearly need to
continue to investigate what is missing in the simulations to properly
reproduce the observed data, but we can note that we have found a
statistical tool sensitive enough to make the difference between
simulations and real datasets.
 
 \section{Conclusion}
 \label{conclusion}
 In this work we make use of a new computable
 multifractal formalism in a microcanonical
 formulation, based on ideas from predictability in complex systems, and we apply it to analyze the complex turbulent
 structure of the ISM for the case of a Musca {\sl Herschel}
 observational map. We confirm that to be fully effective, the use of the multifractal
 formalism must be operated with important checks on an observational
 map: determination of the inertial range for which scaling laws
 apply, scale invariance must be checked by computing singularity
 spectra at different scales and checking their coincidence,
 background noise elimination must be operated with particular care in
 order to preserve low dimensional weak coherent structures.

 Our aim is to present a self-contained study with sufficient detail to help reproduce the experiments. Very importantly, we could make the check of
 scaling laws and the determination of the inertial range, such as usually done in canonical formulations, thanks to moments of the chosen partition functions achieved using a 2D 
structure function methodology.
The microcanonical formulation is based on the direct computation of the
 singularity exponents. The singularity spectra
 are derived, and we check the invariance of the resulting spectra
 with respect to scale. Background noise has the tendency to
 $\log$-normalize the signals, by making their spectra parabolic. We
 propose a $L^1$ denoising algorithm based on edge-aware filtering
 which preserves low-dimensional features as opposed to thresholding
 through wavelet decomposition methods. We use the theory of cumulants
 to determine the presence of a multiplicative cascade using
 $\log$-correlations.

 The results show a clear multiplicative cascade with a significant
 and large inertial range from at least 0.05 to 0.65$\,$pc (larger
 scales might be affected by the $\sim 4\,$ pc width of the observed map, and smaller scale by the 0.012 pc beam size)
 pointing to a dominance of the turbulence originating from scales
 probably larger than a parsec. We show that a precise study of the intermittency
 can be achieved by the methodology introduced in this work. For the
 time in ISM studies, our data and analysis challenge the
 $\log$-normality of the singularity spectrum of the turbulence. This
 is one of the major results of this work. It suggests that the
 turbulence associated with the Musca filament formation exhibits more
 a dynamical intermittency with localized, enhanced energy dissipation
 than a space intermittency. Some hints for different turbulent
 behavior in the region and for some missing physics in simulations
 shows that the work presented in this study is only a first step in
 our long term goal to seek for statistical signatures of different
 turbulent, physical processes at work in the ISM.

\begin{appendix} 
\section{Definition of the measure}
\label{definitionmeasure}
We keep the notations and conventions defined in Sect.~\ref{ss}. Let
us begin by considering the set ${\cal V}_n({\bf x})$ displayed on the
left of Fig.~\ref{voisin}. In that case each ${\cal V}_n({\bf x})$
is made of two points and we introduce the discrete measure noted
$\mu_n$ on the unit square by
 \begin{equation}
 \label{mu-n}
 \mu_n =  \displaystyle \left (\frac{1}{2^n} \right )^2\sum_{{\bf x} \in \Omega_n}  \sum_{{\bf y} \in {\cal V}_n({\bf x})} D({\bf x},{\bf y}) \delta_{{\bf x}}
 \end{equation}
 with $\delta_{{\bf x}}$: Dirac measure at ${{\bf x}}$. For each $n$,
 $\mu_n$ can be written as a sum of local measures
 \begin{equation}
 \label{mu-n2}
 \mu_n =  \displaystyle \left (\frac{1}{2^n} \right )^2\sum_{{\bf x} \in \Omega_n}  \mu_n^{\bf x}
\end{equation}
 with $ \mu_n^{\bf x} = \displaystyle \sum_{{\bf y} \in {\cal
     V}_n({\bf x})} D({\bf x},{\bf y}) \delta_{{\bf x}}$. We examine
 the behavior of the measures $\mu_n$ when $n \rightarrow +\infty$.
 Let us suppose first that $s$ is differentiable. Then
 \begin{equation}
 \label{grad1}
| s({\bf x})- s( {\bf y}) | = | \langle \nabla s({\bf x}) \, | \, ({\bf x} - {\bf y}) \rangle + \| {\bf x} - {\bf y}\| \varepsilon({\bf x} - {\bf y}) |
\end{equation}
so that
 \begin{equation}
 \label{grad2}
\displaystyle  \frac{| s({\bf x} )- s({\bf y}) | }{ \| {\bf x} - {\bf y} \| } = | \langle \nabla s({\bf x}) \, | \, \frac{ ({\bf x} - {\bf y}) }{ \| {\bf x} - {\bf y}\| }\rangle + \varepsilon({\bf x} - {\bf y}) |
\end{equation}
Consequently, if $f$ is any continuous bounded function then 
 \begin{equation}
 \label{vague-convergence1}
\int f\, \mbox{d} \mu_n^{\bf x} \rightarrow f({\bf x}) ( | \nabla s({\bf x})_1 | +  | \nabla s({\bf x})_2 |)
\end{equation}
when $n \rightarrow +\infty$ with $\nabla s({\bf x}) = (\nabla s({\bf x})_1, \nabla s({\bf x})_2)$ because $ \frac{ ({\bf x} - {\bf y} ) }{ \| {\bf x} - {\bf y} \| }$ is a unitary vector and the types of 
neighboring points considered in Fig.~\ref{voisin}. Hence, when $n \rightarrow +\infty$ we have that 
$ \int f\, \mbox{d} \mu_n^{\bf x} \rightarrow f({\bf x}) \| \nabla s({\bf x}) \|_{L_1}$.  If we want to obtain $f({\bf x}) \| \nabla s({\bf x}) \|_{L_2}$, then one must start from the measure:
 \begin{equation}
 \label{mu-n-2}
 \mu_n =  \displaystyle \left (\frac{1}{2^n} \right )^2\sum_{{\bf x} \in \Omega_n}  \sqrt{\sum_{{\bf y} \in {\cal V}_n({\bf x})} D({\bf x},{\bf y})^2} \delta_{{\bf x}}
\end{equation}
If one uses the second set of ${\cal V}({\bf x})$ displayed on the right of Fig.~\ref{voisin}, then one must use
 \begin{equation}
 \label{mu-n-3}
 \mu_n =  \displaystyle \left (\frac{1}{2^n} \right )^2\sum_{{\bf x} \in \Omega_n}  \displaystyle \frac{1}{2} \left (\sum_{{\bf y} \in {\cal V}_n({\bf x})} D({\bf x},{\bf y}) \right )\delta_{{\bf x}}
\end{equation}
to ensure convergence towards $f({\bf x}) \| \nabla s({\bf x}) \|_{L_1}$ for $ \int f\, \mbox{d} \mu_n^{\bf x}$. Then, if $f$ is any continuous bounded function as above then, when $n 
\rightarrow +\infty$, we get
$$
\int f \, \mbox{d}\mu_n \rightarrow \int f \| \nabla s \| \, \mbox{d}\lambda
$$
with $\lambda = $ Lebesgue measure on the unit square because the
points in each lattice set $\Omega_n$ are regularly spaced with
distance $2^{-n}$ and as a classical theorem from Lebesgue
integration. Consequently, the measures $\mu_n$ converge vaguely
towards the measure associated to the gradient's norm when $s$ is
differentiable.

\paragraph*{}When $s$ is any signal, differentiable or not, let us consider ${\bf x} \in \Omega = \displaystyle \bigcup_{n \geq 0} \Omega_n$ so that ${\bf x} \in \Omega_n$ for a certain 
$n$; then ${\bf x} \in \Omega_k$ for all $k \geq n$. Let ${\cal B}({\bf x}, 2^{1-n})$ with $n \geq 1$ a ball centred at $ {\bf x}$ at resolution $\displaystyle \frac{1}{2^{n-1}}$. Then
\begin{equation}
\label{scalemeasure}
\begin{array}{lll}
\mu({\cal B}({\bf x}, 2^{1-n})) & = & \displaystyle \lim_{k \rightarrow +\infty} \mu_k({\cal B}({\bf x}, 2^{1-n})) \\
~&  = & \displaystyle \lim_{k \rightarrow +\infty}  \left ( \frac{1}{2^k} \right )^2 \left (  \sum_{{\bf z} \in \Omega_k \cap {\cal B}({\bf x}, 2^{1-n})} \sum_{{\bf y}_k \in {\cal V}_k({\bf z}), \,k > n} 
D({\bf y}_k , {\bf z})\right ) 
\end{array}
\end{equation}
When $n \rightarrow +\infty$ the last term is a sum of discrete differences taken in the balls ${\cal B}({\bf x}, 2^{1-n})$. Consequently we make the following {\it scaling hypothesis}: 
\begin{equation}
\label{scalinghypothesis}
\mu({\cal B}({\bf x}, 2^{1-n})) \sim \displaystyle A \left ( \frac{1}{2^n} \right)^{{\bf h}(\bf x)} ~~\mbox{as } n \rightarrow +\infty
\end{equation}
which expresses the scaling of the measure as $n \rightarrow
+\infty$. This defines a scalar field of singularities on the dense
subset $\Omega = \displaystyle \bigcup_{n \geq 0} \Omega_n$ of the
unit square; consequently ${\bf h}({\bf x})$ possesses a unique
continuous continuation over the whole unit square as long as the
limit $\displaystyle \lim_{{\bf x} \rightarrow {\bf y}, {\bf x} \in
  \Omega} {\bf h}({\bf x})$ exists for all ${\bf y}$ in the unit
square ([\citeads{dieudonne1969}] prop. (3.15.5)), which is a
reasonable hypothesis also assumed.

To count the sites ${\bf x}$ with same scaling behavior, one introduces the density of the exponents at resolution $ \displaystyle \frac{1}{2^n}$
 \begin{equation}
 \label{historesolm}
 \rho_n ({\bf h}) = \sum_{{\bf x} \in \Omega_n}\delta \left ( \frac{\log \mu({\cal B}({\bf x}, 2^{1-n})) }{\log 2^n}  - {\bf h} \right ).
 \end{equation}
In the limit $ n \rightarrow +\infty$ one has 
 \begin{equation}
 \label{historesolinf}
 \rho_n ({\bf h}) \sim c_n({{\bf h}}) \sqrt{n\log 2 } \,n^{\log (2)D({\bf h})} \\
 \end{equation}
 where the mapping ${\bf h} \mapsto D({\bf h})$ is the singularity
 spectrum of the measure defined by the gradient's norm density
 (i.e. $\mbox{d}\mu = \| \nabla s \| \, \mbox{d} {\bf x}$) (when the
 signal is differentiable).

 Let ${\cal F}_{\bf h} = \{ ~{\bf x} ~| ~{\bf h}({\bf x}) = {\bf h}
 ~\}$. There is a general physical argument showing that, for
 sufficient complex signals $s$ such as the ones in Fully Developed
 Turbulence, the sets ${\cal F}_{\bf h} $ are dense; indeed, if there
 were an ${\bf h}$ such that the corresponding ${\cal F}_{\bf h}$ was
 not dense, then one could find an open ball in the signal's domain
 such that no point in that ball has a singularity exponent ${\bf h}$;
 in the differentiable case, this means a particular gradient's norm
 is forbidden in that ball.

\section{Singularity exponents and predictability} 
\label{h-upm}
In this appendix we describe a different methodology to evaluate the
singularity exponents ${\bf h}({\bf x})$ which leads to a much better
evaluation of the singularity spectrum ${\bf h} \mapsto D({\bf h})$ as
we will see in the experiments. It is based on the notion of
predictability in complex systems. In the theory of dynamical systems,
there is a simple notion of predictability which consists in
evaluating the so-called {\it predictability time}: for an initial
small perturbation of size $\delta$, and an accepted tolerance error
$\Delta$, the predictability time is the quantity
\begin{equation}
\label{predictabilitytime}
T_p  = \displaystyle \frac{1}{\lambda_d} \log \left ( \frac{\Delta}{\delta}\right  )
\end{equation}
with $\lambda_d$: leading Lyapunov exponent. In \citet{Aurell1997} 
this elementary notion is extended to the case of Fully Developed
Turbulence with a formula which involves the singularity spectrum
${\bf h} \mapsto D({\bf h})$. Consequently, the singularity exponents,
through the singularity spectrum, are related to predictability. As pointed 
out in \citet{PhysRevE.74.061110,Turiel2006,Turiel2008,doi:10.1080/00207160.2012.748895} 
it is therefore proposed to relate the computation of singularity
exponents to predictability, information content and
reconstructibility: the most unpredictable points are considered to
encode the information in the system, and they form a set from which
the whole system can be completely reconstructed. But it is necessary
to precise this notion of reconstructibility.

Given an observational map $s({\bf x})$ where ${\bf x}$ designates 2D
spatial coordinates, a subset ${\cal F}$ is said to reconctruct the
signal $s$ if we have
\begin{equation}
\label{reconstruction}
\nabla s({\bf x}) = \displaystyle {\cal G} \left ( \nabla_{\cal F} s({\bf x}) \right )
\end{equation}
where ${\cal G}$ is a reconstruction functional, and $ \nabla_{\cal F}
$ means the gradient operator restricted to the set ${\cal
  F}$. In \citet{Turiel2008} it is shown that, under the
assumption that ${\cal G}$ is deterministic, linear, translationally
invariant and isotropic, eq.~\ref{reconstruction} then implies the
existence of a reconstruction kernel ${\bf g}$, which leads to the
following reconstruction formula in Fourier space:
\begin{equation}
\label{reconstructionfourier}
\hat{s}({\bf k}) = \hat{{\bf g}}({\bf k}) \cdot \widehat{\nabla_{\cal F}  s}({\bf k})
\end{equation}
with ${\bf k}$: frequency vector. This formula has the consequence that if a set ${\cal F}$ satisfies the reconstruction eq.~\ref{reconstructionfourier}, then
\begin{equation}
\label{reconstructiondivergence}
\mbox{div} \displaystyle \left ( \nabla_{{\cal F}^{\mbox{c}}}s \right ) = 0.
\end{equation}
with ${\cal F}^{\mbox{c}}$: complementary set of ${\cal F} $. Since
the gradient and divergence operators are local, this means that the
decision whether a point ${\bf x}$ belongs or not to ${\cal F} $
should be done only locally. From these considerations, 
\citet{PhysRevE.74.061110,Turiel2006,Turiel2008,doi:10.1080/00207160.2012.748895}  
made the following assumption: \linebreak \linebreak {\it The set
  of most unpredictable points ${\cal F}_{\infty}$ is the one which
  gives a perfect reconstruction according to
  equation~\ref{reconstructionfourier}; the decision that a point
  ${\bf x} \in {\cal F}_{\infty}$ can be made locally around ${\bf x}$
  and the set ${\cal F}_{\infty}$ is identical to the set of points
  which have the lowest singularity exponents in the signal
  i.e. ${\cal F}_{\infty} = {\cal F}_{{\bf h}_{\infty} }= \{ {\bf x}
  ~|~{\bf h}({\bf x}) = {\bf h}_{\infty} \}$ with ${\bf h}_{\infty }$
  being the minimum of the singularity exponents.} \\ 

Under the previous assumption, the computation of the values of the
singularity exponents can be done at the lowest scale ${\bf r}_{0} =
2^{-n}$ of the acquisition signal instead of $\log$-regression through
scales according to the formula:
\begin{equation}
\label{hupm}
{\bf h}({\bf x}) = \displaystyle \frac{\log ({\cal H}(\mu_n, {\bf x}, {\bf r}_{0}))/\langle {\cal H}(\mu_n, {\bf x}, {\bf r}_{0}) \rangle}{\log {\bf r}_{0}}
\end{equation}
with $\cal H$ being a function of the input measure $\mu_n$ which
makes uses of local information around point ${\bf x}$ at scale ${\bf
  r}_{0}$. The term $\langle {\cal H}(\mu_n, {\bf x}, {\bf r}_{0})
\rangle$ is the spatial average of ${\cal H}$ over the whole
observational map which lessens the relative amplitudes of $\mu_n$'s
fluctuations at scale ${\bf r}_{0}$. $\cal H$ is defined as one of the
simplest and most generic ways of measuring the local
unpredictability: subtracting the signal value at a given point from
the value inferred from its neighbour points. These values must be
previously detrended to cancel any global offset influence. The result
$\cal H$ then measures local correlation. We now describe how ${\cal
  H}$ is computed.

\begin{figure}[h]
  \centering
  \includegraphics[width=0.40\textwidth]{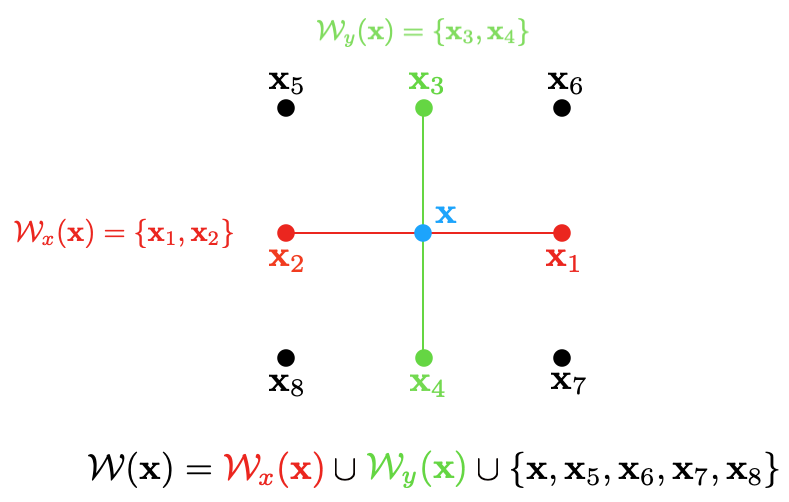}
\caption{The neighborhood system ${\cal W}({\bf x})$ considered at a
  point ${\bf x}$ for the computation of local correlation measure
  ${\cal H}(\mu, {\bf x}, {\bf r}_{0})$. It consists of a "cross" made
  of the horizontal and vertical sets ${\cal W}_x({\bf x})$, ${\cal
    W}_y({\bf x})$ with other neighbors points including ${\bf
    x}$. \label{upmcomputation}}
\end{figure}

We consider a point ${\bf x}$ at which we compute ${\bf h}({\bf x})$
and local neighborhood information around ${\bf x}$ at scale ${\bf
  r}_{0} = 2^{-n}$, ${\cal W}({\bf x})$ as shown in
Fig.~\ref{upmcomputation}.
The vector $(s({\bf x}), s({\bf x}_1),s({\bf x}_2) ,s({\bf x}_3)
,s({\bf x}_4) )$ is first properly detrended by considering the trend
$d = \frac{1}{3}(s({\bf x}) + s({\bf x}_1) + s({\bf x}_2) + s({\bf
  x}_3) + s({\bf x}_4))$ and generate from it the detrended vector
$(s({\bf x}) + d, s({\bf x}_1) -d,s({\bf x}_2)-d ,s({\bf x}_3)-d
,s({\bf x}_4) -d )$. Let us denote this detrended vector by $({
  p}_0({\bf x}),{p}_1({\bf x}),{p}_2({\bf x}),{p}_3({\bf
  x}),{p}_4({\bf x}))$. We define
\begin{equation}
\label{errorupm}
\begin{array}{lcl}
\varepsilon_x({\bf x}) &= &\alpha (p_2({\bf x}) - p_1({\bf x})) \\
\varepsilon_y({\bf x}) &= &\alpha (p_4({\bf x}) - p_3({\bf x})) \\
\end{array}
\end{equation}
with $\alpha > 0$. Note that the values of $\varepsilon_x({\bf x})$
and $\varepsilon_y({\bf x})$ are related to the "cross" neighboring
sets ${\cal W}_x({\bf x})$ and ${\cal W}_y({\bf x})$ shown in red and
green in Fig.~\ref{upmcomputation}. The local correlation measure is
then defined as:
\begin{equation}
 \label{defH}      
 {\cal H}(\mu_n, {\bf x}, {\bf r}_{0})= \displaystyle \left ( \left ( \varepsilon_x({\bf x})^2 + \varepsilon_y({\bf x})^2 \right ) \left ( \frac{{\cal A}({\bf x}, {\bf r}_0)}{ \displaystyle \mu_n({\cal W}({\bf x}))} \right ) \right )^{1/2} 
\end{equation}
with
$$
{\cal A}({\bf x}, {\bf r}_0) = \displaystyle \left | \varepsilon_x({\bf x}) \left (   \sum_{{\bf y} \in {\cal W}({\bf x})}  \varepsilon_x({\bf y} ) \right )  +    \varepsilon_y({\bf x}) \left (   \sum_{{\bf y} \in 
{\cal W}({\bf x})}  \varepsilon_y({\bf y} ) \right )   \right |.
$$ The value ${\bf h}({\bf x})$ is then computed with
equation~\ref{hupm}. Once the singularity exponents are determined,
they define the collection of sets ${\cal F}_{\bf h} = \{ ~{\bf x} ~|
~{\bf h}({\bf x}) = {\bf h} ~\}$ which are of particular importance in
the description of a turbulent system \citep{frisch1995}. In 
\citet{Turiel2008} an argument for the case
of log-Poisson processes is derived, which generalizes the work presented
in \citet{PhysRevLett.72.336}, and which describes the
multiplicative cascade from the geometrical organization of the
level-sets ${\cal F}_{\bf h}$ defined by the local singularity
exponents. However a general justification valid for a large class of
physical processes is still being awaited.

\section{Cumulant Analysis Method}
  \label{cumulant}
In this appendix we go back to the canonical description of
multifractality, considering an advanced and powerful computational
approach based on cumulants, which was introduced
in \citet{Delour2001}; it is used in this work as it provides a
criterion to determine the existence of a multiplicative
cascade \citep{PhysRevLett.80.708,10.2307/55171} and the existence
of long-range correlations. The reader is referred
to \citet{Venugopal2006b} for a detailed and accurate
description. If $\mu$ is a probability measure on the line, its
characteristic function is $f_{\mu}(z) = {\E}_{\mu}(e^{iz})$ which can
be expanded in a power series involving $\mu$'s moments $M_n$:
$f_{\mu}(z) = \displaystyle \sum_{n=0}^{\infty} M_n \frac{(iz)^n}{n!}$
with $M_n = \int x^n \mbox{d}{\mu}(x)$. The cumulant generating
function of $\mu$, $g_{\mu}(z)$, is the $\log$ of $\mu$'s
characteristic function: $g_{\mu}(z) = \log f_{\mu}(z)$. Since $M_0 =
1$, $g_{\mu}(z) =  \displaystyle \log  \left (  1 + \left ( \sum_{n=1}^{\infty}
M_n \frac{(iz)^n}{n!} \right ) \right  )$ which can in turn be expanded into a
power series $g_{\mu}(z) = \displaystyle \sum_{n=0}^{\infty} C_n
\frac{(iz)^n}{n!}$; the $C_n$ are called the cumulants of $\mu$. The first cumulants are given by the following relations:
 \begin{equation}
 \label{calcul-cumulants1}
 \begin{array}{lcl}
 C_1 &= & M_1\\
 C_2 & = &M_2 - M_1^2 \\
  C_3& = & M_3 - 3M_2M_1 + 2M_1^3
 \end{array}
 \end{equation}
 and the general recurrence relation is:
 \begin{equation}
\label{eqn:cumulantrec}
C_n = M_n - \displaystyle \sum_{l=1}^{n-1} \left( \begin{array}{c} n-1 \\ l-1 \end{array}\right)C_lM_{n-l}.
\end{equation}
 The most interesting part of the theory is the relation between
 the $C_n$ and the multifractal spectrum when $\mu$ is a multifractal
 measure. We collect the main results here. \\ As in
 Sect.~\ref{multifractal} let $s$ be a given signal, $\psi$ an
 analyzing wavelet, ${\cal T}_{\psi}({\bf x},{\bf r})$ the
 wavelet-projected signal evaluated at position ${\bf x}$ and scale
 ${\bf r}$, $Z(q,{\bf r})$ a suitably chosen $q$th order partition
 function built out of ${\cal T}_{\psi}$. Let us consider, as the
 analyzing measure $\mu$, the one having density $\log | {\cal
   T}_{\psi}({\bf x},{\bf r}) |$. Let $C_n({\bf r})$ be its
 cumulants. Then the evaluation of the slope of $C_n({\bf r})$ vs
 $\log ({\bf r})$ gives, when ${\bf r} \rightarrow 0$, coefficients:
 \begin{equation}
 \label{calcul-cumulants2}
 C_n \sim (-1)^{n+1}c_n\log({\bf r}) 
 \end{equation}
in such a way that $\tau (q)$ can be retriewed as $ \tau(q) =
-c_0 + c_1q - c_2q^2/2! + c_3q^3/3! + \cdots$. For a 1D $\log$-normal process, $c_n = 0$ for $n > 2$ and one has
 \begin{equation}
 \label{lspectrelognormalcumulants}
 D({\bf h}) = c_0 - \displaystyle \frac{({\bf h}-c_1)^2}{2c_2}
\end{equation}
i.e. we get a parabolic spectrum as stated before, and comparing with
the previous general expression for the singularity spectrum of a
$\log$-normal process (eq.~\ref{lognormalspectrum}
Sect.~\ref{multifractal}), we find in this case: $c_0 = 1$,
(dimension of the support of the measure), $c_1 = {\bf h}_m$ (average
value of the singularity exponents) and $c_2 = \sigma_{{\bf h}}^2$
(variance).

\section{Two-point magnitude statistical analysis and the multiplicative cascade} 
\label{twopoint}

Keeping up with the notation of the last section, we define the
two-point correlation function for a given scale ${\bf r}$ and spatial
interval $\Delta {\bf x}$:
 \begin{equation}
 \label{deflogcorr}
 \begin{array}{lcl}
 {\cal C}({\bf r}, \Delta {\bf x}) & = &\langle \, ( \log | {\cal T}_{\psi}({\bf x},{\bf r}) | - \langle \log | {\cal T}_{\psi}({\bf x},{\bf r}) | \rangle \, )\cdot  \\
 ~&~& ( \log | {\cal T}_{\psi}({\bf x }+ \Delta {\bf x},{\bf r}) | - \langle \log | {\cal T}_{\psi}({\bf x},{\bf r}) | \rangle \, ) \, \rangle.
 \end{array}
 \end{equation}
 It is shown that if ${\cal C}({\bf r}, \Delta {\bf x}) $ is
 logarithmic in $\Delta {\bf x}$ and independent of scale ${\bf r}$
 provided $\Delta{\bf x} > {\bf r}$ i.e. ${\cal C}({\bf r}, \Delta
 {\bf x}) \sim \log \Delta {\bf x}$ then long-range dependence exists
 in the system \citep{PhysRevLett.80.708,Venugopal2006b}. If there
 is a multiplicative cascade, then
 \begin{equation}
 \label{verif-cascade}
 {\cal C}({\bf r}, \Delta {\bf x})  \sim -c_2 \log \Delta {\bf x}. 
 \end{equation}
We note that some random processes can be self-similar
without displaying a multiplicative
cascade \citep{Arneodo1999,10.2307/55171,Venugopal2006b}. This
feature is often neglected in multifractal analysis of signals in
astronomy, although it is of primary importance if one wants to prove
the existence of a multiplicative cascade in the ISM.
\end{appendix}
\begin{acknowledgements}
  This work is supported by the {\it GENESIS} project ( \emph{GENeration and Evolution of Structure in the ISm}), via the french ANR and the german DFG through grant numbers ANR-16-CE92-0035-01 and DFG1591/2-1. 
\end{acknowledgements}
  \bibliographystyle{aa} 
  \bibliography{citations1} 
\end{document}